\documentclass[12pt,preprint]{aastex}
\slugcomment{To be submitted to {\it The Astrophysical Journal}}
\shortauthors{Kirkpatrick}
\shorttitle{WISE Brown Dwarfs}

\begin{document}

\title{The First Hundred Brown Dwarfs Discovered by the Wide-field Infrared Survey
Explorer (WISE)}

\author{J.\ Davy Kirkpatrick\altaffilmark{a},
Michael C.\ Cushing\altaffilmark{b},
Christopher R.\ Gelino\altaffilmark{a},
Roger L.\ Griffith\altaffilmark{a},
Michael F.\ Skrutskie\altaffilmark{d},
Kenneth A.\ Marsh\altaffilmark{a},
Edward L.\ Wright\altaffilmark{c},
Amanda K.\ Mainzer\altaffilmark{b},
Peter R.\ Eisenhardt\altaffilmark{b},
Ian S.\ McLean\altaffilmark{c},
Maggie A.\ Thompson\altaffilmark{j},
James M.\ Bauer\altaffilmark{b},
Dominic J.\ Benford\altaffilmark{l},
Carrie R.\ Bridge\altaffilmark{k},
Sean E.\ Lake\altaffilmark{c},
Sara M.\ Petty\altaffilmark{c},
S.\ Adam Stanford\altaffilmark{m},
Chao-Wei Tsai\altaffilmark{a},
Vanessa Bailey\altaffilmark{t},
Charles A.\ Beichman\altaffilmark{a},
John J.\ Bochanski\altaffilmark{g,u},
Adam J.\ Burgasser\altaffilmark{h},
Peter L.\ Capak\altaffilmark{n},
Kelle L.\ Cruz\altaffilmark{i},
Philip M.\ Hinz\altaffilmark{t},
Jeyhan S.\ Kartaltepe\altaffilmark{o},
Russell P.\ Knox\altaffilmark{t},
Swarnima Manohar\altaffilmark{p},
Daniel Masters\altaffilmark{q},
Maria Morales-Calder{\'o}n\altaffilmark{n},
Lisa A.\ Prato\altaffilmark{e},
Timothy J.\ Rodigas\altaffilmark{t},
Mara Salvato\altaffilmark{r},
Steven D.\ Schurr\altaffilmark{s},
Nicholas Z.\ Scoville\altaffilmark{p},
Robert A.\ Simcoe\altaffilmark{g},
Karl R.\ Stapelfeldt\altaffilmark{b},
Daniel Stern\altaffilmark{b},
Nathan D.\ Stock\altaffilmark{t},
William D.\ Vacca\altaffilmark{f}
}

\altaffiltext{a}{Infrared Processing and Analysis Center, MS 100-22, California 
    Institute of Technology, Pasadena, CA 91125; davy@ipac.caltech.edu}
\altaffiltext{b}{NASA Jet Propulsion Laboratory, 4800 Oak Grove Drive, Pasadena, CA 91109}
\altaffiltext{c}{Department of Physics and Astronomy, UCLA, Los Angeles, CA 90095-1547}
\altaffiltext{d}{Department of Astronomy, University of Virginia, Charlottesville, VA, 22904}
\altaffiltext{e}{Lowell Observatory, 1400 West Mars Hill Road, Flagstaff, AZ, 86001}
\altaffiltext{f}{SOFIA-USRA, NASA Ames Research Center, Moffett Field, CA 94035}
\altaffiltext{g}{Massachusetts Institute of Technology, 77 Massachusetts Avenue, Building 37, Cambridge, MA 02139}
\altaffiltext{h}{Department of Physics, University of California, San Diego, CA 92093}
\altaffiltext{i}{Department of Physics and Astronomy, Hunter College, New York, NY 10065}
\altaffiltext{j}{The Potomac School, 1301 Potomac School Road, McLean, VA 22101}
\altaffiltext{k}{Division of Physics, Mathematics, and Astronomy, MS 220-6, California Institute of Technology, 
  Pasadena, CA 91125}
\altaffiltext{l}{Infrared Astrophysics Branch, NASA-Goddard Space Flight Center, 8800 Greenbelt Road, 
  Greenbelt, MD 20771}
\altaffiltext{m}{University of California, Davis, CA 95616}
\altaffiltext{n}{Spitzer Science Center, California Institute of Technology, Pasadena, CA 91125}
\altaffiltext{o}{National Optical Astronomy Observatory, 950 North Cherry Avenue, Tucson, AZ 85719}
\altaffiltext{p}{California Institute of Technology, MC 104-24, Pasadena, CA 91125}
\altaffiltext{q}{University of California, Riverside}
\altaffiltext{r}{Max-Planck-Institut f\"ur Plasmaphysik, Boltzmanstrasse 2, D-85741 Garching, Germany}
\altaffiltext{s}{Planck Science Center, MS 220-6, California Institute of Technology, Pasadena, CA 91125}
\altaffiltext{t}{Steward Observatory, The University of Arizona, 933 N. Cherry Ave., Tucson, AZ 85721}
\altaffiltext{u}{Astronomy and Astrophysics Department, Pennsylvania State University, 525 Davey Laboratory, University Park, PA 16802}

\begin{abstract}

We present ground-based spectroscopic verification of six Y dwarfs (see also Cushing et al.), 
eighty-nine T dwarfs, eight L dwarfs, and
one M dwarf identified by the Wide-field Infrared Survey Explorer (WISE). Eighty of these are
cold brown dwarfs with spectral types $\ge$T6, six of which have been announced earlier 
in Mainzer et al.\ and Burgasser et al. We present color-color and color-type
diagrams showing the locus of M, L, T, and Y dwarfs in WISE color space. Near-infrared and, in a 
few cases, optical spectra are presented for these discoveries. Near-infrared classifications as late as early Y are 
presented and objects with peculiar spectra are discussed. Using these new discoveries, we are also able
to extend the optical T dwarf classification scheme from T8 to T9. After deriving an absolute WISE 4.6 $\mu$m 
(W2) magnitude vs.\ spectral type relation, we estimate spectrophotometric distances to our discoveries.
We also use available astrometric measurements to provide preliminary trigonometric parallaxes to four
our discoveries, which have types of L9 pec (red), T8, T9, and Y0; all of these lie within 10 pc of 
the Sun. The Y0 dwarf, WISE 1541$-$2250, is the closest at $2.8^{+1.3}_{-0.6}$ pc; if this 2.8 pc value persists
after continued monitoring, WISE 1541$-$2250 will become the seventh closest stellar system to the Sun.
Another ten objects, with types between T6 and $>$Y0,
have spectrophotometric distance estimates also placing them within 10 pc. The closest of these, 
the T6 dwarf WISE 1506+7027, is believed to fall at a distance of $\sim$4.9 pc.
WISE multi-epoch positions supplemented with positional info primarily from {\it Spitzer}/IRAC
allow us to calculate proper motions and tangential velocities for roughly one half of the new
discoveries. This work represents the first step 
by WISE to complete a full-sky, volume-limited census of late-T and
Y dwarfs. Using early results from this census, we present preliminary, lower limits to the space density of these objects
and discuss constraints on both the functional form of the mass function and the low-mass limit
of star formation. 

\end{abstract}

\section{Introduction}

Brown dwarfs, objects whose central temperatures never reach the critical threshold for 
stable thermonuclear burning (\citealt{kumar1963}; \citealt{hayashi1963}), are the lowest mass 
products of star formation. Hundreds of examples are now known\footnote{See DwarfArchives.org.}, enabling the 
study of brown dwarfs as a population in their own right (\citealt{kirkpatrick2005}).
The study of brown dwarfs helps to constrain 
mechanisms for small-object formation, which include turbulent fragmentation (\citealt{padoan2005};
\citealt{boyd2005}), magnetic field confinement (\citealt{boss2004}), stellar embryo 
ejection through dynamical interactions (\citealt{reipurth2001}; \citealt{bate2005}), 
and photo-evaporation of embryos by nearby hot stars (\citealt{whitworth2004}).

Brown dwarfs also represent a ``fossilized'' record of star formation throughout the 
Galaxy's history because their mass is never ejected 
back into the interstellar medium. They therefore preserve information on metallicity enrichment 
over the lifetime of the Milky Way (\citealt{burgasser2008}). Solitary brown dwarfs have also 
proven to be excellent calibrators of the atmospheric models on which our inference 
of the properties of giant exoplanets depends (\citealt{fortney2005}; \citealt{barman2005}; 
\citealt{marois2008}). Their effective temperatures are similar to those of the exoplanets
discovered thus far but their spectra lack the complication of irradiation from a host star.

Despite uncovering hundreds of brown dwarfs, 
previous surveys have allowed us to identify only the warmest 
examples. The latest object currently having a measured spectrum is UGPS J072227.51$-$054031.2,
whose effective temperature is estimated to be 520$\pm$40 K (\citealt{lucas2010}; Bochanski et al., submitted,
find $T_{eff}$ = 500-600K) and whose
spectrum is used as the near-infrared T9 spectral standard (Cushing et al., accepted)\footnote{Previously
published objects with spectral types $\ge$T8.5 have been reclassified now that the end of the
T dwarf sequence and beginning of the Y dwarf sequence has been defined (Cushing et al., accepted).}.
Two other objects -- WD 0806$-$661B (also known as GJ 3483B; \citealt{luhman2011}) and CFBDSIR J145829+101343B (\citealt{liu2011}) 
-- are probably even colder and later in type than UGPS J072227.51$-$054031.2, 
although both currently lack spectroscopic confirmation.  Both of these objects underscore
the fact that the coldest brown dwarfs are extremely faint even at near-infrared wavelengths
where ground-based spectroscopy has its best chance of characterizing the spectra. WD 0806$-$661B,
a common proper motion companion to a white dwarf with a measured distance of 19.2$\pm$0.6 pc,
has yet to be detected in ground-based imaging observations down to J=23.9 mag (Luhman et al., submitted; 
see also \citealt{rodriguez2011}).  CFBDSIR J145829+101343B is the secondary in a system 
with a composite spectral type of T9 and a measured distance of 23.1$\pm$2.4 pc (\citealt{liu2011}).
A combination of its close proximity (0.11 arcsec) to the primary along with a faint magnitude
(J=21.66$\pm$0.34) make the acquisition of a spectrum challenging. Of the two,  
WD 0806$-$661B is less luminous at J band and presumably intrinisically fainter bolometrically 
(see also \citealt{wright2011}). Finding even closer and brighter examples of cold brown dwarfs
will be necessary to maximize our chances of best characterizing them.

Canvassing the immediate Solar Neighborhood for such cold objects is one of the goals of the
all-sky mission performed by the Wide-field Infrared Survey Explorer (\citealt{wright2010}).
Discoveries will directly measure the low-mass cutoff of star formation (Figure
12 of \citealt{burgasser2004}) and provide even colder fiducial atmospheres for 
modeling cold exoplanets and understanding the gas giants of our own Solar System. The
discovery of cold objects raises the question of whether a new spectral class, dubbed ``Y''
(\citealt{kirkpatrick2000}; \citealt{kirkpatrick2008}; see also \citealt{kirkpatrick1999}), 
will be needed beyond the T class. In this paper we present an overview of the our first $\sim$100 
WISE brown dwarf discoveries and show that objects colder than those previously known, including 
Y-class brown dwarfs, are being uncovered.

\section{Brown Dwarf Selection}

WISE is an Earth-orbiting NASA mission that surveyed the entire sky simultaneously at wavelengths of 3.4, 4.6, 
12, and 22 $\mu$m, hereafter referred to as bands W1, W2, W3, and W4, respectively. 
As shown in Figures 6, 7, and 13 of \cite{wright2010} as well as
Figure 2 of \cite{mainzer2011},
the W1 and W2 bands were specifically designed to probe the deep, 
3.3 $\mu$m CH$_4$ absorption band in brown dwarfs and the region relatively free of opacity 
at $\sim$4.6 $\mu$m.  Since the peak of the Planck function at low temperatures is in 
the mid-infrared, a large amount of flux emerges in the 4.6 $\mu$m window, and this makes the W1-W2 
colors of cool brown dwarfs extremely red (see \S2.1).  Such red colors, 
which are almost unique among astronomical sources, make the identification of cool brown 
dwarfs much easier.

WISE launched on 2009 Dec 14 and, 
after an in-orbit checkout, began surveying the sky on 
2010 Jan 14. Its Sun-synchronous polar orbit around the Earth meant that each location along
the ecliptic was observed a minimum of eight times, with larger numbers of re-visits
occurring at locations nearer the ecliptic poles.
WISE completed it first full pass of the sky on 2010 Jul 17 and its second pass on 
2011 Jan 09. During this second pass, the outer, secondary tank depleted its cryogen on 2010 Aug 05, 
rendering the W4 band unusable, and 
the inner, primary tank depleted its cryogen on 2010 Sep 30, rendering the W3 band unusable. 
Thus, this second full sky pass is partly missing bands W3 and W4. Fortunately, the bands most crucial for brown 
dwarf selection -- W1 and W2 --
were little affected by this cryogen exhaustion. WISE continued to collect data on a third, incomplete 
sky pass in bands W1 and W2 until data acquisition was halted on 2011 Jan 31.

Preliminary processing of the data, including single-frame and coadded 
images and photometrically and astrometrically characterized detections, has been used
to search for cold brown dwarf candidates, as described in detail below. This is the 
same version of the pipeline software that produced the WISE Preliminary Data Release, 
details of which can be found in the Explanatory 
Supplement\footnote{See \url{http://wise2.ipac.caltech.edu/docs/release/prelim/expsup/}.}. 
For a more 
detailed description of the WISE mission and data products, see \cite{wright2010}
and the NASA/IPAC Infrared Science Archive (IRSA; http://irsa.ipac.caltech.edu).
Because processing of the data continues as of this writing, our candidate selection
is on-going, and only a fraction of our candidates has been followed up, it is
not possible to estimate the sky coverage or volume surveyed for discoveries presented
herein. However, objects discussed here can be added to the growing census of brown dwarfs 
in the Solar Neighborhood and can be used to place lower limits, as we do in \S5.3 below,
to the brown dwarf space density as a function of type or temperature. As such, this paper
should be regarded as a progress report on the continuing WISE search for previously missed brown dwarfs in the
Sun's immediate vicinity.

\subsection{Comparison to Known M, L, and T Dwarfs}

Before beginning the hunt for brown dwarfs, it is necessary to establish empirically the locus of known
brown dwarfs in WISE color space and to understand what other kinds of astrophysical objects
might fall in the same area. This will not only inform the search of WISE color space itself
but also dictate the kinds of photometric follow-up that need to take place before time-intensive
spectroscopic characterization begins.

We have performed a positional cross-correlation of nearby stars from Dwarf 
Archives\footnote{See http://DwarfArchives.org.} against source lists derived from
the WISE coadded data. Many of these stars are known to have
substantial proper motion, so it was necessary to verify each cross-match by visually 
inspecting the WISE and Two Micron All-Sky Survey (2MASS; \citealt{skrutskie2006}) images. The final cross-identifications are given in 
Table~\ref{photometry_knownMLT}, which lists photometry{\footnote {In this and all subsequent tables,
the errors listed are one-sigma values.}}
from 2MASS (when detected) and WISE 
for 118 previously cataloged T dwarfs, 142 L dwarfs, and 92 M dwarfs.
Figure~\ref{W1W2_vs_type} shows the resulting trend of 
WISE W1-W2 color as a function of spectral type for these objects, ranging from early-M
through late-T. Note that there is a slow increase in the W1-W2 color between early-M and early-L,
with the color stagnating near 0.3 mag between early- and mid-L. The W1-W2 color then rapidly
increases at types later than mid-L, corresponding to the appearance of the methane 
fundamental band at 3.3 $\mu$m (\citealt{noll2000}).
The average W1-W2 color is $\sim$0.6 mag at T0 and $\sim$1.5 mag at T5, with the color 
increasing to above 3.0 mag at late T. 
%
%

The red W1-W2 colors ($>$1.7 mag) of dwarfs of type mid-T and later are almost, but not entirely,
unique among astrophysical sources. Dust-obscured galaxies (DOGs) and asymptotic giant 
branch stars (AGBs) are the major sources of contamination at these red W1-W2 colors,
as analysis of the Spitzer Deep Wide-Field Survey (SDWFS) results of 
\cite{eisenhardt2010} has shown. The three short-wavelength bands of WISE -- which 
are close in wavelength to the 3.6, 4.5, and 8.0 $\mu$m
bands (hereafter denoted as ch1, ch2, and ch4, respectively) of the Infrared Array Camera (IRAC) 
onboard the {\it Spitzer} Space Telescope -- can help to distinguish between these 
populations. As Figure 1 of \cite{eisenhardt2010} shows, most AGBs with very red W1-W2 
(or ch1-ch2) colors can be distinguished by their very red W2-W3 (or ch2-ch4) colors. 
Brown dwarfs with similar W1-W2 colors 
are much {\it bluer} in W2-W3 color than these contaminants. 
Similarly, DOGs should be easily separable from brown dwarfs because, like AGBs, their
W2-W3 (or ch2-ch4) colors tend to the red for objects with very red W1-W2 colors.
This is further demonstrated in Figure 12 of \cite{wright2010}, where the bulk of 
the extragalactic menagerie, including red Active Galactic Nuclei, can be distinguished 
from cold brown dwarfs using a color of W2-W3$\approx$2.5 as the dividing line.

As with any set of generic color cuts, however, one should be ever vigilant for
exceptions. As Figure~\ref{W1W2_vs_type} and Figure~\ref{W2W3_vs_type} show, very 
late-T dwarfs have colors approaching W2-W3$\sim$2.0 mag, near the locus of 
extragalactic sources, but their W1-W2 colors are extreme ($>$3.0 mag). Few
extragalactic sources have W1-W2 colors this red, so the W2-W3 color criterion
can be relaxed for the coldest objects (see Figure~\ref{W1W2_vs_W2W3}). Indeed,
brown dwarfs with $T_{eff}<300K$ are expected to
turn to the red in W2-W3 color (Figure 14 of \citealt{wright2010}). It should also
be noted that 
the location of low-gravity or low-metallicity brown dwarfs may not follow the
general rule set by normal-gravity, solar-metallicity cases, so it is important
to use other data (proper motion, parallax) when possible to identify brown dwarfs independent
of photometric selections. Nonetheless, a number of possibly low-metallicity
T dwarfs have been uncovered using these same photometric selections, as further 
discussed in the Appendix.

\subsection{Search for New Candidates}

We have used two different sets of criteria to search the WISE source lists for
nearby brown dwarfs:

(1) To find the coldest brown dwarfs, we selected 
high signal-to-noise (SNR$>$3 at W2) detections having W1-W2 colors (or limits) 
greater than 1.5 mag, corresponding roughly to types $\ge$T5. (Because of the relative
depths of the W1 and W2 bands, the W1-W2 requirement imposes a more severe W2 SNR limit
of its own, generally W2 SNR$>$7.) In order to assure 
that an object is real, we required it to have been
detected at least eight times in the individual W2 exposures; 
this eliminated spurious sources like
cosmic rays and satellite trails that would otherwise not be eliminated during the
outlier rejection step in coadd image creation (see section IV.5.a.v of the Explanatory Supplement to the WISE
Preliminary Data Release\footnote{Available at http://wise2.ipac.caltech.edu/docs/release/prelim/expsup/ .}).
For our initial candidate selection, we also required W2-W3 $<$ 3.0 mag if the object
has a detection in W3. 

(2) To find bright, nearby (i.e.,
high proper motion) L and T dwarfs that other surveys have missed, we
searched for objects with W1-W2 colors greater than 0.4 mag 
(roughly types $\ge$L5), W2 SNR values greater than 30, and no association with
a 2MASS source (implying that the J-W2 color is either very red or the
object has moved). It should be noted that the WISE source lists report 2MASS associations 
falling within 3 arcseconds of each WISE source\footnote{See section IV.7.a.i of http://wise2.ipac.caltech.edu/docs/release/prelim/expsup/ .}.
As with the first search, we required the object to have been
detected at least eight times in W2 to assure its reliability. To
eliminate extragalactic contaminants, we imposed one additional criterion that
W1-W2 $>$ 0.96(W2-W3) $-$ 0.96 (see dashed line in Figure~\ref{W1W2_vs_W2W3})\footnote{It should be noted that five objects from
Table~\ref{photometry_knownMLT} fall more than one sigma to the right of this line. 
These objects are the M dwarfs CTI 064951.4+280442 and CTI 065950.5+280228 and the L dwarfs
SDSS J082030.12+103737.0, SDSS J102947.68+483412.2, and SDSS J204317.69$-$155103.4. Visual inspection of the WISE
images for each of these shows that the W3 detections for all five sources are almost certainly spurious, the
photometry code having found a low-level W3 ``detection'' at the sky location of the object.
The signal-to-noise values for each of these W3 detections are 3.1, 2.2, 3.1, 2.0, and 4.4, respectively.
Because the W3 band is sensitive to extended structure within the Milky Way as well as background variations
by bright stars and the Moon, low-level W3 detections should be regarded with caution.}.

For both searches, no constraint on galactic latitude was imposed, although
additional constraints were placed on object detections in
order to assure that they were real, particularly for our earliest searches of
the WISE source lists. First, the reduced ${\chi}^2$ value from the
PSF photometric fitting (``rchi2'' in the WISE Preliminary Release Source Catalog)
was required to fall between 0.5 and 3.0 to assure that the 
source was pointlike. Second, the early version of the WISE data processing pipeline automatically
flagged artifacts -- bright star halos, diffraction
spikes, latent images, ghost reflections from bright stars, etc. --
only on individual frames and not on the coadded images. 
For search \#2 above, because those objects
all have high-SNR W2 detections, we are able to use these individual frame
flags to remove objects marked as spurious. For fainter objects found in
search \#1, we created three-color images from the W1 (blue), W2 (green), and
W3 (red) coaded images, which were then inspected by eye to eliminate artifacts.
Third, for all objects passing
the above tests, we created finder charts showing the Digitized Sky Survey (DSS; 
$B$, $R$, $I$), the Sloan Digital Sky Survey (SDSS; \citealt{york2000} -- $u$, $g$, $r$, 
$i$, and $z$, if available), 2MASS ($J$, $H$, $K_s$), and WISE
(W1, W2, W3, W4, + 3-color image made from W1+W2+W3) images. A visual inspection of 
these finder charts allowed us to remove other spurious sources and objects
clearly visible in the short wavelength optical bands, while also
allowing us to check for proper motion between surveys for objects bright 
enough to have been detected at shorter wavelengths, such as $J$ and $H$ for brighter
T dwarf candidates.

Example images are shown in Figure~\ref{finder_chart1} through Figure~\ref{finder_chart19} 
for our spectroscopically confirmed candidates. WISE photometry for these 
same sources is given in Table~\ref{WISE_photometry_discoveries}.

\section{Follow-up Imaging Observations}

\subsection{Ground-based Near-infrared Follow-up}

Follow-up imaging observations of WISE candidates are important not only for verifying that
the source has the characteristics of a brown dwarf at shorter wavelengths 
but also for determining how bright the object is at wavelengths observable from the
ground. This latter knowledge is necessary for determining which facility to use for
spectroscopic confirmation.

Ground-based near-infrared observations at $J$ and $H$ bands are technically the easiest to
acquire. Before discussing specifics, it should be
noted that the two main near-infrared filter systems being used today -- the
2MASS system and the Mauna Kea Observatories Near-infared (MKO-NIR, or just MKO) system -- will yield
somewhat different results for the same objects. The 2MASS
bandpasses are illustrated in Figure 2 of \cite{skrutskie2006}, and the MKO filter profiles
are shown in Figure 1 of \cite{tokunaga2002}. A comparison of the two filter sets is illustrated
in Figure 4 of \cite{bessell2005}. For $J$ and $H$ bands, the main difference
between the two systems lies in the width of the $J$-band filter; the $H$-band filters
are very similar. The 2MASS $J$-band profile extends to bluer
wavelengths than does the MKO $J$-band profile, and the overall shapes of the two $J$ filters are quite different, the 2MASS
$J$ filter lacking the top-hat shape that is characteristic of most filter profiles. As a result, the
measured $J$-band magnitude of a brown dwarf, whose spectral signature is also quite complex 
at these same wavelengths, can be considerably different between the two systems. This is
dramatically illustrated in Figure 3 of \cite{stephens2004}, which shows that although the $H$-band magnitudes
are (as expected) very similar between the two systems for a wide range of brown dwarf types, the $J$-band
magnitudes (and hence, $J-H$ colors) can differ by as much as 0.5 mag for late-type T dwarfs. In the
discussion that follows, we note the systems on which our photometry was obtained and we mark
photometry from the two systems with different colors and symbols on the plots.

As Figure~\ref{JH_vs_type} and
Figure~\ref{JH_vs_W1W2} show, mid- to
late-T dwarfs (W1-W2 $>$ 1.5 mag) have relatively blue colors, $J-H \lesssim 0.4$ 
mag, on the 2MASS filter system. 
The upper righthand panel of Figure 5 of \citealt{leggett2010} shows the trend of $J-H$ color versus
spectral type using the MKO filter 
system, and there we find that $J-H \lesssim -0.1$ mag for dwarfs $\ge$T5. These blue colors in both systems are a consequence of stronger
methane bands and collision-induced H$_2$ absorption at $H$-band compared to $J$-band. 
This color stands in contrast to the
majority of other astrophysical sources, whose $J-H$ colors are much redder than this.
For example, Figure 19 of \cite{skrutskie2006} shows the 2MASS $J-H$ color distribution of detected sources
at high Galactic latitude and confirms that most 2MASS objects 
have colors redder than those of mid- to late-T dwarfs. Nevertheless, there are true astrophysical
sources -- e.g., main sequence stars earlier than type K0 (Table 2 of \citealt{bessell1988}); 
certain AGNs, particularly those with $0.7 < z < 1.1$ (Figure 2 of \citealt{kouzuma2010}) -- that aren't brown dwarfs 
and have $J-H$ colors below 0.4 mag.

Fortunately, other colors like $J$-W2 and 
$H$-W2 can also be used to distinguish populations. 
Figure~\ref{JW2_vs_type} and Figure~\ref{HW2_vs_type} show these colors
as a function of spectral type and demonstrate that the $J$-W2 color of mid- 
to late-T dwarfs runs from $\sim$2.0 mag at T5 to $>$4.0 mag at late-T; $H$-W2 color runs 
from $\sim$1.5 mag to $\sim$5.0 mag for the same range of types. Figure~\ref{JW2_vs_W1W2}
and Figure~\ref{HW2_vs_W1W2}
show the trend of $J$-W2 and $H$-W2 with W1-W2 color. The correlation is very tight for
M and L dwarfs (W1-W2 $<$ 0.6 mag), but at redder W1-W2 colors, corresponding to the L/T transition
and beyond, there is a much larger spread of $J$-W2 and $H$-W2 colors at a given W1-W2 color.
Nonetheless, both $J$-W2 and $H$-W2 color increase dramatically beyond W1-W2 $>$1.5 mag ($>$T5).

With our list of brown dwarf candidates in hand, we have obtained $J$ and $H$ observations -- and
in some cases, $Y$ and $K_s$ as well -- using a variety of different facilities in both 
hemispheres. Details of those observations are given below, and a listing of the resultant
photometry can be found in Table~\ref{followup_photometry_discoveries}.

\subsubsection{Fan Mountain/FanCam}

The Fan Mountain Near-infrared Camera (FanCam) at the University of Virginia's 31-inch telescope
has a 1024$\times$1024-pixel HAWAII-1 array (pixel scale of 0.51 arcsec/pix) that images an 8.7-arcminute square field of view 
(\citealt{kanneganti2009}). Observations of eight of our candidates
were obtained in 2MASS $J$ and $H$ bands, and for some objects $Y$-band observations were also
acquired. Details on observing strategy and data reduction are described further in \cite{mainzer2011}.

\subsubsection{Magellan/PANIC}

Persson's Auxiliary Nasmyth Infrared Camera (PANIC) at the 6.5m Magellan Baade Telescope has
a 1024$\times$1024 HAWAII array with a plate scale of
0$\farcs$125 pixel$^{-1}$, resulting in a 2$\times$2-arcminute field of view (\citealt{martini2004}).
Observations of three of our candidates were obtained in Carnegie (essentially MKO) $J$ and $H$ bands. A series of dithered 
images of short integration times was obtained in each band, as is the standard in near-infrared 
imaging observations.

The data were reduced using custom Interactive Data Language (IDL)
routines written by one of us (M.C.C.). Each image was first corrected for non-linearity
using the relation given in the PANIC Observer's Manual\footnote{See
\url{http://www.lco.cl/telescopes-information/magellan/instruments/panic/panic-online-documentation/panic-manual/panic-observers-s-manual/panic-manual}}.
Sky flat fields were constructed by first subtracting a median-averaged
dark frame from each twilight flat frame.  The dark-subtracted twilight
flats were then scaled to a common flux level and then medianed.  After
each frame was sky subtracted and flat fielded, we corrected for optical
distortions as described in the PANIC Observer's Manual.  Frames
were then aligned and combined using a median average to produce the
final mosaic.  2MASS stars in the field of view were used to both
astrometrically calibrate the mosaic as well as determine the zero
points for the images.

\subsubsection{Mt.\ Bigelow/2MASS}

The 2MASS camera on the 1.5m Kuiper Telescope on Mt.\ Bigelow, Arizona, has three 256$\times$256 
NICMOS3 arrays capable of observing simultaneously in 2MASS $J$, $H$, and $K_s$ filters 
(\citealt{milligan1996}).  The plate scale for all three arrays is
1$\farcs$65 pixel$^{-1}$, resulting in a 7$\times$7 arcmin field of
view. Images for twenty-three of our candidates were obtained
using a 3$\times$3 box dither pattern, and to reduce the
overheads associated with nodding the telescope four images were taken
at each of nine dither positions.  At the conclusion of each dither
sequence, the telescope was offset slightly before the start of the next
sequence.

The data were reduced using custom IDL routines. Flat fields in each band
were constructed using on-source frames. These images were scaled to a
common flux level and combined using an average, the latter process rejecting the
lowest 10 and highest 20 values at each pixel location to eliminate imprinting of 
real objects.  Each raw frame was then flat
fielded with these derived flats.  Sky subtraction was then accomplished in a two-step
process.  First, all of the coadded frames were median averaged
to produce a first-order sky frame.  This frame was then subtracted from
each of the coadded frames so that stars could be easily identified.  A
sky frame was then constructed for each coadded image using the five
previous frames in the sequence and the next five frames in the
sequence.  Stars identified in the previous step were masked out
before the ten frames were median averaged. The sky frame was then scaled
to the same flux level as the image and subtracted.  After all of the
coadded frames had been sky subtracted, the frames were co-registered on the sky and averaged.  
2MASS stars in
the field of view were used to both astrometrically calibrate the mosaic
as well as determine the zero points for the images.

\subsubsection{PAIRITEL}

The Peters Automated Infrared Imaging Telescope (PAIRITEL; \citealt{bloom2006}) on Mt.\ Hopkins, Arizona, 
is a 1.3m telescope equipped with three 256$\times$256 NICMOS3 arrays with 2$''$ pixels that
cover an 8.5$\times$8.5 arcminute field of view 
simultaneously in 2MASS $J$, $H$, and $K_s$ filters (\citealt{milligan1996}). 
This camera is, in fact, one of the original 2MASS cameras (\citealt{skrutskie2006}) and
the telescope is the original 2MASS northern facility, now roboticized for transient
follow-up and other projects. Thirty-three of our candidates were observed here.

Upon acquiring a target field, the system acquires a series of 7.8 s exposures
until the desired integration time (1200 s) is met.  After every
third exposure the telescope shifts to a position offset by about one
tenth of the field of view in order to account for bad pixels on the
arrays.  Within hours of a night's observations an automated data
pipeline delivers Flexible Image Transport System (FITS; \citealt{hanisch2001})
mosaic images of the combined stack
of observations for each object in each of the three wavebands.  The
pipeline subtracts a background from each raw image based on the median
of several images adjacent in time.  The resulting individual frames are
calibrated with a flat field constructed from sky observations spanning
many nights. Astrometric and photometric calibrations were accomplished using
observations of 2MASS-detected stars in the field of view.

\subsubsection{Palomar/WIRC}

The Wide-field Infrared Camera (WIRC) at the 5m Hale Telescope at Palomar Observatory 
has a pixel scale of 0$\farcs$2487 pixel$^{-1}$ and uses
a 2048$\times$2048-pixel HAWAII-2 array to image an 8.7-arcminute square field of view
(\citealt{wilson2003}). Observations of eighteen of our candidates were obtained in MKO $J$ and $H$ bands.

Images were reduced using a suite of IRAF\footnote{The Image Reduction and
  Analysis Facility (\citealt{tody1986}) is distributed
  by the National Optical Astronomy Observatory, which is operated by
  the Association of Universities for Research in Astronomy, Inc., under
  cooperative agreement with the National Science Foundation.}  scripts
and FORTRAN programs kindly provided by T. Jarrett. The images were first
linearized and dark subtracted.
Sky background and flat field images were created from the list of input images, and then these were
subtracted from and divided into, respectively, each input image.  At
this stage, WIRC images still contained a significant bias not
removed by the flat field.  Comparison of 2MASS and WIRC photometric
differences across the array showed that this flux bias had a level of
$\approx$10\% and the pattern was roughly the same for all filters.
Using these 2MASS-WIRC differences for many fields, we created a flux
bias correction image that was then applied to each of the ``reduced''
images.  Finally, we astrometrically calibrated the images using 2MASS
stars in the field.  The images were then mosaicked together.  This
final mosaic was photometrically calibrated using 2MASS stars and a
custom IDL script.  Magnitudes were calculated
using the zero points computed using 2MASS stars.

\subsubsection{Shane/Gemini}

The Gemini Infrared Imaging Camera at the 3m Shane Telescope at Lick Observatory uses two
256$\times$256-pixel arrays (pixel scale of 0.70 arcsec/pix) for simultaneous observations of a 3$\times$3-arcminute field 
of view (\citealt{mclean1994}) over each array. The short-wavelength
channel uses a NICMOS3 HgCdTe array, and the long-wavelength channel uses an InSb array.
Observations were obtained for three of our candidates and used 2MASS $J$, $H$, and $K_s$ filters. Observations were acquired in pairs 
($J$ + $K_s$ or $H$ + $K_s$) so that twice as much integration time could be obtained on the InSb array
at $K_s$ as in either the $J$ or $H$ filter on the HgCdTe array. Observations were obtained in
dithered sequences. Data reduction
was handled similarly to that described for the Mt.\ Bigelow/2MASS data.

\subsubsection{SOAR/SpartanIRC}

The Spartan Infrared Camera (SpartanIRC) at the 4.1m Southern Astrophysics Research 
(SOAR) Telescope on Cerro Pach{\'o}n, Chile, uses a mosaic of two 2048$\times$2048-pixel HAWAII-2 arrays
to cover either a 3-arcminute square (0.04 arcsec/pix) or 5-arcminute square (0.07 arcsec/pix)
field of view (\citealt{loh2004}). For seven of our candidates,
we used the 5$\times$5-arcminute mode and acquired observations in the MKO $J$ and $H$ (and for one object object, 
$K$) filters. Observing strategy and data reductions followed the same prescription discussed
in \cite{burgasser2011}.

\subsubsection{Keck/NIRC2}

The second-generation Near-infrared Camera (NIRC2) at the 10m W.~M.~Keck 
Observatory atop Mauna Kea, Hawai'i, employs a 1024$\times$1024 Aladdin-3 array. Used with
the Keck II laser guide star adaptive optics system (\citealt{wizinowich2006}; \citealt{vandam2006}),
it can provide deep, high-resolution, imaging for our faintest targets. In the wide-field mode,
the camera has a plate scale of 0.039686 arcsec/pix resulting in a field of view of 40$\times$40 arcsec.
WISEPA J182831.08+265037.8 and WISEPC J205628.90+145953.3 are the only targets in this paper whose photometry comes solely from the 
NIRC2 data because they were undetected in $J$- and/or $H$-band at other facilities. 
The camera employs MKO $J$ and $H$ filters. 

For WISEPA J182831.08+265037.8 we used the $R$=16 star USNO-B 1168-0346313
(\citealt{monet2003}), located 41$\arcsec$ from the target, for
the tip-tilt reference star.  The tip-tilt reference star for
WISEPC J205628.90+145953.3 was USNO-B 1050-0583683 ($R$=13,
separation=13$\arcsec$).  A three-position dither pattern was used to avoid
the noisy lower-left quadrant. Each position of the dither pattern
consisted of a 120 s exposure; the pattern was repeated two or three
times with a different offset for each repeat in order to build up
deeper exposures.

The images were reduced using a custom set of IDL scripts written by one
of us (C.R.G.).  The raw images were first sky-subtracted using a sky
frame constructed from all of the images.  Next, a dome flat was used to
correct for pixel-to-pixel sensitivity variations.  In order to shift
the reduced images to a common astrometric grid for the creation of the
mosaic, we used the header keywords AOTSX and AOTSY, which record the
position of the AO tip-tilt sensor stage.  As the telescope is dithered,
this sensor must move so that the tip-tilt star stays properly centered.
Although this method of computing image offsets is more precise than
using the right ascension and declination offsets in the header, it can
be prone to small positional errors.  To account for this, we computed
the minimum of the residuals of the shifted image relative to the
reference image (the first image of the stack) over a 5$\times$5 pixel
grid centered on the AOTSX/AOTSY-computed offset.  This correction was
generally $<$2 pixels.  The aligned images were then medianed to form
the final mosaic.

There are typically very few, if any, 2MASS stars in the NIRC2 field of
view that can be used to determine the photometric zero point.  We
therefore bootstrapped a photometric zero point using faint stars in the
NIRC2 images that are also present in the much wider field-of-view Palomar/WIRC
images that have been calibrated using 2MASS stars.  For this ensemble
we computed the average difference 
between the calibrated WIRC magnitudes and the NIRC2 instrumental
magnitudes.  We used the standard deviation of the differences as the
error in the photometric calibration and note that it is the dominant
source of uncertainty.

\subsubsection{CTIO-4m/NEWFIRM}

The NOAO Extremely Wide Field Infrared Imager (NEWFIRM) at the 4m Victor M. Blanco Telescope on Cerro Tololo, Chile,
consists of four 2048$\times$2048 InSb arrays arranged in a 2$\times$2 grid.  With a pixel scale of 0.40 arcsec/pixel,
this grid covers a total field of view of 27.6$\times$27.6 arcmin.   (\citealt{swaters2009}).  
The only source in this paper with NEWFIRM photometry is WISEPA J1541521.66$-$225025.2, so the discussion that 
follows is specific to this set of observations.

At $J$-band, ten sets of images were obtained with integration times of 30 s and two coadds; at $H$-band ten sets with 
exposure times of
5 s and twelve coadds were obtained. Thus, the total integration time was 600 s in each
filter, and these are on the MKO system.  Because of the large field of view and the 35$\arcsec$ gap
between the arrays, we positioned the target near the center of the
North-East array, designated SN013 in the Proposal Preparation
Guide\footnote{\url{http://www.noao.edu/ets/newfirm/documents/Quick\_Guide\_for\_Proposal\_Preparation.pdf}}.

Photometry was performed on the mosaics produced by the NEWFIRM pipeline
(\citealt{swaters2009}).  Although the pipeline computed a
magnitude zero-point based on the photometry of 2MASS stars in the
mosaic, the default aperture used was so large ($\approx$ 6 pixel radius) that  WISEPA J1541521.66$-$225025.2 overlapped
with a close neighbor. Furthermore,  WISEPA J1541521.66$-$225025.2 was
barely visible on the mosaics, meaning that a smaller aperture was warranted to
increase the S/N of our measurement. We therefore re-calibrated the mosaics using
an aperture of 2-pixel radius.  

\subsubsection{2MASS and UKIDSS}

Some of the WISE brown dwarf candidates were detected by 2MASS or by
the United Kingdom Infrared Deep Sky Survey (UKIDSS; \citealt{lawrence2007}). For 2MASS,
we have searched the All-Sky Point Source Catalog and the Reject 
Table\footnote{See http://www.ipac.caltech.edu/2mass/releases/allsky/doc/explsup.html for details.}
and found photometry for forty-five of our sources, the other fifty-one being too faint for 2MASS detection.
For UKIDSS, we searched Data Release 6 (or 5 for those mini-surveys not released in 6) and 
found five of our objects -- three in the Large Area Survey and two in the Galactic Clusters
Survey.

\subsection{Follow-up Using Spitzer/IRAC}

Some of the coldest brown dwarf candidates detected by WISE have W1-W2 color {\it limits}
because these sources are too faint to be detected in the W1 band. Acquiring deeper images at
similar bandpasses would therefore be extremely beneficial in further deciding which of
the candidates are the most interesting for spectroscopic
follow-up. Fortunately, the two shortest wavelength bandpasses of the Infrared Array Camera 
(IRAC; \citealt{fazio2004}) onboard the {\it Spitzer} Space Telescope (\citealt{werner2004}) 
are operating during the Warm {\it Spitzer}
mission and can be used to provide this missing info. Although these bandpasses -- at 3.6 and 4.5 $\mu$m
(aka ch1 and ch2, respectively) -- are similar to the two shortest wavelength bands of WISE
(see Figure 2 of \citealt{mainzer2011}), they were not designed specifically for brown dwarf detection 
and therefore yield less extreme colors (ch1-ch2 vs.\ W1-W2) for
objects of the same spectral type. Nonetheless, IRAC photometry in the 
Warm Spitzer era provides a powerful verification tool for WISE brown dwarf candidates.

Figure~\ref{ch1ch2_vs_type} shows the trend of ch1-ch2 color versus spectral type for early-M
through late-T dwarfs taken from \cite{patten2006}, \cite{leggett2009}, and \cite{leggett2010}.
As noted by these authors, there is a clear trend of increasing ch1-ch2 color with later
spectral type, as expected. Figure~\ref{ch1ch2_vs_W1W2} maps the ch1-ch2 color onto W1-W2 color
for these same sources. Two additional plots, Figure~\ref{Jch2_vs_type} and Figure~\ref{Hch2_vs_type},
show composite colors made from ground-based near-infrared and {\it Spitzer} follow-up and
demonstrate that both $J$-ch2 and $H$-ch2 colors can serve as proxies for spectral type
for objects of type mid-T and later.
Plots of $H$-ch2 color versus type have been presented in earlier papers (the upper lefthand 
panel of Figure 6 of \citealt{leggett2010}; Figure 5 of \citealt{eisenhardt2010}). Even earlier,
\cite{warren2007} noted the tight correlation
between $H$-ch2 color and effective temperature for later T dwarfs and suggested
that the color is relatively insensitive to gravity and metallicity, which makes it a reliable
indicator of T$_{eff}$ or type (see also \citealt{burningham2008}).

We have therefore obtained IRAC ch1 and ch2 photometry of many of our WISE brown dwarf
candidates (Program ID=70062; J.D.\ Kirkpatrick, PI).  Data were collected in both channels
using a 5-position, medium-scale ``cycling'' dither script.  Single images were 30s per position, 
for a total on-source exposure time of 150s per channel.  

Although the $Spitzer$ Science Center (SSC) suggests performing photometry on custom-built mosaics from the 
basic calibrated data (BCD) rather than the post-BCD mosaics, we have found that the photometric differences 
between the two mosaics are negligible for our data sets.  Therefore, photometry of our sources was measured on the 
post-BCD mosaics as produced by the SSC IRAC Pipeline, software version S18.18.  The post-BCD mosaics 
have a pixel scale of 0.6$\arcsec$/pixel, which is half of the native pixel scale.  We used a 4-pixel 
aperture, a sky annulus of 24-40 pixels, and applied the aperture corrections listed in Table 4.7 of the 
IRAC Instrument Handbook$\footnote{http://ssc.spitzer.caltech.edu/irac/iracinstrumenthandbook}$ 
to account for our non-standard aperture size. Resulting IRAC photometry for those confirmed candidates observed
so far is listed in Table~\ref{followup_photometry_discoveries}.

\section{Follow-up Spectroscopic Observations}

For candidates whose follow-up imaging strengthened their credibility as a bona fide brown dwarf,
we obtained spectroscopic confirmation at near-infrared wavelengths. The facilities and instruments
we have used, along with the data reductions employed, are discussed below.

\subsection{Keck/LRIS}

The Low Resolution Imaging Spectrometer (LRIS, \citealt{oke1995}) at the 10m W.~M.~Keck 
Observatory atop Mauna Kea, Hawai'i, uses two channels to simultaneously observe blue
and red optical wavelengths. Our observational setup employed only the two
2k$\times$4k CCDs in the red channel. When used with the 400 lines/mm grating
blazed at 8500 \AA\ and a 1$\arcsec$ slit, the red channel produces 10-{\AA}-resolution spectra covering the range
from 6300 to 10100 \AA.  The OG570 order-blocking filter was used
to eliminate second-order light.  Flat-field exposures of the interior of the 
telescope dome were used to normalize the response of the detector, and HgNeAr
arc lamps were taken after each program object to obtain the wavelength calibration.
Because our targets were late-T dwarfs, for which the 9300-\AA\ band of
H$_2$O is the most important spectral diagnostic, observations of G0 dwarf stars near in airmass and
near in time were needed to correct for telluric H$_2$O at these same wavelengths. 
Observations were typically
acquired with the slit oriented to the parallactic angle to minimize wavelength-dependent
slit losses. Once per night, a standard star from the list of \citet{hamuy1994} was observed 
to provide flux calibration. The data were reduced and calibrated 
using standard IRAF routines as described in \cite{kirkpatrick1999} and \cite{kirkpatrick2006}.

\subsection{IRTF/SpeX}
\label{IRTF-section}

SpeX, a medium-resolution spectrograph and imager at NASA's 3m Infrared Telescope Facility (IRTF)
on Mauna Kea, Hawai'i, uses a 1024$\times$1024 InSb array for its spectroscopic observations
(\citealt{rayner2003}). We used the prism mode with a
0$\farcs$5 wide slit to achieve a resolving power of $R\equiv
\lambda / \Delta \lambda \approx 150$ over the range 0.8-2.5 $\mu$m.  A series of 120s or 180s
exposures were typically obtained at two different nod positions along the 15$\arcsec$
long slit.  A0 dwarf stars at similar airmass to the target were observed near in time 
for telluric correction and flux calibration.  Observations were typically obtained 
with the slit oriented to
the parallactic angle to minimize slit loses and spectral slope
variations due to differential atmospheric refraction.  Finally, a set
of exposures of internal flat field and argon arc lamps were obtained for
flat fielding and wavelength calibration.

The data were reduced using Spextool (\citealt{cushing2004}) the
IDL-based data reduction package for SpeX.  The raw images were first
corrected for non-linearity, pair subtracted, and then flat fielded.
For some of the fainter sources, multiple pair-subtracted images were
averaged in order to facilitate tracing.  The spectra
were then optimally extracted (e.g., \citealt{horne1986}) and
wavelength calibrated using the argon lamp exposures.  Multiple spectra
were then averaged and the resulting spectrum was corrected for
telluric absorption and flux calibrated using observations of an A0
V star using the technique described in \cite{vacca2003}.

\subsection{Keck/NIRSPEC}

The Near-Infrared Spectrometer (NIRSPEC, \citealt{mclean1998,mclean2000}) at the 10m 
W.~M.~Keck Observatory on Mauna Kea, Hawai'i, uses a 1024$\times$1024 InSb array for
spectroscopy. In low-resolution mode,
use of the $42{\times}0{\farcs}38$ slit results in a resolving power of 
R~$\equiv~{\lambda}/{\Delta}{\lambda}~{\approx}~2500$. Our brown dwarf candidates
were observed in either or both of the N3 and N5
configurations (see \citealt{mclean2003})
that cover the portion of the $J$-band window from 1.15 to 1.35 $\mu$m and
the portion of the $H$-band window from 1.5 to 1.8 $\mu$m.   

Data were typically obtained in four or more sets of 
dithered pairs, with a 300-second exposure obtained at each dither position. 
To measure telluric absorption and to calibrate the flux levels, A0 dwarf stars
were observed near in time and airmass to the target. Other calibrations 
consisted of neon and argon arc lamp spectra, dark frames, and
spectra of a flat-field lamp.  We employed standard reductions using 
the REDSPEC package, as described in \cite{mclean2003}. 

\subsection{Magellan/FIRE}

The Folded-port Infrared Echellette (FIRE; \citealt{simcoe2008}, \citealt{simcoe2010}) 
at the 6.5m Magellan Baade Telescope uses a 2048$\times$2048 HAWAII-2RG array.
In its high-throughput prism mode, it covers the wavelength range from 0.8 to 2.5 $\mu$m
at a resolution ranging from R=500 at $J$-band to R=300 at $K$-band for a slit
width of 0$\farcs$6. Typically, each observation
used the 50$''$-long slit aligned to the parallactic angle, and consisted of a series of nod pairs 
taken with exposure times not exceeding 120s per position. 
The spectrograph detector was read out using the 4-amplifier mode at ``high gain'' (1.2 counts per
e$^-$) with the Fowler-8 sampling mode.  We also obtained exposures of a
variable voltage quartz lamp (set at 1.2 V and 2.2 V) for flat fielding purposes and 
neon/argon arc lamps were used for wavelength calibration.
A0 dwarf stars were used for 
telluric and flux calibration. 

Data were reduced using a modified version of Spextool. In contrast to SpeX, the spatial axis of the
FIRE slit is not aligned with the columns of the detector so that
the wavelength solution is not only a function of the column number but
also of the row number.  We therefore derived a two-dimensional
wavelength solution in two steps.  First, a one-dimensional wavelength
solution applicable to the center of the slit was determined using the
night-sky OH emission lines (\citealt{cushing2004}).  Second, the OH
emission lines were used to map the optical distortions in the spatial
and dispersion axes within each order.  The 1D wavelength solution and
distortion maps were then combined to assign a wavelength and spatial
position to each pixel in each order.

With the wavelength calibration completed, the remainder of the
reduction steps could proceed.  Pairs of images taken at different
positions along the slit were subtracted in order to remove the bias and
dark current as well as to perform a first-order sky subtraction.  The
resulting pair-subtracted image was then flat fielded using a normalized
flat constructed from dome flats taken at the start of each night.  The
spectral extraction followed the technique described in
\cite{smith2007}.  A set of ``pseudorectangles'' was defined
spanning each order to map out positions of constant wavelength on the
detector.  These rectangles are themselves composed of pseudopixels with
a width and height of $\sim$1 detector pixel.  The pseudorectangles were
then extracted using a polygon clipping algorithm
(\citealt{sutherland1974}), producing 1D profiles of the slit at each
wavelength.  Spectral extraction including residual background
subtraction could then proceed in the standard way.  The raw spectra were
then combined and corrected for telluric absorption and flux
calibrated using the observations of an A0 V star and the technique
described in \cite{vacca2003}.

\subsection{Palomar/TSpec}

The Triple Spectrograph (TSpec) at the 5m Hale Telescope at Palomar Observatory
uses a 1024$\times$2048 HAWAII-2 array to cover simultaneously the range from 1.0 
to 2.45 $\mu$m (\citealt{herter2008}).
With a 1$\times$30-arcsecond slit, it achieves a resolution of $\sim$2700.
Observations were acquired in an ABBA nod sequence with an exposure time per
nod position not exceeding 300s to mitigate problems with changing OH background
levels. Observations of A0 dwarf stars were taken near in time and near in airmass
to the target objects and were used for telluric correction and flux calibration.
Dome flats were taken to calibrate the pixel-to-pixel response. Data reduction
is identical to that discussed above for FIRE because Spextool required the
same changes for TSpec reductions as it did for FIRE reductions.

\subsection{SOAR/OSIRIS}

The Ohio State Infrared Imager/Spectrometer
(OSIRIS) mounted at the 4.1m Southern Astrophysical Research Telescope
(SOAR) located at Cerro Pach{\'o}n, Chile, uses a 1024$\times$1024 HgCdTe
array. The 1$\farcs$0-wide slit yields a 
resolving power of $R \approx 1400$ across the 1.18-2.35 $\mu$m wavelength 
range in three spectral orders. Short exposures (180s) were taken at
six or seven positions nodded along the 24$\arcsec$-long slit. A0 dwarf stars
were observed for telluric correction and flux calibration.  Wavelength calibration 
was based on the OH airglow lines. The data were reduced using a
modified version of the Spextool data reduction package. (See \S~\ref{IRTF-section} for
a description of Spextool.)  The three spectral
orders were then merged into a single spectrum covering the entire
wavelength range.

\subsection{HST/WFC3}

The Wide Field Camera 3 (WFC3) onboard the {\it Hubble} Space Telescope employs
a 1024$\times$1024 HgCdTe detector
with a plate scale of 0$\farcs$13 pixel$^{-1}$ to image a field
of view of 123$\times$126 arcseconds.
We used the G141 grism to acquire slitless spectroscopy over the 1.1-1.7 $\mu$m range for faint
targets not observable from ground-based facilities (Program 12330, PI Kirkpatrick). The
resulting resolving power, $R$$\approx$130, is ideal for the broad characterization of faint
sources. Three spectra -- those
of WISEPC J145018.40+553421.4, WISEPA J173835.53+273258.9, and WISEPA J182831.08+265037.8 -- were
obtained with this setup.
For each target, we first obtained four direct images through the F140W
filter in MULTIACCUM mode using the SPARS25
sampling sequence.  The telescope was offset slightly between each exposure.
We then obtained four dispersed images with the G141 grism using MUTLIACCUM mode and
a SPARS50 sequence. These dispersed images were acquired at the same positions/dithers
as used in the direct images.
 
Bias levels and dark current were first subtracted from the data images
using version 2.3 of the CALWFC3 pipeline. CALWFC3 also
flat fields the direct images, but the grism images are flat fielded during
the extraction process described below.  Spectra were then
extracted using the aXe software (\citealt{kummel2009}).
Because aXe requires knowledge of the position and brightness of the
targets in the field of view, we combined the four direct
images using MULTIDRIZZLE (\citealt{koekemoer2002}) and the latest
Instrument Distortion Coefficient Table (IDCTAB).  SExtractor
(\citealt{bertin1996}) was then used to produce a catalog of
objects in the field. For each cataloged object, two-dimensional subimages 
centered on the first-order spectra of each
object were combined using AXEDRIZZLE to produce a final spectral image.  
These subimages were used to extract flux-calibrated spectra. Additional details 
on the data reduction process are discussed in Cushing et al.\ (accepted).

\section{Analysis}

\subsection{Deriving Spectral Types}

The list of spectroscopically confirmed WISE brown dwarfs is given in Table~\ref{spectral_types}.
Abbreviated source names\footnote{Throughout the rest of the paper,
we will abbreviate WISE source names to the form ``WISE hhmm$\pm$ddmm''.
Full designations can be found in Table~\ref{WISE_photometry_discoveries}.} are shown in column 1; optical 
spectral types are shown in column 2; near-infrared types are shown in column 3; and the source
of the spectrum, integration time, telluric corrector star (for ground-based observations), and 
observation date are shown in columns 4 through 7. The optical and 
near-infrared classifications of these sources are discussed further below.

\subsubsection{Optical Spectral Types}

Keck/LRIS spectra were obtained for seven of our candidates. Reduced spectra
from 8000 to 10000 \AA\ are shown in Figure~\ref{optical_spectra}, all of which
have been corrected for telluric absorption over the regions 
6867--7000 \AA~(the Fraunhofer B band, caused by O$_2$ absorption),
7594--7685 \AA~(the Fraunhofer A band, again caused by O$_2$), and
7186--7273, 8162--8282, and $\sim$8950--9650 \AA~(all caused by H$_2$O absorption).
These spectra show the hallmarks
of T dwarf optical spectra: strong H$_2$O absorption with a bandhead at 9250 \AA,
along with \ion{Cs}{1} absorption at 8521 and 8943 \AA\ in the earlier objects and
CH$_4$ absorption between 8800 and 9200 \AA\ in the later objects.

By-eye comparisons to the T dwarf optical spectral standards of \cite{burgasser2003} show
that these spectra range in type from T$_o$5 for WISE 1841+7000 to later than T$_o$8 for WISE 1741+2553. 
(The ``o'' subscript is used to denote spectral types assigned based on optical spectra.)
This latter spectrum is unusual in that it has stronger 8800-9200
\AA\ CH$_4$, stronger 9250-9400 \AA\ H$_2$O bands of the 3(${\nu}_1$,${\nu}_3$) transition,
and stronger 9450-9800 \AA\ H$_2$O bands of the 2(${\nu}_1$,${\nu}_3$) + 2(${\nu}_2$) transition
than the latest T dwarf optical standard, the T$_o$8 dwarf 2MASS J04151954$-$0935066
(\citealt{burgasser2003}). We therefore propose that WISE 1741+2553 be the 
spectral standard for a newly adopted T$_o$9 spectral class. Figure~\ref{optical_spectral_sequence} 
illustrates the entire sequence of T dwarf optical standards from \cite{burgasser2003}
appended with the proposed T$_o$0 standard SDSSp J083717.22$-$000018.3 from 
\cite{kirkpatrick2008} and this newly proposed T$_o$9 standard.

Further evidence in support of a new optical standard is shown in 
Figure~\ref{optical_T9s}. Shown here is a comparison of the spectrum of
WISE 1741+2553 with our LRIS spectrum of the bright, late-T dwarf
UGPS J072227.51$-$054031.2, which Cushing et al.\ (accepted)
have proposed as the infrared spectral standard for type T9. As our comparison shows,
the CH$_4$ and H$_2$O depths are very similar between these two objects and both are distinctly
different from the T$_o$8 standard. Thus, identifying WISE 1741+2553 as the new, T$_o$9
anchor point would help link the optical and near-infrared sequences, especially since
WISE 1741+2553 is also classified as a T9 on the
near-infrared scheme. See further discussion in the Appendix.  

\subsubsection{Near-infrared Spectral Types}

All of the confirmed brown dwarfs listed in Table~\ref{spectral_types} have near-infrared
follow-up spectra. These spectra were classified using the the near-infrared M and L
sequences of \cite{kirkpatrick2010} and the near-infrared T0-to-T8 dwarf sequence of \cite{burgasser2006}.
Cushing et al.\ (accepted) have extended classifications to T9 and Y0 and have reclassified previously
discovered $>$T8 dwarfs from the literature on this system. This extension of the classification system uses
UGPS 0722$-$0540 as the near-infrared T9 standard and WISE 1738+2732 as the Y0 standard.

Assignment of spectral types
was done by overplotting spectra of these standards onto the candidate spectra and determining by-eye which standard
provided the best match. In some cases two adjacent standards, such as T7 and T8, provided
an equally good match, so the candidate spectrum was assigned an intermediate type, in this case, of T7.5. For L dwarfs, the comparison
was done at $J$-band and, following the prescription discussed in \cite{kirkpatrick2010},
any anomalies at $H$- and $K$-bands were noted. Spectra that did not match any of the standards 
well are marked with a ``pec'' suffix to
indicate that they are peculiar. As a further example, an object that best
fit the L9 spectral standard at $J$-band but failed to provide a good match to the L9 standard at longer
wavelengths because it was considerably redder than the standard was assigned a type of ``L9 pec (red).'' 
See \cite{kirkpatrick2010} for examples of similar classifications.

In Figure~\ref{other_spectra_lt_T0} through Figure~\ref{other_spectra_gt_T8}, we show the 
near-infrared spectra for each of our sources.
Because of the narrow wavelength ranges covered by the Keck/NIRSPEC and SOAR/OSIRIS spectra,
those data are plotted separately in Figure~\ref{NIRSPEC_spectra}
and Figure~\ref{OSIRIS_spectra}.

\subsubsection{Discussion}

Near-infrared spectral types (and optical spectral types, for those with Keck/LRIS spectra) are listed in 
Table~\ref{spectral_types} for all WISE brown dwarf discoveries. For objects with near-infared spectral
types of T0 or later, Figure~\ref{spec_type_histogram} shows the number of newly discovered objects per spectral type bin 
compared to the number of objects previously published. Whereas there 
were 16 objects known previously with types of T8 or later (\citealt{burgasser2002}, \citealt{tinney2005},
\citealt{looper2007}, \citealt{warren2007}, \citealt{delorme2008}, \citealt{burningham2008}, 
\citealt{burningham2009}, \citealt{burningham2010mf}, \citealt{goldman2010}, \citealt{lucas2010}, \citealt{delorme2010}, 
\citealt{burningham2011ugps0521}), the tally now stands at 58 once our objects
are added. WISE has already identified seventeen new objects with types equal to or later than the T9
UGPS J072227.51$-$054031.2, the previous record holder for latest measured spectral type, and six of these belong to the
Y dwarf class (Cushing et al., accepted).

Figure~\ref{W1W2_vs_type} through Figure~\ref{W1W2_vs_W2W3} and Figure~\ref{JH_vs_type} through 
Figure~\ref{Hch2_vs_type}, discussed previously, show the locations of these newly discovered WISE brown dwarfs 
(black symbols) in color space. The T9, T9.5, and early-Y dwarfs continue the trend toward redder 
W1-W2, ch1-ch2, $J$-W2, $H$-W2, $J$-ch2, and $H$-ch2 colors, with the reddest object being the $>$Y0 dwarf
WISE 1828+2650 (J-W2 = 9.39$\pm$0.35 mag; $J$ on the MKO filter system). The blueward trend in $J-H$ color seen 
for later T dwarfs, however, begins to reverse 
near a spectral type of Y0. In particular, the $J-H$ color of WISE 1828+1650 is dramatically redder ($J-H = 0.72{\pm}0.42$ mag;
MKO filter system) than any of the late-T or Y0 dwarfs, the latter of which show a large scatter in $J-H$ colors themselves.
Cushing et al.\ (accepted) explore the trend of $J-H$ colors in more detail and show that the synthetic photometry derived from
our observed spectra generally agree with photometry measured from direct imaging. Given the large spread in $J-H$ color
observed for the six Y dwarfs already identified, $JHK_s$ colors alone cannot be used to confirm or deny objects as cold as Y dwarfs.

\subsection{Distances and Proper Motions}

Distances to the new WISE brown dwarf discoveries can be estimated based on their W2 magnitudes and
measured spectral types. First, however, the relation between absolute W2 magnitude and spectral
type needs to be established using objects with measured trigonometric parallaxes and WISE W2 
detections. Figure~\ref{MW2_vs_type} shows the trend of absolute W2 magnitude as a function of
near-infrared spectral type for previously published objects whose measured parallaxes are at least three times
the measurement error (Table~\ref{abs_mags}).
A third-order least squares relation, weighted by the errors on the $M_{W2}$ values, is shown by the black
curve in Figure~\ref{MW2_vs_type}. For this fit, objects known to be binary (red points) have been
omitted. The resulting relation is

$$M_{W2} = 9.692 + 0.3602(Type) - 0.02660(Type)^2 + 0.001020(Type)^3$$

\noindent where $Type$ is the near-infrared spectral type on the system where L0=0, L5=5, T0=10, T5=15, and Y0=20.

Using this relation, we have estimated distances to our WISE discoveries. These are given in column 2 of
Table~\ref{astrometry}. (The distance to the lone M dwarf, WISE 0106+1518, was estimated using 2MASS
magnitudes and the near-infrared absolute magnitudes listed in Table 3 of \citealt{kirkpatrick1994}.)
These distance estimates for the late-T dwarfs and Y dwarfs are shown graphically in Figure~\ref{distances}.
Also shown in the figure are previously published late-T dwarfs from other surveys. WISE has sufficient
sensitivity to detect the latest T dwarfs out to 15 to 20 pc and because of its all-sky
coverage can complete the census of the Solar Neighborhood for these objects. As the figure shows, 
twelve of our objects have estimated distances placing them within 10 pc of the Sun, and
two of these have estimates placing them within 5 pc. It should also be noted that the fitted 
relation shown in Figure~\ref{MW2_vs_type} may lead to overestimated distances for objects at the latest types
because the relation falls above all four of the previously published T8.5 and T9 dwarfs on that plot. Furthermore, the
extrapolation of this relation to even later Y dwarf types may lead to even more discrepant distance overestimates,
as discussed further in the caption to Figure~\ref{MW2_vs_type}.
Measuring trigonometric parallaxes for more of these latest T dwarfs and early Y dwarfs will be an
important, early step in characterizing the physical nature of these objects.

Because these objects should all lie very close to the Sun, their observed parallaxes will be large.
Thanks to its survey strategy, WISE performed its two passes of the sky with observations always near 90$^\circ$ 
solar elongation, thus capitalizing on the maximum parallactic angle at both epochs. 
(For objects observed during the final $\sim$2 weeks of WISE operations,
three epochs of WISE data are available.) Objects will, of course,
also show displacements due to proper motion, so observations at other epochs and/or from other surveys
are necessary to disentangle the two effects.  Hence, ancillary astrometry from 2MASS and SDSS and our own follow-up
observations from the ground and from space ({\it Spitzer} and {\it HST}) are invaluable. Currently
available astrometric data points\footnote{Future papers will include astrometry taken from our various
ground-based imaging campaigns, once data over a longer time baseline have been acquired.}
are shown in columns 3 through 8 of Table~\ref{astrometry}.

It should be noted here that positions of objects in the WISE preliminary data release may be
offset from their true positions by many times the quoted positional
uncertainty.  Approximately 20\% of the sources fainter than W1$\approx$14.5
mag in the Preliminary Release Source Catalog suffer from a
pipeline coding error that biases the reported position by $\sim$0.2-1.0 arcsec
in the declination direction while an increasingly smaller fraction
of the sources suffer this effect to magnitudes as bright as W1$\approx$13.0 mag.
The Cautionary Notes section of the WISE
Preliminary Release Explanatory Supplement describes the origin and
nature of this effect in detail. For this paper, we have rerun the WISE images
for our sources through a version of the WISE pipeline that eliminates this source 
of systematic error, and we list those remeasured positions in 
Table~\ref{astrometry}. This version of the pipeline is essentially the same one
used to process data for the WISE Final Data Release.

Astrometric fits were made to the multiple observations of each source. 
These fits solved for five parameters: initial (time = $t_i$) positional offsets of
$\Delta\alpha$ and $\Delta\delta$ in Right Ascension ($\alpha$) and in Delination ($\delta$), the Right Ascension
component of proper motion ($\mu_\alpha = (\cos\delta) d\alpha/dt$), the Declination component
of proper motion ($\mu_\delta = d\delta/dt$), and the parallax ($\pi_{trig}$).  For all but four
sources, the data were not sufficient to find an accurate distance,
so the distance was forced to equal the spectrophotometric
estimate, and the fit only solved for the first four parameters; the four sources with a
preliminary parallax measurement are listed in Table~\ref{prelim_parallaxes} and are discussed
individually in the Appendix. The 
equations used are

$$\cos\delta_1 (\alpha_i-\alpha_1)  =  \Delta\alpha + \mu_\alpha (t_i-t_1) + \pi_{trig} \vec{R}_i \cdot \hat{W}, {\rm and}$$

$$\delta_i-\delta_1 =  \Delta\delta +\mu_\delta (t_i-t_1) - \pi_{trig} \vec{R}_i \cdot \hat{N}.$$

\noindent Subscript $i$ refers to the individual astrometric measurements, where $t_i$ 
is the observation time in years, and $R_i$ is the vector
position of the observer relative to the Sun in celestial
coordinates and astronomical units. $\hat{N}$ and $\hat{W}$
are unit vectors pointing North and West from the position of the source.
$R_i$ is the position of the Earth for 2MASS, SDSS, WISE, and {\it HST} observations;
for {\it Spitzer} observations, $R_i$ is the position of the spacecraft.
The observed positional difference on the left hand side is in arcseconds,
the parameters $\Delta\alpha$ and $\Delta\delta$ are in arcseconds,
the proper motion $\mu_\alpha$ and $\mu_\delta$ are in arcseconds/year,
and the parallax $\pi_{trig}$ is in arcseconds.

These equations are solved using standard weighted least-squares techniques,
which also provide the uncertainties in the parameters. These uncertainties
come from propagating the uncertainties in the input. The $\chi^2$ and
number of degrees of freedom are also given and can be used to assess the 
quality of the fit. The resulting proper motions in RA and Dec are listed
in columns 9 and 10 of Table~\ref{astrometry}. For motions with a significance
of $>3\sigma$, the total proper motion is listed in column 11 along with
the tangential velocity in column 12.

\subsection{Space Density of late-T dwarfs}

The brown dwarf discoveries presented here represent only a fraction of the brown
dwarf candidates identified so far from WISE data. WISE
coadded data are not available across the entire sky, and many of those coadds
do not reach the full survey
depth. Nonetheless, we can use these preliminary results to assess our progress
toward completing the tally of cold brown dwarfs in the Solar Neighborhood and gauging the 
functional form of the mass function for these objects by using lower limits to their 
space densities.

Our goal is to complete an all-sky census of objects out to a specified maximum
distance for each spectral subtype of T6 or later.
Table~\ref{space_density_numbers} divides our discoveries into six spectral type
bins (column 1) from T6 through $>$Y0. The approximate range in effective temperature is given
for each bin in column 2. These temperature bins are assigned as follows. We took the
values of $T_{eff}$ for T dwarfs of type T6 and later as computed by
\cite{warren2007},
\cite{delorme2008},
\cite{burningham2008},
\cite{burgasser2010-ross458c},
\cite{lucas2010},
\cite{burgasser2011},
\cite{burningham2011},
\cite{burningham2011ugps0521}, 
Bochanski et al.\ (submitted), 
and Cushing et al.\ (accepted)
or compiled by \cite{kirkpatrick2005}. Then, when necessary, we re-assigned spectral types to these objects
so that they matched the near-infrared spectral classification scheme of 
\cite{burgasser2006} or its extension beyond T8 by Cushing et al.\ (accepted).
We then examined the distribution of temperature within each integral spectral type
bin and found that a 150K width for each bin was enough to encompass the $T_{eff}$ values for
most of the objects. The final assignments are given in column 2 of Table~\ref{space_density_numbers}.

As shown in Figure~\ref{distances}, the depth of our current search
translates to different distances for each bin. In the spirit of determining
the space density using a well defined census of the Solar Neighborhood, we limit 
our sample to those objects falling within 20 pc of the Sun
even if WISE can sample that spectral type to larger distances. Only in the last
three bins -- the T9-T9.5 bin and the Y dwarf bins -- is WISE incomplete at this distance,
so those bins are limited to volumes with smaller radii.
These values, called $d_{max}$, are listed in column 3.

Next, the number of objects per spectral type bin lying closer than the value of
$d_{max}$ is tabulated for previously published objects (column 4), for our new WISE
discoveries (column 5), and in total (column 6). Distances are determined using 
trigonometric parallaxes, if available, or 
spectrophotometric estimates if no parallax has been measured. The resulting space density
in each bin is given in column 7. 

This simple calculation of the space densities can be overestimated for the following 
reasons:

\noindent (1) Spectrophotometric distance estimates have an inherent bias. The absolute
magnitude versus spectral type relation is based on parallaxes, and those parallax
measurements lead to a bias in estimated distances because a parallax value of $\pi_{trig} \pm \sigma$
is more likely to represent an object further away ($\pi_{trig} - \sigma$) than an object
closer ($\pi_{trig} + \sigma$) because the volume of space between parallax values of $\pi_{trig}$
and $\pi_{trig} - \sigma$ is larger than that sampled between parallax values of $\pi_{trig}$
and $\pi_{trig} + \sigma$. Thus, the observed values of $\pi_{trig}$ are larger than the true
values and the measured absolute values will be systematically too large. A correction can
be applied that depends only on the value of $\sigma/\pi_{trig}$ (see Table 1 of \citealt{lutz1973}).
Most of the parallaxes in Table~\ref{abs_mags} have $\sigma/\pi_{trig}$ values of less than
5\% where the correction to the absolute magnitude is $\le$0.02 mag, and those 
with larger errors have already been downweighted in our fit. We therefore
conclude that the Lutz-Kelker effect is negligible here.

\noindent (2) Both the Malmquist bias and the Eddington bias can be largely accounted for by
limiting the sample over which we derive our space densities. Malmquist bias (\citealt{malmquist1920}), in which
more luminous objects can preferentially bias statistics in a magnitude limited sample,
can be eliminated by calculating space densities in narrow spectral type bins in which
all objects have the same (or nearly the same) intrinisic luminosity. The Eddington bias
(\citealt{eddington1913}; \citealt{eddington1940}), in which random errors will bias
magnitude measures to brighter values due to the fact that there are more objects
in the more distant (fainter) population than in the closer (brighter) one, can be
reduced by operating at magnitudes where the random errors are still small. By 
using the brighter and better measured W2 values to estimate distances, we can 
reduce the effects of Eddington bias on our derived densities. (For further discussion, see also \citealt{teerikorpi2004}.)

\noindent (3) Unresolved binarity will cause distances to be underestimated. This may
cause a more distant object to appear closer than it really is and falsely inflate the space density.
Empirical data presented earlier can be used to estimate this degree of binary
contamination. Figure~\ref{MW2_vs_type} shows a well-defined binary sequence (red) overlying
the sequence of single objects on the $M_{W2}$ versus spectral type diagram. If we compute the ratio of known binaries to total objects
between L0 and T4, we find that 12/35, or 34\%, are binary. (See also section 7.4 of \citealt{burgasser2006} for
an in-depth discussion of instrinic binarity, which varies from $\sim$20\% at early-L to $\sim$42\% at the L/T transition.)
While this is a sizable percentage, it does not
mean that all of the unresolved binaries fall outside of the sample considered. Some small fraction will still be contained within the 
distance limit and will have been undercounted by a factor of two. Nonetheless, binarity is likely the largest
contributor to inflating density estimates.
 
On the other hand, other biases discussed below lead to an underestimate in the space densities. These
effects are believed to overwhelm the effects detailed above, and hence our simple space density 
calculations, although preliminary, can be considered as lower limits to the true densities:

\noindent (1) Although WISE has taken data covering the entire sky at multiple epochs,
the available coadded data cover less than 75\% of the entire sky. Also none of the sky has been
coadded to its full depth using all available frames and the detection threshold for first-pass
processing was set higher, in units of signal-to-noise ratio, than it will be for final
processing. The latter points are particularly important
as they will, in the future, enable  more robust colors or color limits for potential Y dwarf candidates.
Moreover, we have followed up less than 50\% of the brown dwarf candidates already 
culled from sections of the sky for which we have access. Thus, we believe that our 
current space density estimates are gross underestimates.

\noindent (2) Except for 2MASS, other surveys providing data in column 4 of Table~\ref{space_density_numbers}
do not have all-sky coverage and can only provide limited help in completing this nearby sample. 
Moreover, none of the current or planned ground-based surveys canvassing the sky for brown dwarfs can reach sizable populations
of the coldest objects because Y dwarfs are intriniscally dim at ground-observable wavelengths. 
This is highlighted in Figure~\ref{MH_vs_type}, which shows the absolute $H$-band magnitude as
a function of spectral type. Note that the Y0 dwarf WISE 1541$-$2250
has $M_H = 23.7{\pm}0.9$ mag; its absolute magnitude is $M_J = 23.9{\pm}0.8$ mag and it is presumably even fainter 
than this shortward of $J$ band. 
WISE operates at wavelengths where these objects are their brightest -- five thousand 
times brighter at W2 than at
$J$ and $H$ bands, in the case of WISE 1828+2650 -- so it is the only survey capable of detecting the coldest brown dwarfs 
in significant numbers. 

\noindent (3) As mentioned earlier, the $M_{W2}$ vs. spectral type relation of Figure~\ref{MW2_vs_type} likely 
overpredicts distances to dwarfs of type $\ge$T9. This means that our surveyed volume may be 
overestimated, leading to an underestimate of the space density.

\noindent (4) Despite the all-sky covereage of WISE, the galactic plane will restrict our ability 
to probe to the same depths as other parts of the sky due to higher backgrounds and confusion.
This loss of coverage is not currently accounted for in our density estimates.

We have checked the distance distribution of objects in each spectral type bin by performing the
$V/V_{max}$ test (column 8 of Table~\ref{space_density_numbers}). This test
was first proposed by \cite{schmidt1968} to check the uniformity of a distribution
of objects in space. The quantity $V$ is the volume of space interior to object $i$ at distance
$d_i$, and $V_{max}$ is the full volume of space contained within the distance limit, $d_{max}$, of 
the sample. For a uniform sample, the average value, $<V/V_{max}>$, should be 0.5 because
half of the sample should lie in the nearer half of the volume and the rest should lie in the farther half.
If this number is not near 0.5, then the sample is either non-isotropic or incomplete. For our sample
we find that the T dwarf bins have $<V/V_{max}>$ values considerably less than 0.5. This points to 
incompleteness in the sample -- our low $<V/V_{max}>$ values are almost certainly a consequence of the 
fact that the brighter (closer) candidates tend to be followed up first. This gives further credence to
the assertion that our number densities are lower limits only. For the Y dwarf bins, however, the values of 
$<V/V_{max}>$ are above 0.5, which is further suggestive that our assumed distances to
these objects are overestimates.

Figure~\ref{space_density} shows these preliminary results (column 9 of Table~\ref{space_density_numbers})
relative to measurements made by other 
surveys and relative to predictions based on different forms of the underlying mass function. Previous
results by \cite{metchev2008}, \cite{burningham2010mf}, and \cite{reyle2010} are shown by the open
symbols and, with the possible exception of the \cite{burningham2010mf} point, support values of $\alpha$ 
of zero or greater, where the functional form is given as $dN/dM \propto M^{-\alpha}$. Our incomplete, volume-limited
sample so far fails to put tighter constraints on the mass function at warmer temperatures than previously
published work, but the
preliminary lower limit to the Y0 space density already rules out the $\alpha = -1.0$ model and may soon,
with additional Y0 discoveries, be able to distinguish between the $\alpha = 0.0$ and $\alpha = +1.0$
models. Results for the early-Y dwarfs already suggest that the low-mass cutoff of star formation must be
below 10 $M_{Jup}$ if $\alpha \le 0$. This result is in accordance with findings in young star formation regions 
(e.g., \citealt{luhman2000}, \citealt{muench2002}, \citealt{lucas2006},
\citealt{luhman2007-cham}),
but the derived masses for those objects should be considered cautiously, as discussed in \cite{baraffe2003},
because models with ages of $\sim$1 Myr or younger are highly sensitive to untested assumptions about initial 
conditions. Using models of older ages is far less sensitive to these assumptions, so field brown dwarfs
provide a better check of star formation's low-mass cutoff. Model fits by Cushing et al.\ (accepted) suggest that our
Y dwarf discoveries have masses as high as 30 $M_{Jup}$ or as low as 3 $M_{Jup}$ or less, which agrees roughly
with the mass values inferred from Figure~\ref{space_density}.

\section{Conclusions}

This paper represents the culmination of a year's worth of effort following up the first batch
of brown dwarf candidates identified by WISE. There are many hundreds more candidates still 
being scrutinized, and there are still areas of sky not yet searched. It is therefore
clear that these first hundred brown dwarf discoveries are harbingers of a much larger
trove of brown dwarfs yet to be uncovered by WISE. Not only is the WISE data archive uniquely 
suited to finding even colder objects than the current batch of early-Y dwarfs, the all-sky and 
multi-epoch nature
of the mission will enable many other brown dwarf studies -- the search for the lowest mass objects in 
nearby moving groups, hunting for low-metallicity objects via their high proper motions, etc. -- that
are well beyond the scope of the photometric search presented here.

\section{Acknowledgments}

This publication makes use of data products from the Wide-field Infrared Survey Explorer, which is a 
joint project of the University of California, Los Angeles, and the Jet Propulsion Laboratory/California 
Institute of Technology, funded by the National Aeronautics and Space Administration. We 
acknowledge fruitful discussions with Tim Conrow, Roc Cutri, and Frank Masci, and
acknowledge assistance with Magellan/FIRE observations by Emily Bowsher.
This publication 
also makes use of data products from 2MASS, SDSS, and UKIDSS. 2MASS
is a joint project of the University of Massachusetts and the Infrared Processing and Analysis 
Center/California Institute of Technology, funded by the National Aeronautics and Space Administration 
and the National Science Foundation. SDSS is funded by the Alfred P. Sloan Foundation, the Participating 
Institutions, the National Science Foundation, the U.S. Department of Energy, the National Aeronautics 
and Space Administration, the Japanese Monbukagakusho, the Max Planck Society, and the Higher Education 
Funding Council for England.
UKIDSS uses the Wide Field Camera at the United Kingdom Infrared Telescope atop Mauna Kea, Hawai'i. We
are grateful for the efforts of the instrument, calibration, and pipeline teams that have made the 
UKIDSS data possible.
We acknowledge use of the DSS, which were produced at the Space Telescope Science Institute under
U.S.\ Government grant NAG W-2166. The images of these surveys are based on photographic data
obtained using the Oschin Schmidt Telescope on Palomar Mountain and the UK Schmidt Telescope.
This research has made use of the NASA/IPAC Infrared Science Archive (IRSA),
which is operated by the Jet Propulsion Laboratory, California Institute of Technology, under contract
with the National Aeronautics and Space Administration. Our research has benefitted from the M, L, and
T dwarf compendium housed at DwarfArchives.org, whose server was funded by a NASA Small Research Grant, 
administered by the American Astronomical Society. We are also indebted to the SIMBAD database,
operated at CDS, Strasbourg, France. 
This work is based in part on observations made with the {\it Spitzer} Space Telescope, which is
operated by the Jet Propulsion Laboratory, California Institute of Technology, under a contract with
NASA. Support for this work was provided by NASA through an award issued to program 70062 by JPL/Caltech. This work
is also based in part on observations made with the NASA/ESA {\it Hubble} Space Telescope, obtained
at the Space Telescope Science Institute, which is operated by the Association of Universities for
Research in Astronomy, Inc., under NASA contract NAS 5-26555. These observations are associated with 
program \#12330. Support for program \#12330 was provided by NASA through a grant from the Space
Telescope Science Institute.
Some of the spectroscopic data presented herein were obtained at 
the W.M. Keck Observatory, which is operated as a scientific partnership among 
the California Institute of Technology, the University of California and the 
National Aeronautics and Space Administration. The Observatory was made 
possible by the generous financial support of the W.M. Keck Foundation.
In acknowledgement of our observing time at Keck and the IRTF, 
we further wish to recognize the very significant 
cultural role and reverence that the summit of Mauna Kea has always had within the indigenous Hawai'ian 
community. We are most fortunate to have the opportunity to conduct observations from this mountain.  
We acknowledge use of PAIRITEL, which is operated by the Smithsonian Astrophysical Observatory (SAO) 
and was made possible by a grant from the Harvard University Milton Fund, the camera loaned from the 
University of Virginia, and the continued support of the SAO and UC Berkeley. The PAIRITEL project is 
supported by NASA Grant NNG06GH50G. This paper also includes data gathered with the 6.5 m Magellan
Telescopes located at Las Campanas Observatory, Chile.

\vfill\eject

\appendix{\bf{Appendix:} Notes on Special Objects}

Notes are given below for objects with unusual spectra, spectrophotometric distance estimates placing them 
within 10 parsecs of the Sun, spectral types later than T9, or possible companionship with a nearby object
previously cataloged. In the sections below, the spectra of objects are assumed to be normal unless
peculiarities are specifically mentioned.

\section{WISEPC J003119.76$-$384036.4 ($J$=14.1 mag, W2=12.0 mag)}

The $J$-band spectrum of this object, which was earlier cataloged as SIPS J0031$-$3840 and
identified to be nearby star via its high proper motion by \cite{deacon2005},
is a good match to the spectrum of the L2 standard
(Figure~\ref{0031m3840}), but the spectrum is much bluer than the standard at $H$ and
$K$ bands. Because this source does not exhibit other telltale signs of low 
metallity, such as strong hydride bands, we classify it as an ``L2 pec (blue)''.
\cite{martin2010} give an optical spectral type of L2.5 and do not note any peculiarities
that might be attributable to low metallicity, either.
Its near-infrared spectral morphology and discrepancy relative to its nearest standard is most
similar to the ``L1 pec (sl.\ blue)'' object 2MASS J14403186$-$1303263 shown in 
Figure 32 of \cite{kirkpatrick2010}. Furthermore, this object has a larger W1-W2 color than a typical L2 
(see Figure~\ref{W1W2_vs_type}) and a bluer $J-H$ color (see Figure~\ref{JH_vs_type}).
As discussed in \cite{kirkpatrick2010}, the physical interpretation of these ``blue L
dwarfs'' is not fully known and may differ from object to object. Some blue L dwarfs
are blue, for example, because they are composte L + T dwarf binaries (e.g., \citealt{burgasser2007}).
Model fitting suggests that others have thin cloud decks and/or large grains in their
atmosphere, though neither seems to be directly attributable to gravity or metallicity
effects (\citealt{burgasser2008-2M1126}). 
Kinematic analysis by \cite{faherty2009} has shown that the blue L dwarfs
have kinematics older than the field L dwarf population, but not nearly as old as that
of low-metallicity M dwarfs. It is possible that some of the blue L dwarfs may be
slightly metal poor, and that even a subtle lowering of the metal abundance in these
objects may result in the directly measurable effects on the spectral energy distribution seen here.

\section{WISEPC J010637.07+151852.8 ($J$=14.4 mag, W2=12.7 mag)}

Despite the fact that the spectrum of this object matches very well to the $J$-band
spectrum of the M8 standard, the $H$-band spectrum is more peaked than that seen in a
normal M8, with the H$_2$O bands on either side of the $H$ peak being stronger than in
the standard (Figure~\ref{0106p1518}). This object, which we classify as ``M8 pec'', is similar to the peculiar late-M dwarf 
2MASS J18284076+1229207 shown in Figure 38 of \cite{kirkpatrick2010}. The cause of the peculiarity 
is not known, but appears not to be due to low gravity, as there are no peculiarities in the
strength of the FeH bands between 0.9 and 1.3 $\mu$m when compared to the M8 standard.
(See, for example, near-IR spectra of low-gravity late-M dwarfs in Figure 14 of 
\citealt{kirkpatrick2010}.) This object also has a larger W1-W2 color than a typical M8 
(see Figure~\ref{W1W2_vs_type}) and a bluer $J-H$ color (see Figure~\ref{JH_vs_type}).
Curiously, this object has a sizable motion -- $\mu$ = 0.412$\pm$0.006 arcsec/yr -- and
a tangential velocity of 85.5$\pm$1.3 km/s, suggesting that it may belong to an old population.
The peculiar spectroscopic features may be caused in part by a slightly subsolar metallicity.
\cite{deacon2009} also identified this object as the high motion source ULAS2MASS J0106+1518, and their proper motion
determination ($\mu = 0.407$ arcsec/yr)
agrees with the one we derive here.

\section{WISEPC J014807.25$-$720258.7 ($J$=19.0 mag, W2=14.6 mag)}

The near-infrared spectrum of this object, discussed in Cushing et al.\ (accepted), 
is distinctly later in type than the T9 near-infrared standard,
UGPS J072227.51$-$054031.2, and is therefore
classified as T9.5. Our spectrophotometric distance places it at 12.1 pc
(Table~\ref{astrometry}). This is the only one of our $\ge$T9.5 discoveries not detected in W3 and
along with WISE 1738+2732 is one of only two $\ge$T9.5 dwarfs detected in W1 (Table~\ref{WISE_photometry_discoveries}).

\section{WISEPA J020625.26+264023.6 ($J$=16.5 mag, W2=12.8 mag)}

The $J$-band spectrum of this object closely matches that of the L9 spectral standard, but
the $H$- and $K$-band portions are much redder than those of the standard L9 (Figure~\ref{0206p2640}).
This extremely red color is supported by independent photometry, namely $J-K_s$=2.007$\pm$0.137 mag, from the 
2MASS All-Sky Point Source Catalog. This color is somewhat redder than the mean $J-K_s$ color,
$\sim$1.78 mag, of very late L's (Figure 14 of \citealt{kirkpatrick2008}). Because of its spectral
peculiarity, we classify this object as ``L9 pec (red)'', and add it to the growing list of L dwarfs
that appear red for reasons not obviously attributable to low gravity (see Table 6 of \citealt{kirkpatrick2010}).
The underlying physical cause for these ``red L dwarfs'' is not known, although two have been studied in
detail by \cite{looper2008}. Using the small sample of red L dwarfs then known, \cite{kirkpatrick2010} 
found that these objects, unlike young, low-gravity
L dwarfs that are also redder than spectral standards of the same type, appear to have older kinematics than that of the field
L dwarf population.

\section{WISEPA J025409.45+022359.1 ($J$=15.9 mag, W2=12.7 mag)}

This object is a nearby T8 dwarf at a spectrophotometric distance of d=6.9 pc. Our astrometry (Table~\ref{astrometry})
over a 10.4yr baseline indicates a high motion of $\mu$ = 2.546$\pm$0.046 arcsec/yr and a large tangential velocity of 83.3$\pm$1.5 km/s.
Our preliminary trigonometic parallax measurement (Table~\ref{prelim_parallaxes}) places this object at
$6.1^{+2.3}_{-1.4}$ pc, in excellent agreement with the spectrophotomeric estimate.

\section{WISEPA J031325.96+780744.2 ($J$=17.7 mag, W2=13.2 mag)}

This T8.5 dwarf has a spectrophotometric distance estimate of only 8.1 pc.

\section{WISEPA J041022.71+150248.5 ($J$=19.3 mag, W2=14.2 mag)}

The near-infrared spectrum of this object, discussed in Cushing et al.\ (accepted), is 
classified as Y0. Our distance estimate places it 9.0 pc from the Sun, and our measurement of
the proper motion indicates that it may also be a high mover -- $\mu$ = 2.429$\pm$0.334 arcsec/yr -- 
although the error bar is large (Table~\ref{astrometry}). As with most of the other Y dwarf discoveries, this
object is detected by WISE only in bands W2 and W3 and not in W1 or W4 (Table~\ref{WISE_photometry_discoveries}).

\section{WISEPA J044853.29$-$193548.5 ($J$=17.0 mag, W2=14.2 mag)}

The depths of the H$_2$O and CH$_4$ absorption bands in the $J$- and $H$-band spectra of this
object best fit the T5 standard; however, the spectrum shows excess flux in the $Y$ band around
0.95 to 1.10 $\mu$m and a flattening of the entire $K$-band spectrum (Figure~\ref{0448m1935}). 
We therefore classify this object as a ``T5 pec''. Excess flux at $Y$ band and a flattening at $K$ 
have also been noted in
\cite{burgasser2006BBK} for the T6 pec dwarf 2MASS J09373487+2931409, which may be slightly 
metal-poor ($[M/H] \approx -0.5$ to $-0.1$) based on fits to model spectra.
\cite{burgasser2011} and \cite{burgasser2010} have noted the same
peculiarities in the spectrum of the T7.5 dwarf ULAS J141623.94+134836.3 (\citealt{scholz2010}; \citealt{burningham2010}), 
which is a common proper motion companion to the nearby, late-L dwarf SDSS J141624.08+134826.7 (\citealt{schmidt2010};
\citealt{bowler2010}). The latter is classified by \cite{kirkpatrick2010} as sdL7 and by \cite{burningham2010} 
as d/sdL7. Given that the two objects are presumably coeval, it can be assumed that they have the same metallicity and
that the peculiar features in the spectrum of ULAS J141623.94+134836.3 -- and by extension, WISE 0448$-$1935 --
are caused by a metal content below solar. The high motion of WISE 0448$-$1935 -- $\mu$ = 1.168$\pm$0.029 arcsec/yr,
which translates into a tangential velocity of 118.5$\pm$2.9 km/s for an estimated distance of 21.4 pc -- also indicates
that this object belongs to an old kinematic population.

\section{WISEPA J045853.89+643452.9 ($J$=18.3 mag, W2=13.0 mag)}

This object was discussed in detail by \cite{mainzer2011}. Our distance estimate of 7.3 pc assumes a single
source, but analysis of laser guide star adaptive optics data from \cite{gelino2011} indicates that the source is a double
in which the components have ${\Delta}J \approx {\Delta}H \approx 1$ mag. Individual magnitudes are measured
as $J_A = 17.50{\pm}0.09$, $J_B = 18.48{\pm}0.12$ and $H_A = 17.81{\pm}0.13$, $H_B = 18.81{\pm}0.17$ on
the MKO filter system. \cite{gelino2011}
suggest an actual distance to the system of 12.3$\pm$2.3 pc and individual spectral types of T8.5 and T9.

\section{WISEPA J052536.33+673952.3 ($J$=17.5 mag, W2=14.9 mag)}

The $J$-band spectrum of this object best fits the T6 standard, but there are
discrepancies at $Y$ and $K$ bands. At $Y$ band the spectrum of WISE 0525+6739 shows
excess flux relative to the standard, and at $K$ band the spectrum shows less flux
relative to the standard (Figure~\ref{0525p6739}). We therefore classify this object as a ``T6 pec''.
The physical cause, as discussed above for WISE 0448$-$1935, may be low metal content.

\section{WISEPA J052844.51$-$330823.9 ($J$=16.7 mag, W2=14.5 mag)}

The $J$-band spectrum of this object best fits the T7 standard, but there are
discrepancies at $Y$ and $K$ bands. At $Y$ band the spectrum of WISE 0528$-$3308 shows
excess flux relative to the standard, and at $K$ band the spectrum shows less flux
relative to the standard (Figure~\ref{0528m3308}). We therefore classify this object as a ``T7 pec''.
The physical cause, as discussed above for both WISE 0448$-$1935 and WISE 0525+6739, may be low metal content.

\section{WISEPC J083641.12$-$185947.2 ($J$=unknown, W2=15.0 mag)}

The $J$-band spectrum of this object best fits the T8 standard, but there are major
discrepancies at $Y$ and $K$ bands. At $Y$ band the spectrum of WISE 0836$-$1859 shows
excess flux relative to the standard, and at $K$ band the spectrum shows less flux
relative to the standard (Figure~\ref{0836m1859}). 
We therefore classify this object as a ``T8 pec''.
Several other objects discussed in this section -- WISE 0448$-$1935, WISE 0525+6739, 
WISE 0528$-$3308, WISE 1436$-$1814, WISE 2134$-$7137, and WISE 2325$-$4105 --
have similar $Y$- and $K$-band discrepancies, which may result from low metal content, but none are as severe as the discrepancies in
this object. Unlike in those objects, the $H$-band flux in WISE 0836$-$1859 is markedly lower than the closest matching standard at $J$,
making this object the most peculiar one of the group, and perhaps also the most metal poor.

Ideally, we would like to study the frequency of these metal-poor T dwarfs to see if the
numbers found are what star formation theory would predict for an old, field population.
Unfortunately, we do not have cold models across a large grid of metallicities with which
to determine the [M/H] values of our spectra. We suspect that in cool objects such as 
these, slight changes in the metal content can have profound effects on the emergent
spectra. As a result, we may be able to detect via spectroscopy smaller changes in
[M/H] for T dwarfs than are possible in hotter stars due to the richness of molecular
species and the important role of condensation in determining the absorbing species of
colder objects. In other words, the metal content of these T dwarfs may not be 
too different from solar. As discussed above for WISE 0448$-$1935, the T7.5 dwarf ULAS J141623.94+134836.3
shows peculiarities in its spectrum that are similar to the ones seen in these unusual WISE T dwarfs,
yet models with a subsolar abundance of only [M/H] = $-$0.3 provide good fits to the emergent spectrum
of that object (\citealt{burgasser2010}).

\section{WISEPC J112254.73+255021.5 ($J$=16.7 mag, W2=14.0 mag)}

This object, a normal T6 dwarf, lies 265 arcseconds away from the nearby M5 V star LHS 302 (GJ 3657). Our
spectrophotometric distance estimate for WISE 1122+2550 (16.9 pc; Table~\ref{astrometry}) is very similar
to the distance of 17.2 pc obtained via trigonometric parallax (0.0581$\pm$0.0039 arcsec) for LHS 302 (\citealt{dahn1988}).
Moreover, the Right Ascension and Declination components of the proper motion of WISE 1122+2550 are measured by us to
be $-$0.954$\pm$0.016 and $-$0.276$\pm$0.018 arcsec/yr, respectively, which are only $\sim$3$\sigma$ different from those measured for
LHS 302 ($-$1.002$\pm$0.001 arcsec/yr and $-$0.330$\pm$0.001 arcsec/yr; \citealt{dahn1988}).
If these two objects are a common proper motion binary, the projected separation between them is $\sim$4500 AU.

Other stellar + substellar binaries of large separation are known. Examples are the Gliese 570 system comprised of K4 V, M1.5 V + M3 V, and T7.5
componenets with the latter having a projected separation of 1500 AU from the K star;
the Gliese 584 system comprised of 
G1 V + G3 V and L8 components
with a projected separation of 3600 AU (\citealt{kirkpatrick2001}); the Gliese 417 system comprised of G0 V and L4.5 components
with a projected separation of 2000 AU (\citealt{kirkpatrick2001}); the Gliese 618.1 system comprised of M0 V and L2.5 components
with a projected separation of 1000 AU (\citealt{wilson2001}); the HD 89744 system comprised of F7 IV-V and L0 components
with a projected separation of 2500 AU (\citealt{wilson2001}); and the HD 2057 system comprised of F8 and L4 components
with a projected separation of 7000-9000 AU (\citealt{cruz2007}). Each of these systems, however, has a more massive
primary than the one discussed here. Assuming that LHS 302 has a mass of $\sim$0.2 $M_{\odot}$ in concert with other M5 dwarfs (see
\citealt{lopez-morales2007}), then our projected separation of 4500 AU falls well outside the 
$\Delta_{max} = 10^{3.33M_{tot} + 1.1} = 58$ AU stability limit suggested empirically by \cite{reid2001}. This suggests that
WISE 1122+2550 and LHS 302 are either physically unbound while sharing a common proper motion or are totally unassociated.

\section{WISEPC J115013.88+630240.7 ($J$=17.7 mag, W2=13.4 mag)}

Our distance estimate places this T8 dwarf 9.6 pc from the Sun.

\section{WISEPC J121756.91+162640.2 ($J$=17.8 mag, W2=13.1 mag)}

Our distance estimate places this T9 dwarf 6.7 pc from the Sun. Our measurement of
$\mu$=1.765$\pm$0.388 arcsec/yr, based on astrometry covering only 0.7 yr, may also indicate a high proper motion, but the 
uncertainty in this value is very large.

\section{WISEPC J131141.91+362925.2 ($J$=15.5 mag, W2=13.1 mag)}

The near-infrared spectrum of this source is an excellent match to the L5 standard at
$J$ band. At longer wavelengths, however, the spectrum is considerably bluer, as shown
in Figure~\ref{1311p3629}. There is no evidence in the $J$-band that this source has a
low-metallicity, and therefore that the $H$ and $K$ bands are being suppressed by the
relatively stronger collision-induced absorption by H$_2$ one would expect in a metal-starved atmosphere.
The notch near 1.62 $\mu$m in the top of the $H$ band peak is very similar to the
interesting feature noted by \cite{burgasser2007} in the spectrum of SDSS J080531.84+481233.0,
which those authors claim is an unresolved mid-L + mid-T binary. The $H$-band notch is also seen by \cite{burgasser2011b}
in the spectrum of 2MASS J1315$-$2649, which those authors have successfully split, via
high-resolution imaging and spectroscopy, into an L5 + T7 double. We
classify WISE 1311+3629 as an ``L5 blue'' and note that its peculiar features may be caused by
unresolved binarity as well. Our formal fits to synthetic binaries (see \citealt{burgasser2007} for details)
show that the most likely spectral types of the two components are L3.5$\pm$0.7 and T2$\pm$0.5 
with estimated relative magnitudes of $\Delta{J} = 1.4{\pm}0.2$, $\Delta{H} = 1.7{\pm}0.3$, and 
$\Delta{K} = 2.2{\pm}0.3$ mag.
This object would be an excellent target for high-resolution imaging.
We note that this object was also identified as a brown dwarf candidate by \cite{zhang2009} and
given the designation SDSS J131142.11+362923.9.

\section{WISEPC J140518.40+553421.4 ($J$=20.2 mag, W2=14.1 mag)}

This object is tentatively classfied as Y0 (pec?) by Cushing et al.\ (accepted), who describe its spectral features and 
derived physical parameters. We estimate a distance of 8.6 pc and
find a high proper motion of 2.693$\pm$0.398 arcsec/yr and high tangential
velocity of 109.8$\pm$16.2 km/s. As with most of the other Y dwarf discoveries, this
object is detected by WISE only in bands W2 and W3 and not in W1 or W4 (Table~\ref{WISE_photometry_discoveries}).

\section{WISEPA J143602.19$-$181421.8 ($J$=unknown, W2=14.7 mag)}

The $J$-band spectrum of this object best fits the T8 standard, but there are
discrepancies at $Y$ and $K$ bands. At $Y$ band the spectrum of WISE 1436$-$1814 shows
excess flux relative to the standard, and at $K$ band the spectrum shows less flux
relative to the standard (Figure~\ref{1436m1814}). We therefore classify this object as a ``T8 pec''.
The physical cause, as discussed above for several other objects, may be low metal content.

\section{WISEPC J150649.97+702736.0 ($J$=13.6 mag, W2=11.3 mag)}

This T6 dwarf is estimated to fall only 4.9 pc from the Sun. Its large motion ($\mu$ =
1.388$\pm$0.131 arcsec/yr) placed it nearly in front of a star, now to the southeast, of similar near-infrared
brightness during the 2MASS survey (see Figure~\ref{finder_chart12}), and this confusion led to the source having been
missed in photometric searches of the 2MASS Point Source Catalog.

\section{WISEPA J154151.66$-$225025.2 ($J$=21.2 mag, W2=14.0 mag)}

The near-infrared spectrum of this object, discussed in Cushing et al.\ (accepted), is 
classified as Y0 because of its similarity to the spectrum of the Y0 near-infrared standard,
WISE 1738+2732. As with most of the other Y dwarf discoveries, this
object is detected by WISE only in bands W2 and W3 and not in W1 or W4 (Table~\ref{WISE_photometry_discoveries}).
In contrast to our crude spectrophotometric distance estimate from Table~\ref{astrometry} of
8.2 pc,
we measure a trignometric parallax
placing it at $2.8^{+1.3}_{-0.6}$ pc (Table~\ref{prelim_parallaxes}), along with
a proper motion of $\mu = 0.81{\pm}0.34$ arcsec/yr. 
This parallax result is
significant only at the 3$\sigma$ level and is measured only over a 1.2yr baseline, so it should be treated 
as preliminary only. Nonetheless, if
confirmed, this distance would place WISE 1541$-$2250 as the seventh closest stellar system to the Sun 
after the $\alpha$ Centauri system (d=1.3 pc, \citealt{vanleeuwen2007}),
Barnard's Star (d=1.8 pc, \citealt{vanleeuwen2007}),
Wolf 359 (d=2.4 pc, \citealt{vanaltena2001}),
Lalande 21185 (d=2.5 pc, \citealt{vanleeuwen2007}),
Sirius AB (d=2.6 pc, \citealt{vanleeuwen2007}),
and L 726-8 AB (also known as BL Ceti and UV Ceti; d=2.7 pc, \citealt{vanaltena2001}).
The measured distance implies
absolute magnitudes of $M_J = 23.9{\pm}0.8$ and $M_H = 23.8{\pm}0.9$ mag on the MKO filter system
and $M_{W2} = 16.7{\pm}0.7$ mag. This indicates a rapid dimming at those wavelengths in just a single 
spectral subclass from T9 to Y0 (see, e.g., Figure~\ref{MW2_vs_type} and Figure~\ref{MH_vs_type}).

\section{WISEPA J164715.59+563208.2 ($J$=16.6 mag, W2=13.1 mag)}

The $J$-band spectrum of this object best fits the L9 standard, but there is excess
flux at $H$ and particularly $K$ relative to the standard itself (Figure~\ref{1647p5632}).
We therefore classify this object as an ``L9 pec (red)''. This object adds to a growing
list of red L dwarfs whose red colors cannot obviously  be attributed to low gravity.
It becomes the seventh example of this class, which now includes 2MASS J21481633+4003594
and 2MASS J18212815+1414010 from \cite{looper2008}; 2MASS J13313310+3407583,
2MASS J23174712$-$4838501, and 2MASS J23512200+3010540 from \cite{kirkpatrick2010};
and WISE 0206+2640 from above. As mentioned in \cite{kirkpatrick2010}, the kinematics of the
first five examples suggest that these objects derive from an old population, making them
distinct from the red L dwarfs that have low-gravity spectral signatures and young kinematics.
We note, however, that this object has a low tangential velocity of only 28.1$\pm$1.2 km/s.
However, this velocity assumes our spectrophotometric distance estimate of 20.2 pc, and
our preliminary astrometric measurements over an 11.8yr baseline indicate a closer distance of 
$8.6^{+2.9}_{-1.7}$ pc, along with a motion of $\mu = 0.293{\pm}0.012$ arcsec/yr. 
Continued astrometric monitoring of this object is needed to
see if this closer distance is confirmed, as this would provide another clue in
deciphering the physical nature of this rare class of red (non low-$g$) L dwarfs.

\section{WISEPA J173835.53+273258.9 ($J$=19.5 mag, W2=14.5 mag)}

Cushing et al.\ (accepted) propose this object as the Y0 spectroscopic standard.
It is the only one of our Y
dwarfs detected in all three short-wavelength bands of WISE (W1, W2, and W3).
Our spectrophotometric distance estimate places it at 10.5 pc; the available
astrometry for this Y0 dwarf spans barely six months (Table~\ref{astrometry}), so we
are not yet able to derive proper motion or parallax. 

\section{WISEPA J174124.26+255319.5 ($J$=16.5 mag, W2=12.3 mag)} 

This nearby dwarf of near-infrared spectral type T9 is detected in the 2MASS and SDSS surveys but was overlooked
because of its weak detection in both. Using a 10.4yr baseline, we find a high proper motion of $\mu$ = 1.555$\pm$0.023
arcsec/yr and estimate a distance of 4.7 pc. Our preliminary trigonometric parallax measurement places
it at $5.5^{+1.4}_{-1.0}$ pc (Table~\ref{prelim_parallaxes}). This object has identical optical and near-infrared
types to another nearby object, UGPS J072227.51$-$054031.2, whose trigonometric parallax from \cite{lucas2010}
places it at a distance of 4.1 pc. \cite{gelino2011} note that WISE 1741+2553 appears single in 
near-infared imaging observations with laser guide star adaptive optics.

\section{WISEPA J180435.40+311706.1 ($J$=18.7 mag, W2=14.7 mag)}

The near-infrared spectrum of this object is 
classified as T9.5: because of its similiarity to the spectrum of the T9.5 dwarf
WISE 0148$-$7202. Although noisy, the narrowness of the J-band peaks falls intermediate
between that of the T9 standard UGPS J072227.51$-$054031.2 and the Y0 standard
WISE 1738+2732. We estimate that this object falls at a distance
of 13.0 pc. The WISE detections for this object are very similar to those seen for Y
dwarfs; namely, the object is detected only in bands W2 and W3 and not in W1 or W4
(Table~\ref{WISE_photometry_discoveries}).

\section{WISEPA J182831.08+265037.8 ($J$=23.6 mag, W2=14.3 mag)}

This object, with a tentative classification of $>$Y0 from Cushing et al.\ (accepted), is the
latest object so far found with WISE. The spectrum is unique among late-T and Y dwarfs in
that the $J$- and $H$-band peaks, in units of $f_\lambda$, are nearly the same height (Figure~\ref{other_spectra_gt_T8}).
This reddening of the near-infrared colors ($J-H = 0.72{\pm}0.42$ mag; Table~\ref{followup_photometry_discoveries})
is predicted by model atmosphere calculations to occur
at effective temperatures below 300-400K (\citealt{burrows2003}). This effect is due to the fact that the Wien tail of the
spectral energy distribution becomes the overwhelming effect shaping the
spectrum at those wavelengths, and this may be even more dramatically illustrated by the extremely red $J-W2$ and 
$H-W2$ colors measured for this object (Figure~\ref{JW2_vs_type} and Figure~\ref{HW2_vs_type}). Our spectrophotometric
distance estimate of $<$9.4 pc implies exceedingly dim absolute magnitudes of $M_J > 23.7$ 
and $M_H > 23.0$ mag, on the MKO filter system. These values agree with expectations that WISE 1828+2650 should
be dimmer at these wavelengths than the presumably warmer WISE 1541$-$2250, which also has exceedingly dim $J$
and $H$ magnitudes (Table~\ref{prelim_parallaxes}).

\section{WISEPC J205628.90+145953.3 ($J$=19.2 mag, W2=13.9 mag)}

This Y0 dwarf, discussed in Cushing et al. (accepted), is estimated to lie at a distance of 7.7 pc.
As with several other Y dwarf discoveries, this
object is detected by WISE only in bands W2 and W3 and not in W1 or W4 (Table~\ref{WISE_photometry_discoveries}).

\noindent{\bf{AA. WISEPA J213456.73$-$713743.6 ($J$=19.8 mag, W2=13.9 mag)}}

The $J$-band spectrum of this object best fits the T9 standard, but there are
discrepancies at $Y$ and $K$ bands. At $Y$ band the spectrum of WISE 2134$-$7137 shows
excess flux relative to the standard, and at $K$ band the spectrum shows less flux
relative to the standard (Figure~\ref{2134m7137}). We therefore classify this object as a ``T9 pec''.
The physical cause, as discussed above for several other objects, may be low metal content.

\noindent{\bf{BB. WISEPC J232519.54$-$410534.9 ($J$=19.7 mag, W2=14.1 mag)}}

As with the previous object, the $J$-band spectrum best fits the T9 standard, but there are
discrepancies at $Y$ and $K$ bands. At $Y$ band the spectrum of WISE 2325$-$4105 shows
excess flux relative to the T9 standard, and at $K$ band the spectrum shows less flux
relative to the T9 standard (Figure~\ref{2325m4105}). We therefore classify this object as a ``T9 pec''.
The physical cause, as discussed above for other objects, may be low metal content.

\noindent {\bf{CC. WISEPC J232728.75$-$273056.5 ($J$=16.7 mag, W2=13.2 mag)}}

\begin{figure}
\epsscale{0.8}
\figurenum{1}
\plotone{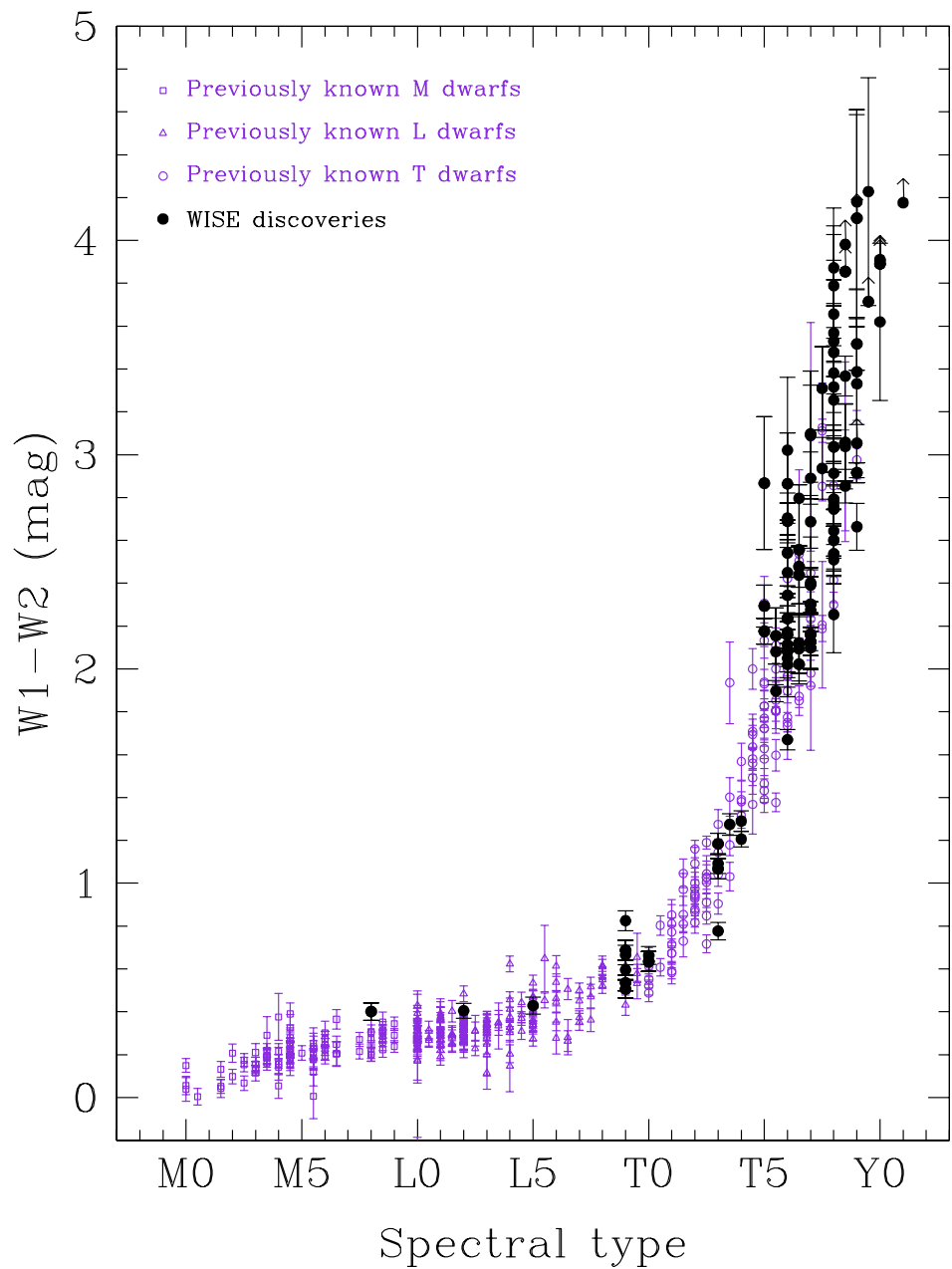}
\caption{WISE W1-W2 color versus spectral type. The W1-W2 colors for a sample of previously known M 
(open squares), L (open triangles) and T (open circles) dwarfs from Table 1 are shown in blue violet. 
The colors of new WISE discoveries from Table 2 are shown by the solid, black circles. W1-W2 color 
limits are indicated by arrows. 
\label{W1W2_vs_type}}
\end{figure}

\begin{figure}
\epsscale{1.0}
\figurenum{2}
\plotone{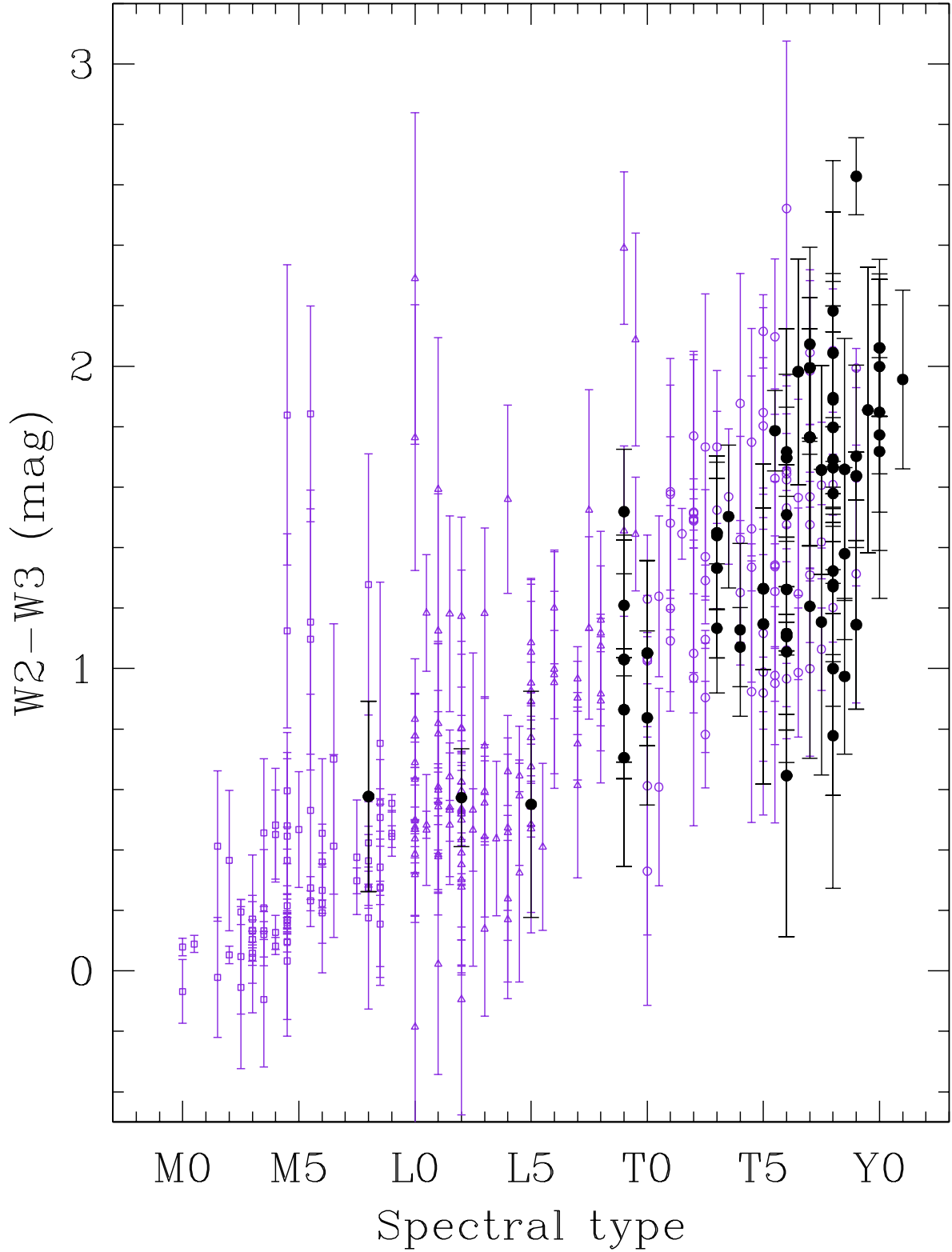}
\caption{WISE W2-W3 color versus spectral type. The color scheme is identical to that of 
Figure~\ref{W1W2_vs_type}. For clarity, only those objects with detections in both W2 and W3 are shown.
\label{W2W3_vs_type}}
\end{figure}

\begin{figure}
\epsscale{0.9}
\figurenum{3}
\plotone{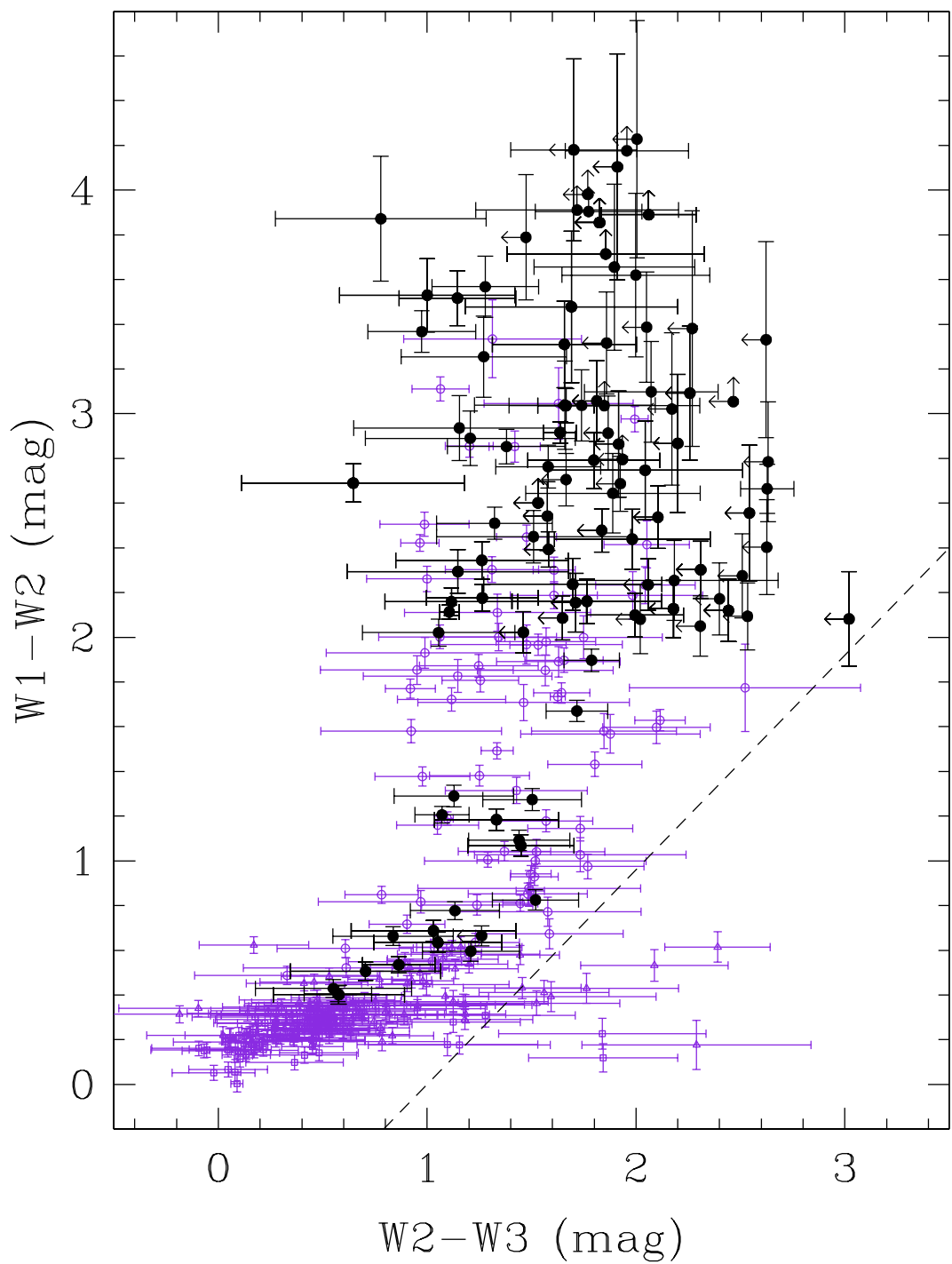}
\caption{WISE color-color plot showing W1-W2 versus W2-W3. Color coding is the same as in
Figure~\ref{W1W2_vs_type}. The dashed line indicates the criterion used, in search \#2
(see section 2.2), to eliminate extragalactic sources to the right of the line from the bulk of the M, L, and T dwarfs to the left.
\label{W1W2_vs_W2W3}}
\end{figure}

This object has the near-infrared spectrum of a normal L9 dwarf, but its WISE color of 
W1-W2 = 0.825$\pm$0.046 is markedly redder than the other L9 dwarfs on Figure~\ref{W1W2_vs_type}.
The {\it Spitzer}/IRAC color of ch1-ch2 = 0.247$\pm$0.025
is redder than all other L and early-T dwarfs in Figure~\ref{ch1ch2_vs_type}. 
Effects such as low-gravity or low-metallicity would cause the near-infrared spectrum of this
object to appear unusually red or unusually blue, respectively, in the near-infrared (\citealt{kirkpatrick2010}),
and this is not seen.
One hypothesis is that this object is an unresolved L+T binary, but not one with such a warm T dwarf
that the near-infrared spectrum of the composite shows itself to be peculiar. We can test this
as follows. The W1-W2 color of WISE 2327$-$2730 is $\sim$0.23 mag redder than the mean W1-W2 for other L9 dwarfs.
Using the absolute W2 versus spectral type plot of
Figure~\ref{MW2_vs_type} along with the trend of W1-W2 color with spectral
type in Figure~\ref{W1W2_vs_type}, we estimate that the type of the hypothesized companion would have to 
be roughly T6.5. However, as Figure 4 of \cite{burgasser2007} shows, an object with a composite type
of $\sim$L9 and secondary of $\sim$T6.5 would have a noticeably peculiar near-infrared spectrum
which would distinguish it from a normal L9. Hence, binarity appears not to be the cause of
the redder W1-W2 and ch1-ch2 colors. The reason for this color peculiarity remains unexplained.

\clearpage

\begin{deluxetable}{lllcccccccc}
\tabletypesize{\tiny}
\rotate
\tablewidth{9in}
\tablenum{1}
\tablecaption{WISE and Near-infrared Photometry for Known M, L, and T Dwarfs\label{photometry_knownMLT}}
\tablehead{
\colhead{WISE Designation\tablenotemark{a}} &
\colhead{Other Designation} &
\colhead{Dis.} &
\colhead{W1} &
\colhead{W2} &
\colhead{W3} &
\colhead{W4} &
\colhead{J} &
\colhead{H} &
\colhead{K$_s$} &
\colhead{Spec.} \\
\colhead{} &
\colhead{} &
\colhead{Ref.} &
\colhead{(mag)} &
\colhead{(mag)} &
\colhead{(mag)} &
\colhead{(mag)} &
\colhead{(mag)} &
\colhead{(mag)} &
\colhead{(mag)} &
\colhead{Typ.\tablenotemark{b}} \\
\colhead{(1)} &
\colhead{(2)} &
\colhead{(3)} &
\colhead{(4)} &
\colhead{(5)} &
\colhead{(6)} &
\colhead{(7)} &
\colhead{(8)} &
\colhead{(9)} &
\colhead{(10)} &
\colhead{(11)}
}
\startdata
\multicolumn{10}{c}{T Dwarfs:} \\
WISEPC J003402.80-005207.4& ULAS J003402.77-005206.7  &  1 & 17.320$\pm$0.249&  14.465$\pm$0.076&  $>$11.801       &   $>$9.224       &    18.150$\pm$0.030&     18.490$\pm$0.040&     18.480$\pm$0.050&            T8.5  \\      
WISEPC J005021.03-332229.2& 2MASS J00501994-3322402   &  2 & 15.506$\pm$0.050&  13.526$\pm$0.036&  11.957$\pm$0.236&   $>$8.989       &    15.928$\pm$0.070&     15.838$\pm$0.191&     15.241$\pm$0.185&             T7   \\   
WISEPC J005911.09-011400.6& CFBDS J005910.90-011401.3 &  3 & 17.003$\pm$0.169&  13.668$\pm$0.044&  12.355$\pm$0.424&   $>$9.290       &    18.060$\pm$0.030&     18.270$\pm$0.050&     18.630$\pm$0.050&             T9   \\   
WISEPC J013657.45+093347.0& IPMS J013656.57+093347.3  &  4 & 11.967$\pm$0.025&  10.962$\pm$0.022&   9.671$\pm$0.047&   9.002$\pm$0.442&    13.455$\pm$0.030&     12.771$\pm$0.032&     12.562$\pm$0.024&            T2.5  \\      
WISEPA J015024.39+135924.3& ULAS J015024.37+135924.0  &  5 & 17.392$\pm$0.265&  15.186$\pm$0.131&  $>$12.108       &   $>$8.942       &    17.730$\pm$0.020&     18.110$\pm$0.020&     17.840$\pm$0.160&            T7.5  \\   
WISEPA J015142.21+124429.8& SDSS J015141.69+124429.6  &  6 & 14.595$\pm$0.039&  13.823$\pm$0.053&  12.246$\pm$0.445&   8.563$\pm$0.383&    16.566$\pm$0.129&     15.603$\pm$0.112&     15.183$\pm$0.189&             T1   \\      
WISEPC J020742.96+000056.9& SDSS J020742.48+000056.2  &  6 & 16.403$\pm$0.097&  15.035$\pm$0.100&  $>$12.618       &   $>$9.231       &    16.799$\pm$0.156&     $>$16.396       &     $>$15.412       &            T4.5  \\
\enddata
\end{deluxetable}

\clearpage

\begin{deluxetable}{cccrccccccc}
\tabletypesize{\tiny}
\rotate
\tablewidth{8in}
\tablenum{2}
\tablecaption{WISE Photometry for WISE Brown Dwarf Discoveries\label{WISE_photometry_discoveries}}
\tablehead{
\colhead{Object Name\tablenotemark{a}} &
\colhead{Disc.} &
\colhead{\it{b}} &
\colhead{W1} &
\colhead{W2} &
\colhead{W3} &
\colhead{W4} &
\colhead{W1-W2} &
\colhead{W2-W3} &
\colhead{No.\ of WISE} \\
\colhead{} &
\colhead{Ref.} &
\colhead{(deg)} &
\colhead{(mag)} &
\colhead{(mag)} &
\colhead{(mag)} &
\colhead{(mag)} &
\colhead{(mag)} &
\colhead{(mag)} &
\colhead{Coverages} \\
\colhead{(1)} &
\colhead{(2)} &
\colhead{(3)} &
\colhead{(4)} &
\colhead{(5)} &
\colhead{(6)} &
\colhead{(7)} &
\colhead{(8)} &
\colhead{(9)} &
\colhead{(10)}  
}
\startdata
WISEPC J000849.76$-$173922.6&   1&  $-$76.3&  16.593$\pm$0.114&  14.543$\pm$0.072&  $>$12.235       &   8.928$\pm$0.381&  2.050$\pm$0.135& $<$2.308       & 13 \\
WISEPC J003119.76$-$384036.4\tablenotemark{b}&   1&  $-$77.7&  12.433$\pm$0.026&  12.028$\pm$0.024&  11.455$\pm$0.160&   $>$8.763       &  0.405$\pm$0.035& 0.574$\pm$0.162& 12 \\
WISEPC J004928.48+044100.1  &   1&  $-$58.1&  13.448$\pm$0.028&  12.942$\pm$0.030&  12.237$\pm$0.359&   8.677$\pm$0.286&  0.506$\pm$0.041& 0.705$\pm$0.360& 14 \\
WISEPC J010637.07+151852.8\tablenotemark{c}  &   1&  $-$47.3&  13.088$\pm$0.030&  12.687$\pm$0.028&  12.110$\pm$0.313&   9.105$\pm$0.465&  0.401$\pm$0.041& 0.577$\pm$0.314& 13 \\
WISEPA J012333.21+414203.9  &   1&  $-$20.7&  17.123$\pm$0.168&  14.848$\pm$0.086&  $>$12.339       &   $>$8.839       &  2.275$\pm$0.189& $<$2.509       & 13 \\
WISEPC J013836.59$-$032221.2&   1&  $-$63.7&  14.427$\pm$0.034&  13.359$\pm$0.034&  11.910$\pm$0.252&   $>$8.957       &  1.068$\pm$0.048& 1.449$\pm$0.254& 14 \\
WISEPC J014807.25$-$720258.7& 1,5&  $-$44.4&  18.812$\pm$0.529&  14.584$\pm$0.051&  $>$12.579       &   $>$9.521       &  4.228$\pm$0.531& $<$2.005       & 24 \\
WISEPA J015010.86+382724.3  &   1&  $-$22.9&  13.619$\pm$0.028&  12.984$\pm$0.036&  11.933$\pm$0.305&   $>$8.779       &  0.635$\pm$0.046& 1.051$\pm$0.307& 10 \\
\enddata
\tablecomments{Discovery Reference: (1) this paper, (2) \citealt{mainzer2011}, (3) \citealt{burgasser2011}, (4) Gelino et al., submitted, (5) Cushing et al., accepted}
\tablenotetext{a}{WISE sources are given designations as follows. The prefix is ``WISE'' followed by either ``PC'' for sources taken from the first-pass precessing operations coadd Source
Working Database, or ``PA'' for objects drawn from the preliminary release Atlas Tile Source Working Database. The suffix is the J2000 position of the source in the format Jhhmmss.ss$\pm$ddmmss.s.
As stated in section 5.2, the positions measured in first-pass WISE processing and used to derive these designations should not be used for astrometric purposes. Instead, refer to the
re-measured astrometry given in Table~\ref{astrometry}.}
\tablenotetext{b}{Identified as a high-motion object by \cite{deacon2005} and classified in the optical as an early-L dwarf by \cite{martin2010}. Alternate name is SIPS J0031$-$3840.}
\tablenotetext{c}{Identified as a high-motion object by \cite{deacon2009}. Alternate name is ULAS2MASS J0106+1518.}
\tablenotetext{d}{Identified as a brown dwarf candidate by \cite{zhang2009}. Alternate name is SDSS J131142.11+362923.9.}
\end{deluxetable}

\clearpage

\begin{deluxetable}{lcllllllllll}
\tabletypesize{\tiny}
\rotate
\tablewidth{9in}
\tablenum{3}
\tablecaption{Follow-up Photometry of WISE Brown Dwarf Discoveries\label{followup_photometry_discoveries}}
\tablehead{
\colhead{Object} &
\colhead{Obs.} &
\colhead{Y} &
\multicolumn{3}{c}{Two Micron All-Sky Survey filter system} &
\multicolumn{3}{c}{Mauna Kea Observatories filter system} &
\multicolumn{3}{c}{Spitzer/IRAC observations} \\
\colhead{Name} &
\colhead{Ref.} &
\colhead{(mag)} &
\colhead{J} &
\colhead{H} &
\colhead{K$_s$} &
\colhead{J} &
\colhead{H} &
\colhead{K$_s$} &
\colhead{ch1} &
\colhead{ch2} &
\colhead{ch1-ch2} \\
\colhead{} &
\colhead{} &
\colhead{} &
\colhead{(mag)} &
\colhead{(mag)} &
\colhead{(mag)} &
\colhead{(mag)} &
\colhead{(mag)} &
\colhead{(mag)} &
\colhead{(mag)} &
\colhead{(mag)} &
\colhead{(mag)} \\
\colhead{(1)} &
\colhead{(2)} &
\colhead{(3)} &
\colhead{(4)} &
\colhead{(5)} &
\colhead{(6)} &
\colhead{(7)} &
\colhead{(8)} &
\colhead{(9)} &
\colhead{(10)}&
\colhead{(11)}&
\colhead{(12)} 
}
\startdata
WISE J0008$-$1739&  4 &      ---            &    17.03$\pm$0.06  &    17.77$\pm$0.13  &    $>$17.0          &    ---             &    ---             &    ---              &    16.044$\pm$0.030 & 14.867$\pm$0.021 & 1.177$\pm$0.037 \\
WISE J0031$-$3840&  1 &      ---            &    14.101$\pm$0.032&    13.399$\pm$0.023&    12.924$\pm$0.034 &    ---             &    ---             &    ---              &    extant           & ---              & ---             \\
WISE J0049+0441  &  1 &      ---            &    15.854$\pm$0.067&    14.674$\pm$0.068&    14.170$\pm$0.066 &    ---             &    ---             &    ---              &    12.977$\pm$0.017 & 12.950$\pm$0.017 & 0.027$\pm$0.024 \\
                 &  3 &      ---            &    15.85$\pm$0.05  &    14.77$\pm$0.07  &    14.17$\pm$0.04   &    ---             &    ---             &    ---              &    ---              & ---              & ---             \\
WISE J0106+1518  &  1 &      ---            &    14.358$\pm$0.031&    13.815$\pm$0.033&    13.434$\pm$0.053 &    ---             &    ---             &    ---              &    12.862$\pm$0.017 & 12.716$\pm$0.017 & 0.146$\pm$0.024 \\
                 & 10 &    15.095$\pm$0.004 &    ---             &    ---             &    ---              &    14.277$\pm$0.003&    13.867$\pm$0.004&   13.363$\pm$0.004  &    ---              & ---              & ---             \\
WISE J0123+4142  &  3 &      ---            &    17.38$\pm$0.11  &    17.20$\pm$0.13  &    18.37$\pm$0.37   &    ---             &    ---             &    ---              &    16.125$\pm$0.032 & 14.845$\pm$0.021 & 1.280$\pm$0.038 \\
WISE J0138$-$0322&  1 &      ---            &    16.389$\pm$0.096&    15.801$\pm$0.147&    15.198$\pm$0.129 &    ---             &    ---             &    ---              &    13.888$\pm$0.018 & 13.426$\pm$0.018 & 0.461$\pm$0.026 \\
\enddata
\tablecomments{Ref.\ for JHK instrument or catalog: 
2MASS filter system: 
(1) 2MASS All-Sky Point Source Catalog,
(2) 2MASS Reject Table,
(3) Bigelow/2MASS,
(4) PAIRITEL,
(5) Shane/Gemini,
(6) FanMt./FanCam;
MKO filter system:
(7) Palomar/WIRC,  
(8) Magellan/PANIC,
(9) SOAR/SpartanIRC,
(10) UKIDSS,
(11) Keck/NIRC2.
(12) CTIO-4m/NEWFIRM}                       
\end{deluxetable}

\clearpage

\begin{deluxetable}{lcccccc}
\tabletypesize{\scriptsize}
\tablewidth{6in}
\tablenum{4}
\tablecaption{Follow-up Spectroscopy of WISE Brown Dwarf Discoveries\label{spectral_types}}
\tablehead{
\colhead{Object Name} &
\colhead{Opt.\ Sp.} &
\colhead{NIR Sp.} &
\colhead{Spectrograph} &
\colhead{Int.\ Time} &
\colhead{Tell.\ Corr.} &
\colhead{Obs.\ Date} \\
\colhead{and J2000 Coordinates} &
\colhead{Type\tablenotemark{c}} &
\colhead{Type\tablenotemark{c}} &
\colhead{} &
\colhead{(s)} &
\colhead{Star\tablenotemark{e}} &
\colhead{(UT)} \\
\colhead{(1)} &
\colhead{(2)} &
\colhead{(3)} &
\colhead{(4)} &
\colhead{(5)} &
\colhead{(6)} &
\colhead{(7)} 
}
\startdata
WISE J0008$-$1739 & ---  &  T6            & Keck/NIRSPEC-J& 1200 & HD 13936    & 2010 Dec 18 \\
WISE J0031$-$3840 & ---  &  L2 pec (blue) & IRTF/SpeX     &  720 & HD 4065     & 2010 Nov 17 \\
WISE J0049+0441   & ---  &  L9            & IRTF/SpeX     & 1200 & HD 7215     & 2010 Sep 14 \\
WISE J0106+1518   & ---  &  M8 pec        & IRTF/SpeX     &  960 & HD 7215     & 2010 Sep 14 \\
WISE J0123+4142   & ---  &  T7            & Keck/NIRSPEC-J&  600 & HD 13936    & 2010 Dec 19 \\
WISE J0138$-$0322 & ---  &  T3            & IRTF/SpeX     &  960 & HD 13936    & 2010 Aug 17 \\
WISE J0148$-$7202 & ---  &  T9.5          & Magellan/FIRE &  960 & HD 1881     & 2010 Sep 18 \\
WISE J0150+3827   & ---  &  T0            & IRTF/SpeX     & 1920 & HD 19600    & 2010 Sep 12 \\
\enddata
\tablenotetext{a}{See Mainzer et al.\ (2010) for details on spectroscopic follow-up.}
\tablenotetext{b}{See Burgasser et al.\ (submitted) for details on spectroscopic follow-up.
A newer spectrum of this object from Magellan/FIRE supports a type of T6.5 rather than
T5.5 as originally assigned by Burgasser et al.\ (submitted).}
\tablenotetext{c}{Spectral types are accurate to $\pm$0.5 type unless indicated by a ``:'' symbol to indicate a more uncertain assignment.}
\tablenotetext{d}{Due to closing of the telescope due to fog, no A0 V star was taken on this night 
so an A0 V observation from an earlier run on 2010 Jul 18 was used.}
\tablenotetext{e}{Telluric corrector stars for optical spectral are G0 dwarfs; those for near-infrared
spectra are A0 dwarfs.}
\end{deluxetable}

\clearpage

\begin{deluxetable}{lccccc}
\tabletypesize{\scriptsize}
\tablewidth{6in}
\tablenum{5}
\tablecaption{L and T Dwarfs with Measured Trigonometric Parallaxes\label{abs_mags}}
\tablehead{
\colhead{Object Name} &
\colhead{$\pi_{trig}$} &
\colhead{Ref.} &
\colhead{NIR Sp.} &
\colhead{W2} &
\colhead{$H$}\\
\colhead{} &
\colhead{(mas)} &
\colhead{} &
\colhead{Type} &
\colhead{(mag)} &
\colhead{(mag)} \\
\colhead{(1)} &
\colhead{(2)} &
\colhead{(3)} &
\colhead{(4)} &
\colhead{(5)} &
\colhead{(6)} 
}
\startdata
a) Objects not known to be binary \\
UGPS J072227.51$-$054031.2  &       246$\pm$33      &   2  &  T9   &  12.197$\pm$0.027&  16.147$\pm$0.205 \\
DENIS J081730.0$-$615520    &       203$\pm$13      &   3  &  T6   &  11.237$\pm$0.017&  13.526$\pm$0.031 \\
DENIS-P J0255$-$4700        &       201.37$\pm$3.89 &   4  &  L8   &  10.190$\pm$0.021&  12.204$\pm$0.024 \\
2MASSI J0415195$-$093506    &       174.34$\pm$2.76 &   5  &  T8   &  12.232$\pm$0.026&  15.537$\pm$0.113 \\
Gliese 570D                 &       169.3$\pm$1.7   &   6  &  T7.5 &  12.105$\pm$0.026&  15.268$\pm$0.089 \\
2MASSI J0937347+293142      &       163.39$\pm$1.76 &   7  &  T6   &  11.652$\pm$0.023&  14.703$\pm$0.068 \\
2MASSW J1507476$-$162738    &       136.4$\pm$0.6   &   8  &  L5   &  10.367$\pm$0.022&  11.895$\pm$0.024 \\
\enddata
\tablecomments{References: (1) \cite{marocco2010}; (2) \cite{lucas2010}; (3) \cite{artigau2010}; (4) \cite{costa2006}; (5) \cite{vrba2004}; (6) \cite{perryman1997}; 
(7) \cite{schilbach2009}; (8) \cite{dahn2002}; (9) \cite{tinney2003}; (10) \cite{vanaltena1995}; }
\tablenotetext{a}{It is assumed that most of the W2 flux comes from the L dwarf companion and not the white dwarf primary.}
\end{deluxetable}

\clearpage

\begin{deluxetable}{ccllccclrrrr}
\tabletypesize{\tiny}
\rotate
\tablewidth{8in}
\tablenum{6}
\tablecaption{Astrometry for WISE Brown Dwarf Discoveries\label{astrometry}}
\tablehead{
\colhead{Object Name} &
\colhead{Dist.\ Est.} &
\colhead{RA (J2000)} &
\colhead{Dec (J2000)} &
\colhead{RA err} &
\colhead{Dec err} &
\colhead{Reference} &
\colhead{MJD} &
\colhead{$\mu_{\alpha}$} &
\colhead{$\mu_{\delta}$} &
\colhead{$\mu_{total}$} &
\colhead{$v_{tan}$} \\
\colhead{} &
\colhead{(pc)} &
\colhead{(deg)} &
\colhead{(deg)} &
\colhead{(arcsec)} &
\colhead{(arcsec)} &
\colhead{} &
\colhead{} &
\colhead{arcsec/yr} &
\colhead{arcsec/yr} &
\colhead{arcsec/yr} &
\colhead{km/s} \\
\colhead{(1)} &
\colhead{(2)} &
\colhead{(3)} &
\colhead{(4)} &
\colhead{(5)} &
\colhead{(6)} &
\colhead{(7)} &
\colhead{(8)} &
\colhead{(9)} &
\colhead{(10)} &
\colhead{(11)} &
\colhead{(12)}
}
\startdata
WISE 0008-1739&  22.0&     2.2073557&  -17.6563363&   0.179&   0.185&  WISE epoch 1&     55363.39   & -0.116$\pm$0.589&  -1.108$\pm$0.532\\
              &      &     2.2073061&  -17.6565639&   0.279&   0.293&  WISE epoch 2&     55542.36   \\
              &      &     2.2073379&  -17.6564659&   0.485&   0.309&  Spitzer     &     55571      \\
\\   
WISE 0031-3840&  22.0&     7.830238 &  -38.676575 &   0.06 &   0.07 &  2MASS PSC   &     51391.3054 &  0.548$\pm$0.006&  -0.049$\pm$0.007&  0.550$\pm$0.009  & 57.4$\pm$1.0 \\
              &      &     7.8323524&  -38.6766992&   0.045&   0.044&  WISE epoch 1&     55357.83   \\ 
              &      &     7.8324281&  -38.6767498&   0.046&   0.044&  WISE epoch 2&     55535.75   \\
\\
\enddata
\tablenotetext{a}{Gelino et al.\ (submitted) find this object to be a binary. Their revised distance to this system is 12.3$\pm$2.3 pc.}
\tablenotetext{b}{Gelino et al.\ (submitted) find this object to be a binary. Their revised distance to this system is 40.1$\pm$3.0 pc.}
\end{deluxetable}

\clearpage

\begin{deluxetable}{cccccccc}
\tabletypesize{\scriptsize}
\tablewidth{6in}
\tablenum{7}
\tablecaption{Preliminary Parallaxes and Absolute Magnitudes for WISE Brown Dwarf Discoveries\label{prelim_parallaxes}}
\tablehead{
\colhead{Object} &
\colhead{Near-infrared} &
\colhead{$\pi_{trig}$} &
\colhead{Dist.\ Range (${\pm}1\sigma$)} &
\colhead{$M_J$\tablenotemark{a}} &
\colhead{$M_H$\tablenotemark{a}} &
\colhead{$M_{W2}$} \\
\colhead{Name} &
\colhead{Spectral Type} &
\colhead{(arcsec)} &
\colhead{(pc)} &
\colhead{(mag)} &
\colhead{(mag)} &
\colhead{(mag)} \\
\colhead{(1)} &
\colhead{(2)} &
\colhead{(3)} &
\colhead{(4)} &
\colhead{(5)} &
\colhead{(6)} &
\colhead{(7)}  
}
\startdata
WISE 0254+0223&  T8          & 0.165$\pm$0.046 & 4.7-8.4   & 17.0$\pm$0.6 & 17.4$\pm$0.6 & 13.8$\pm$0.6 \\
WISE 1541-2250&  Y0          & 0.351$\pm$0.108 & 2.2-4.1   & 23.9$\pm$0.8 & 23.7$\pm$0.9 & 16.7$\pm$0.7 \\
WISE 1647+5632&  L9 pec (red)& 0.116$\pm$0.029 & 6.9-11.5  & 16.9$\pm$0.5 & 15.7$\pm$0.5 & 13.4$\pm$0.5 \\
WISE 1741+2732&  T9          & 0.182$\pm$0.038 & 4.5-6.9   & 17.8$\pm$0.5 & 17.9$\pm$0.5 & 13.6$\pm$0.5 \\
\enddata
\tablenotetext{a}{Absolute $J$ and $H$ magnitudes for WISE 0254+0223 
and WISE 1541$-$2250 are on the MKO filter system; magnitudes
for WISE 1647+5632 and WISE 1741+2732 are on the 2MASS filter system.}
\end{deluxetable}

\clearpage

\begin{deluxetable}{ccccccccc}
\tabletypesize{\tiny}
\rotate
\tablewidth{8in}
\tablenum{8}
\tablecaption{Preliminary Space Densities for an All-sky, Volume-limited Sample of Late-T and Y Dwarfs\label{space_density_numbers}}
\tablehead{
\colhead{Spectral Type} &
\colhead{Approx.\ $T_{eff}$} &
\colhead{$d_{max}$} &
\colhead{\# from Previous} &
\colhead{\# from WISE} &
\colhead{Total \#}&
\colhead{Obs.\ Space Density} &
\colhead{$<V/V_{max}>$} &
\colhead{\# per 100K Bin} \\
\colhead{Range} &
\colhead{Range (K)} &
\colhead{(pc)} &
\colhead{Surveys} &
\colhead{to date} &
\colhead{to date}&
\colhead{(\#/$pc^3$)} &
\colhead{} &
\colhead{w/in 10 pc} \\
\colhead{(1)} &
\colhead{(2)} &
\colhead{(3)} &
\colhead{(4)} &
\colhead{(5)} &
\colhead{(6)} &
\colhead{(7)} &
\colhead{(8)} &
\colhead{(9)}
}
\startdata
T6-T6.5 &  900-1050 & 20 & 14 & 16 & 30 & $>$9.0e-4 & 0.36 & $>$2.5 \\
T7-T7.5 &   750-900 & 20 & 13 & 12 & 25 & $>$7.5e-4 & 0.34 & $>$2.1 \\
T8-T8.5 &   600-750 & 20 &  9 & 26 & 35 & $>$1.0e-3 & 0.30 & $>$2.9 \\
T9-T9.5 &   450-600 & 15 &  1 &  9 & 10 & $>$7.1e-4 & 0.40 & $>$2.0 \\
Y0      &   300-450 & 10 &  0 &  4 &  4 & $>$9.5e-4 & 0.59 & $>$2.7 \\
$>$Y0   &    $<$300 & 10 &  0 &  1 &  1 & $>$2.4e-4 & 0.83 & $>$0.7 \\
\enddata
\end{deluxetable}

\clearpage

\clearpage

\clearpage

\clearpage

\begin{figure}
\epsscale{0.65}
\figurenum{4.1}
\plotone{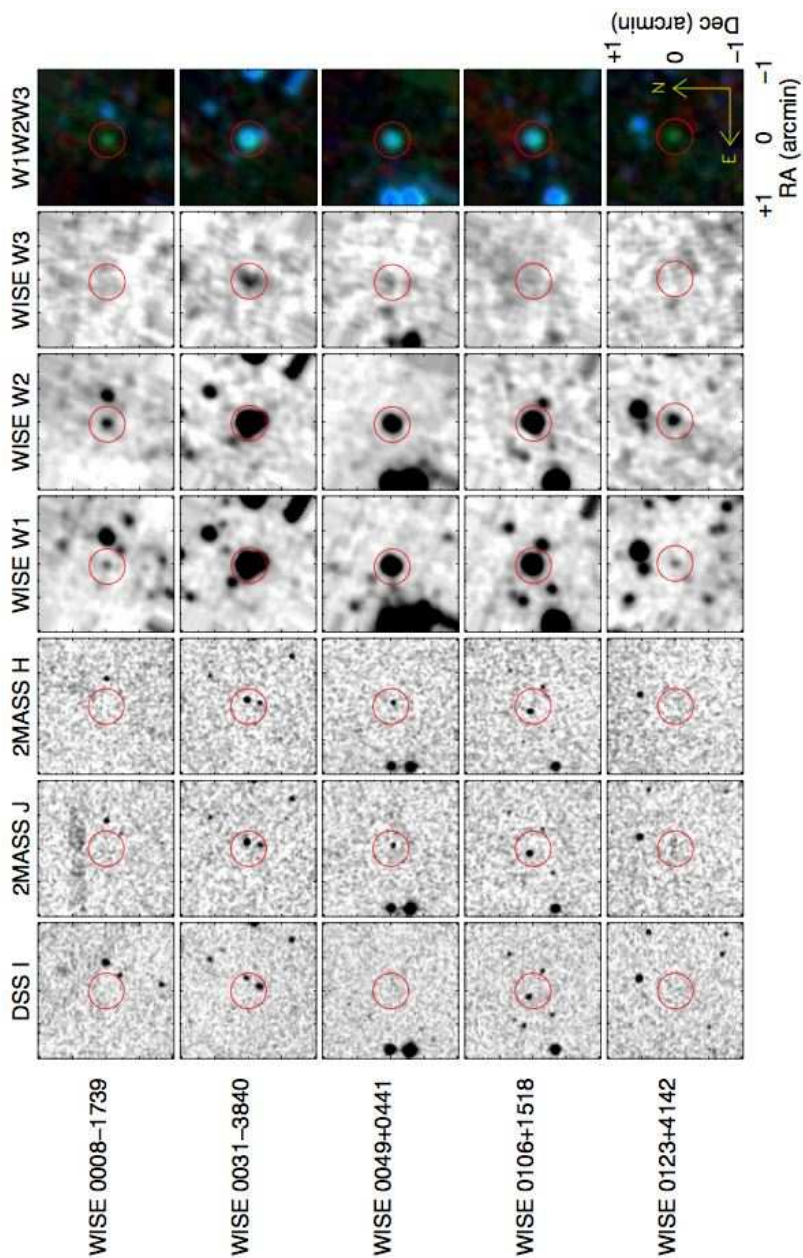}
\caption{Finder charts for new WISE brown dwarf discoveries. 
Each row represents one object and shows a selection of 2$\times$2 arcmin cutouts from 
various all-sky surveys centered at the position (red circle) of the WISE source. From 
left to right, these cutouts are from the Digitized Sky Survey I band, the Two Micron 
All-Sky Survey J and H bands, the three shortest wavelength bands of WISE (W1, W2, and W3), 
and a three-color image made from these same three WISE bands (W1 encoded as blue, W2 as 
green, and W3 as red). In each cutout, north is oriented up and east is to the left. 
Images are not shown for sources already presented elsewhere -- WISE 0458+6434 from 
\cite{mainzer2011}; WISE 1617+1807, WISE 1812+2721, WISE 2018-6423, WISE 2313-8037, and 
WISE 2359-7335 from \cite{burgasser2011}; and WISE 0148-7202, WISE 0410+1502, WISE 1405+5534, 
WISE 1541-2250, and WISE 2056+1459 from Cushing et al. (accepted).
\label{finder_chart1}}
\end{figure}

\clearpage

\begin{figure}
\epsscale{0.85}
\figurenum{4.2}
\plotone{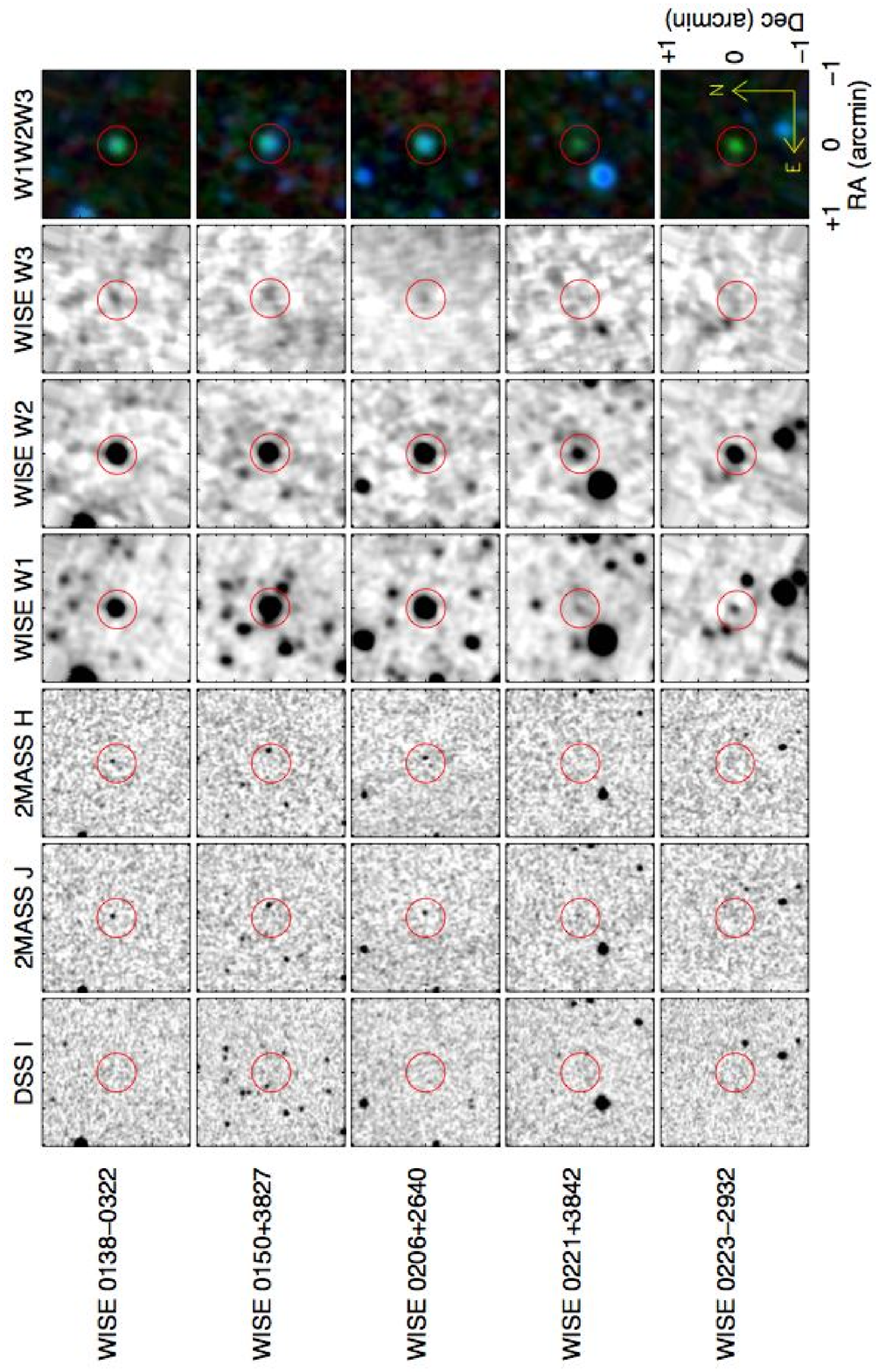}
\caption{Continued.
\label{finder_chart2}}
\end{figure}

\clearpage

\begin{figure}
\epsscale{0.85}
\figurenum{4.3}
\plotone{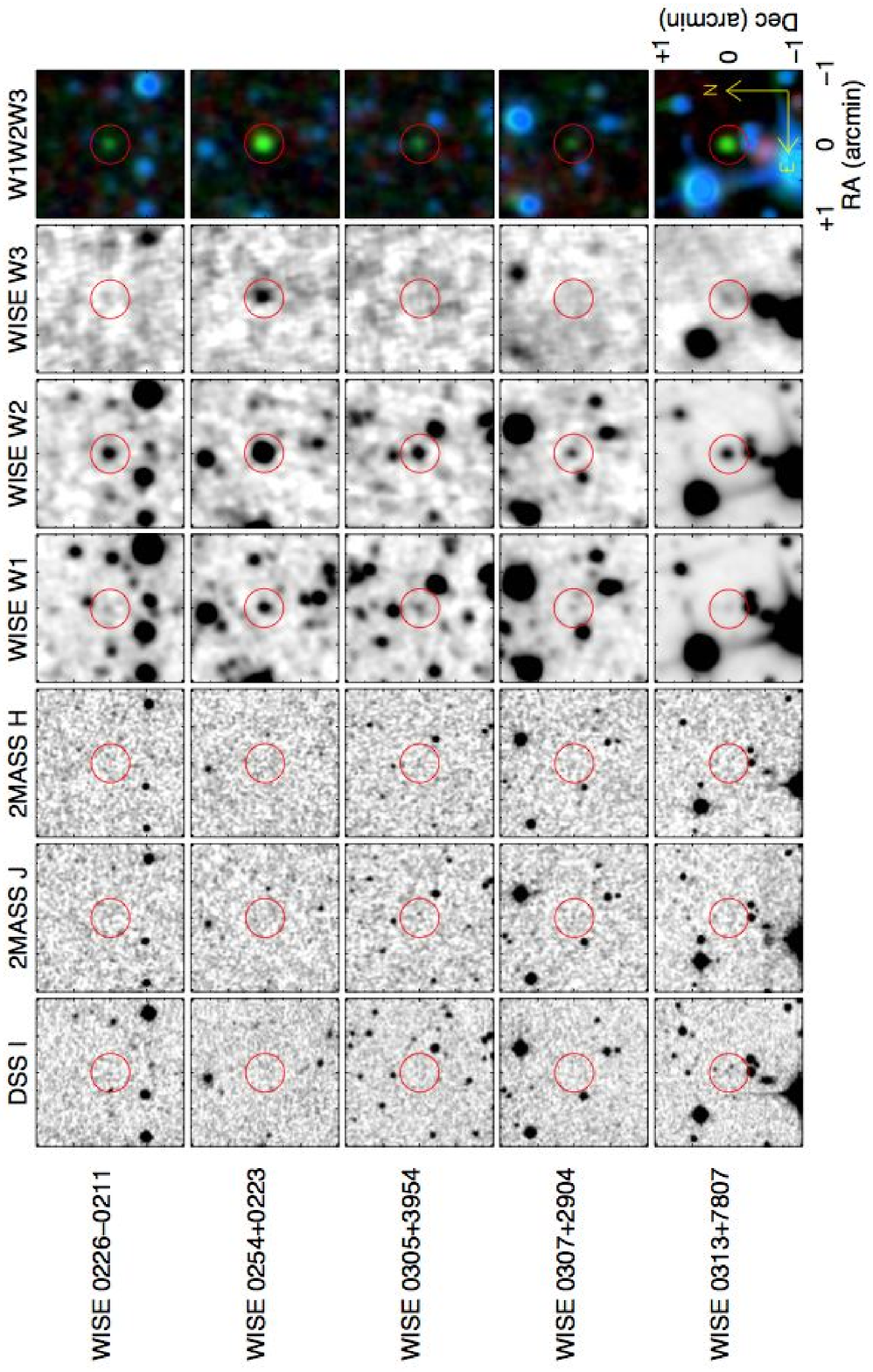}
\caption{Continued.
\label{finder_chart3}}
\end{figure}

\clearpage

\begin{figure}
\epsscale{0.85}
\figurenum{4.4}
\plotone{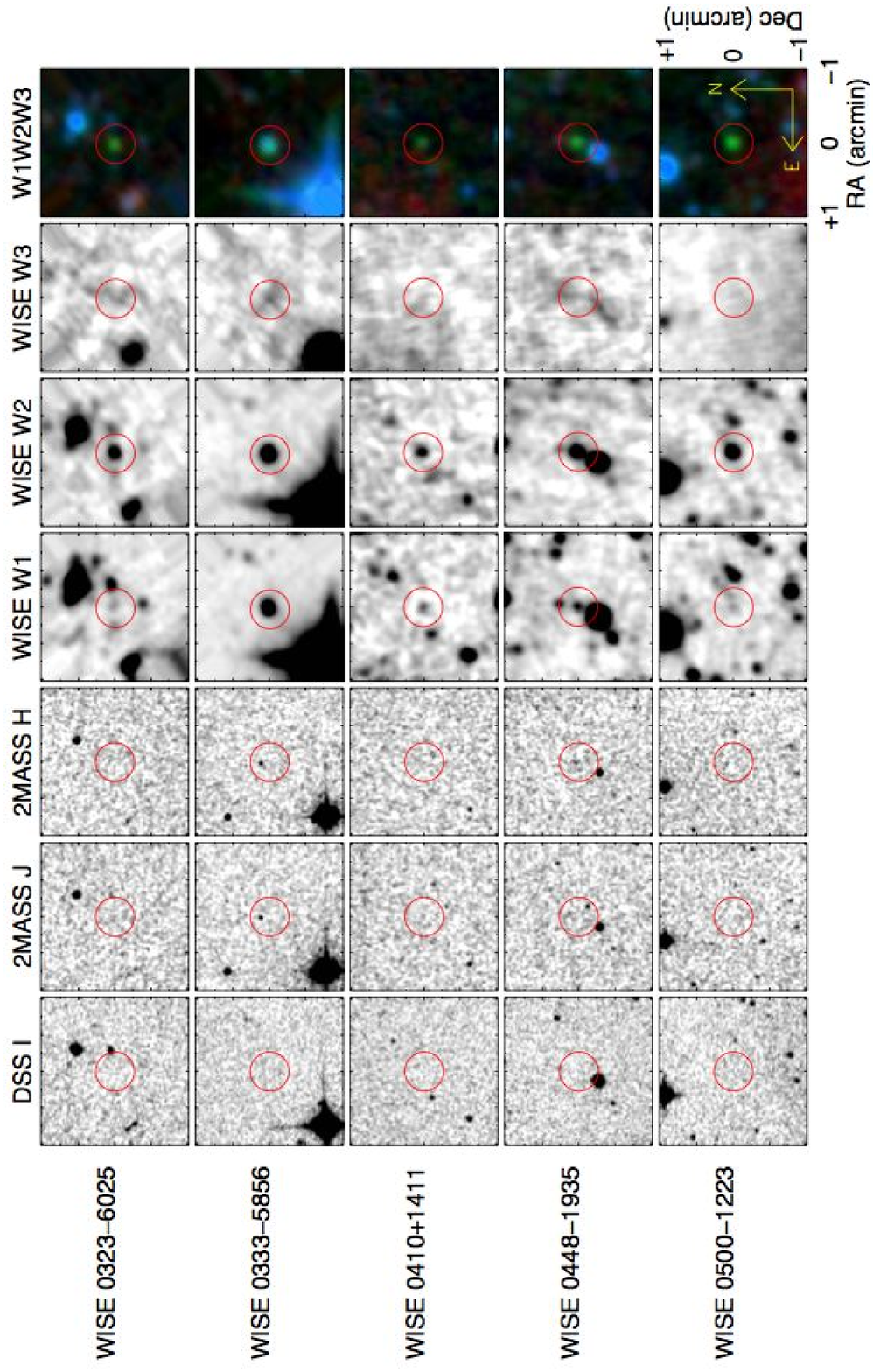}
\caption{Continued.
\label{finder_chart4}}
\end{figure}

\clearpage

\begin{figure}
\epsscale{0.85}
\figurenum{4.5}
\plotone{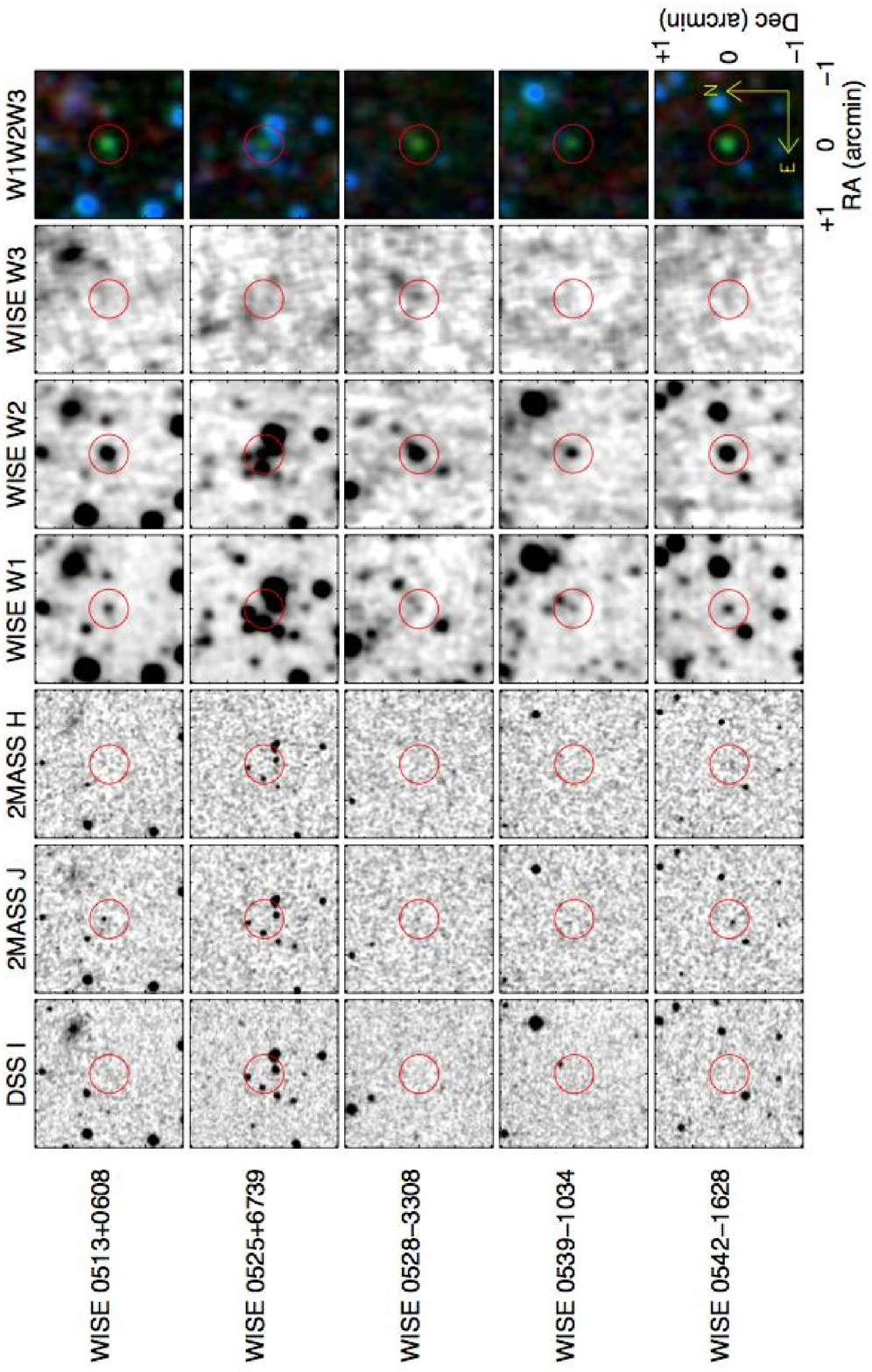}
\caption{Continued.
\label{finder_chart5}}
\end{figure}

\clearpage

\begin{figure}
\epsscale{0.85}
\figurenum{4.6}
\plotone{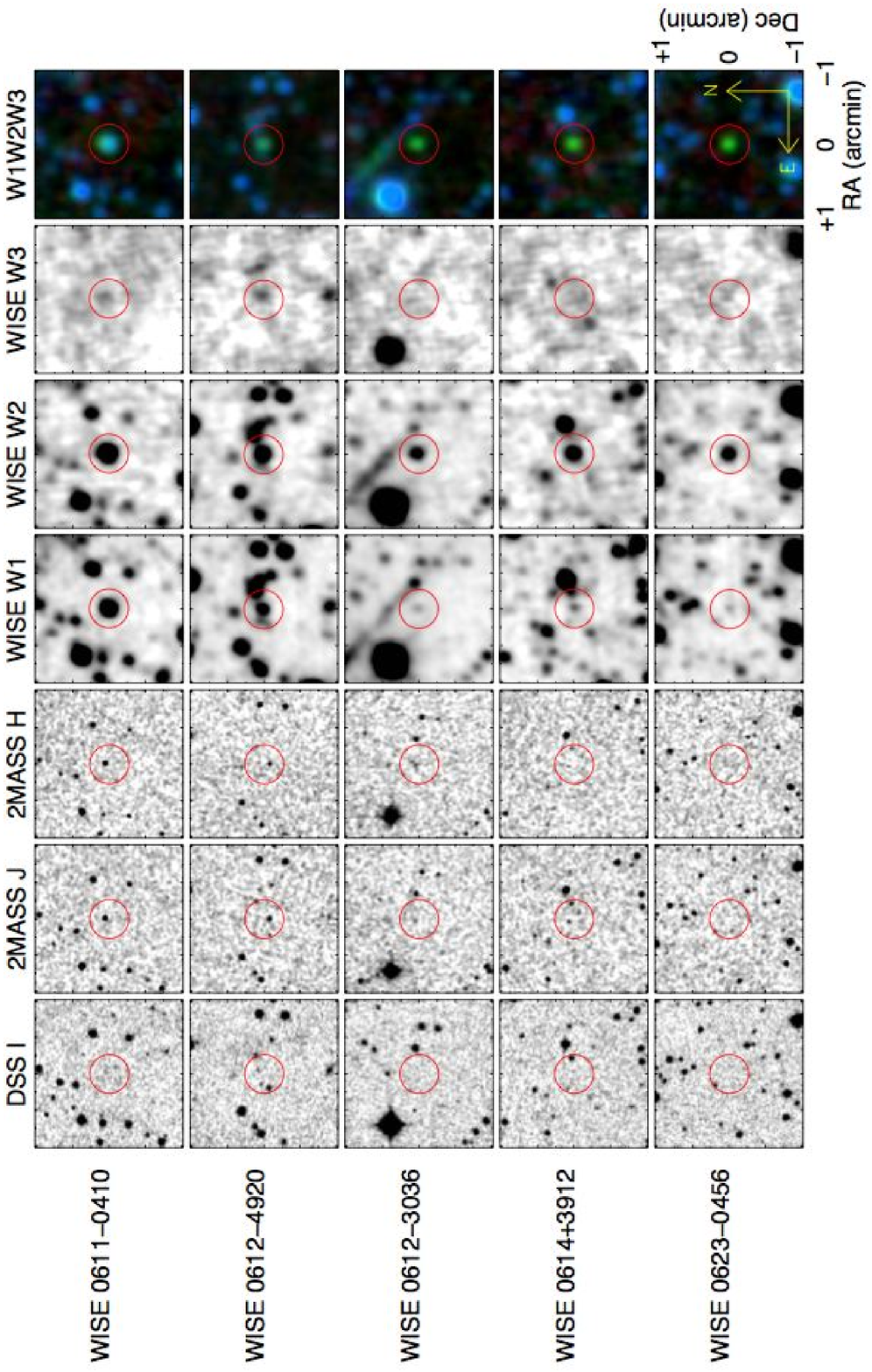}
\caption{Continued.
\label{finder_chart6}}
\end{figure}

\clearpage

\begin{figure}
\epsscale{0.85}
\figurenum{4.7}
\plotone{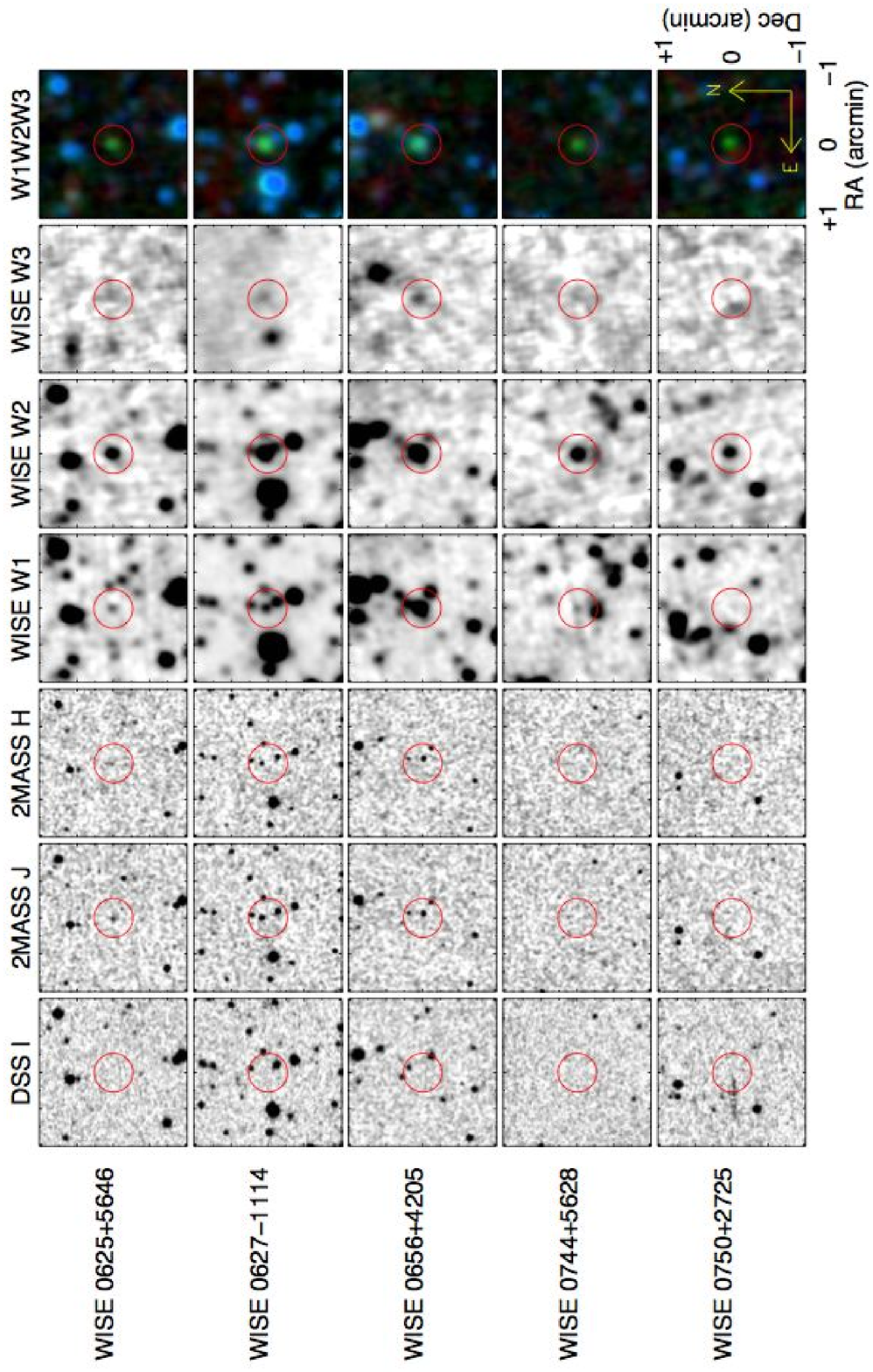}
\caption{Continued.
\label{finder_chart7}}
\end{figure}

\clearpage

\begin{figure}
\epsscale{0.85}
\figurenum{4.8}
\plotone{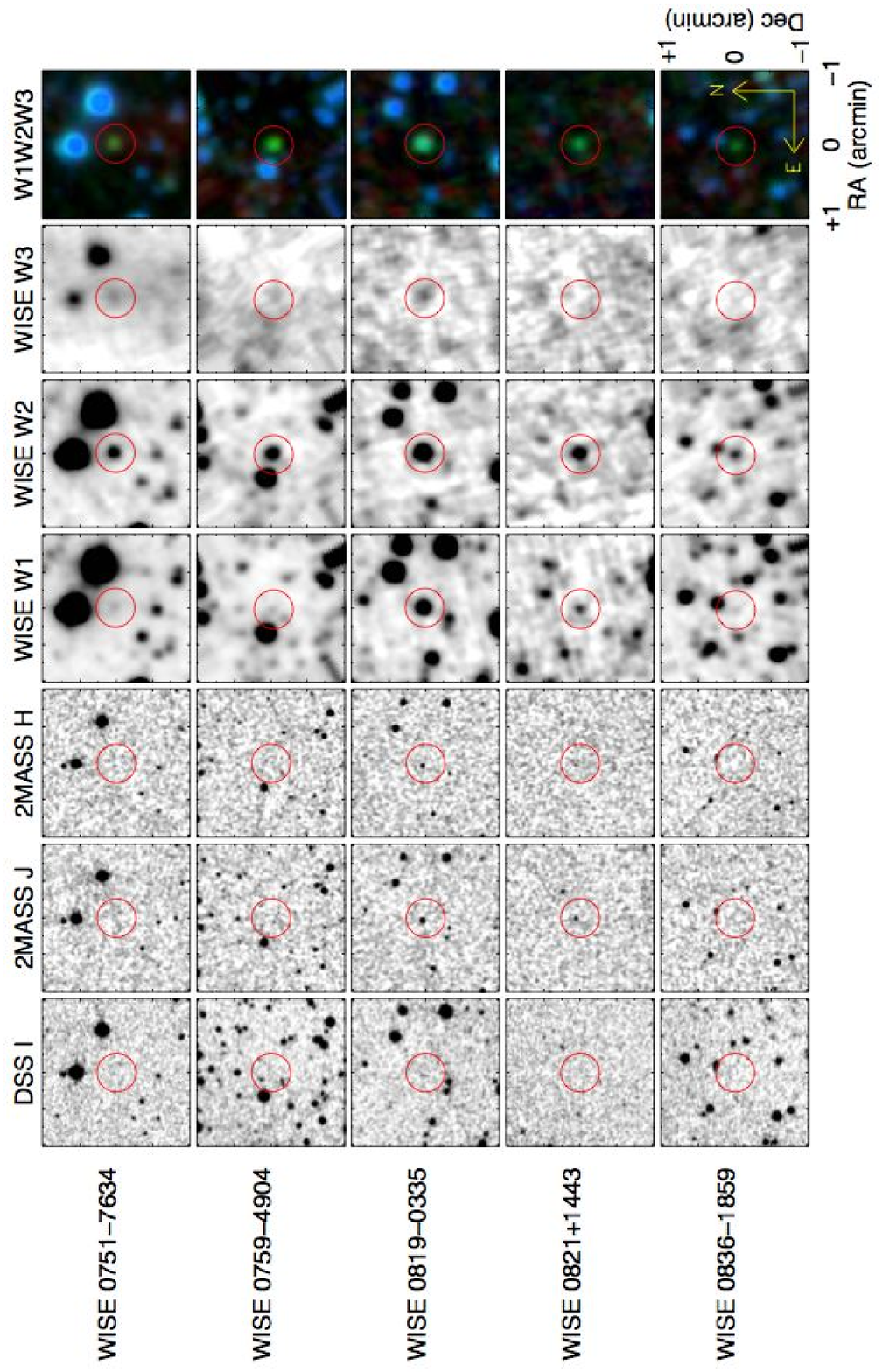}
\caption{Continued.
\label{finder_chart8}}
\end{figure}

\clearpage

\begin{figure}
\epsscale{0.85}
\figurenum{4.9}
\plotone{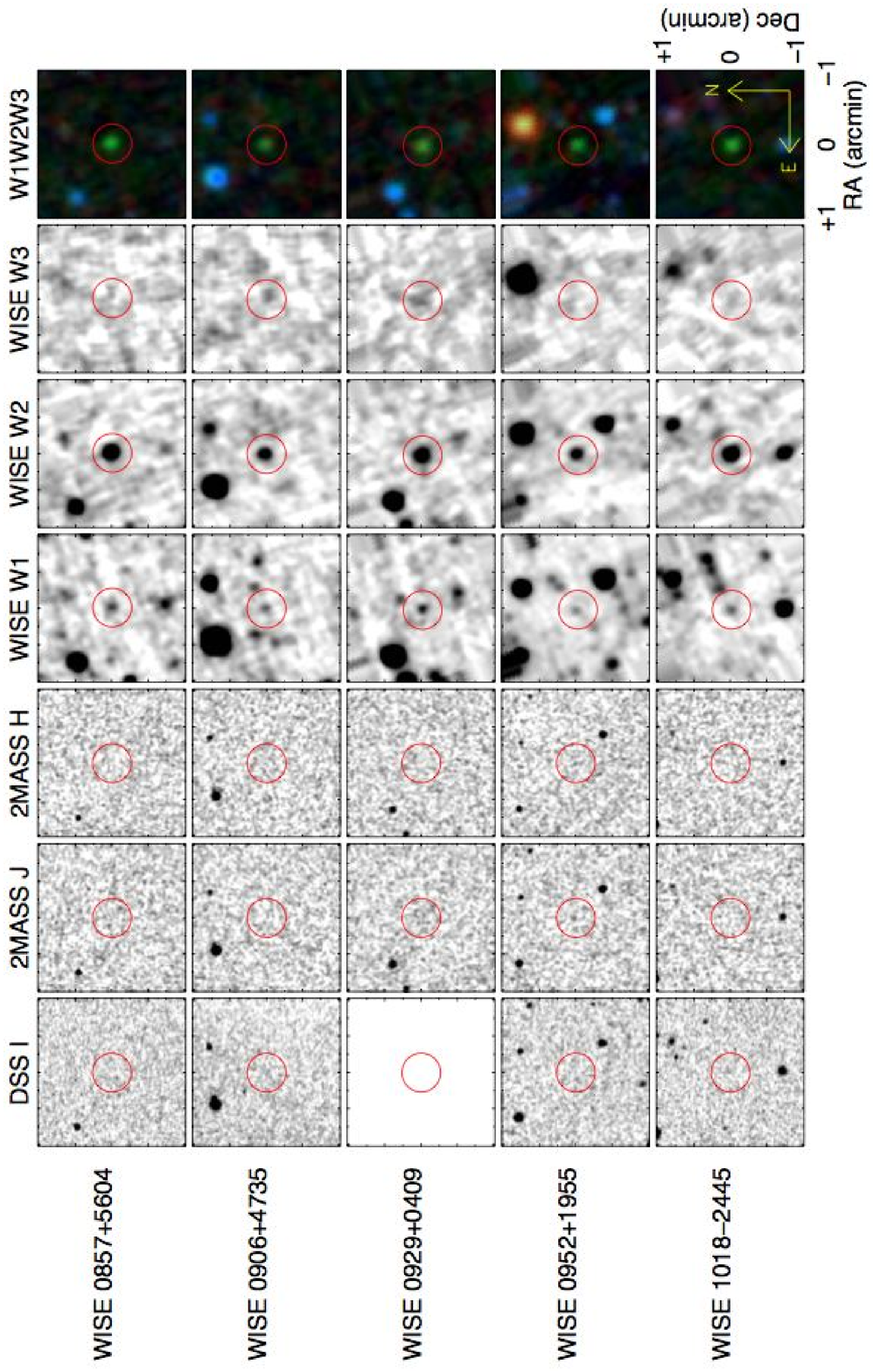}
\caption{Continued.
\label{finder_chart9}}
\end{figure}

\clearpage

\begin{figure}
\epsscale{0.85}
\figurenum{4.10}
\plotone{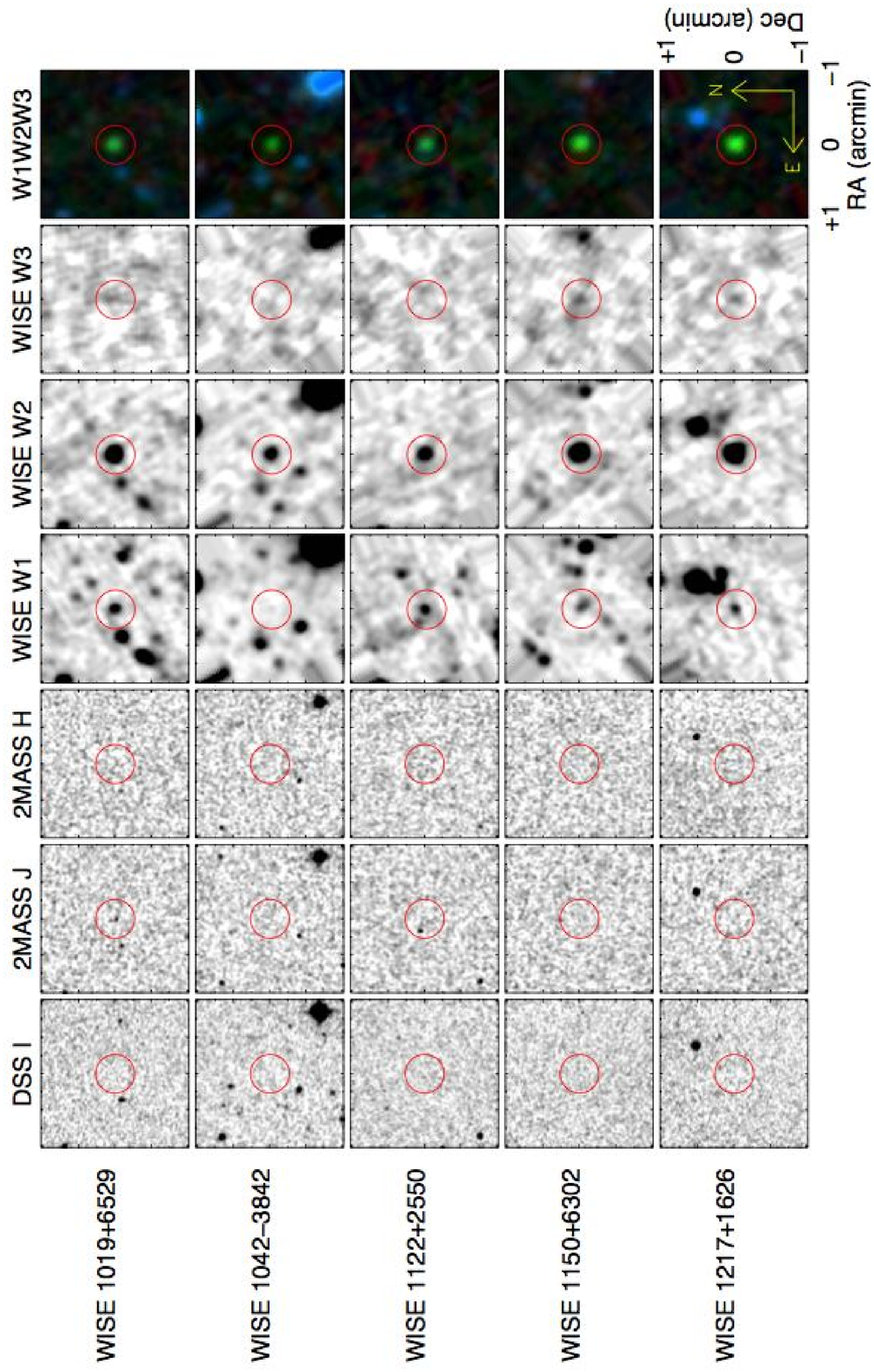}
\caption{Continued.
\label{finder_chart10}}
\end{figure}

\clearpage

\begin{figure}
\epsscale{0.85}
\figurenum{4.11}
\plotone{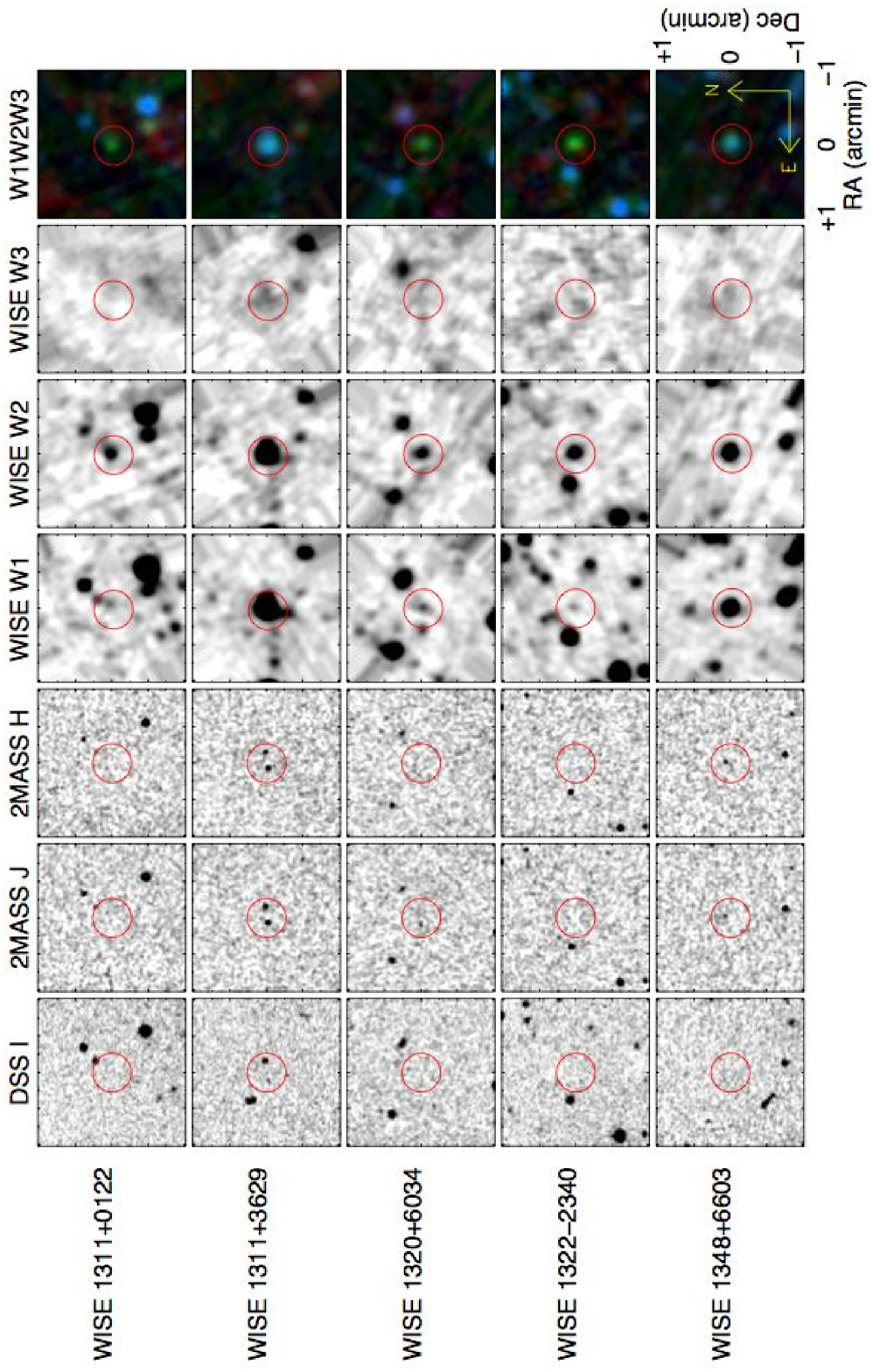}
\caption{Continued.
\label{finder_chart11}}
\end{figure}

\clearpage

\begin{figure}
\epsscale{0.85}
\figurenum{4.12}
\plotone{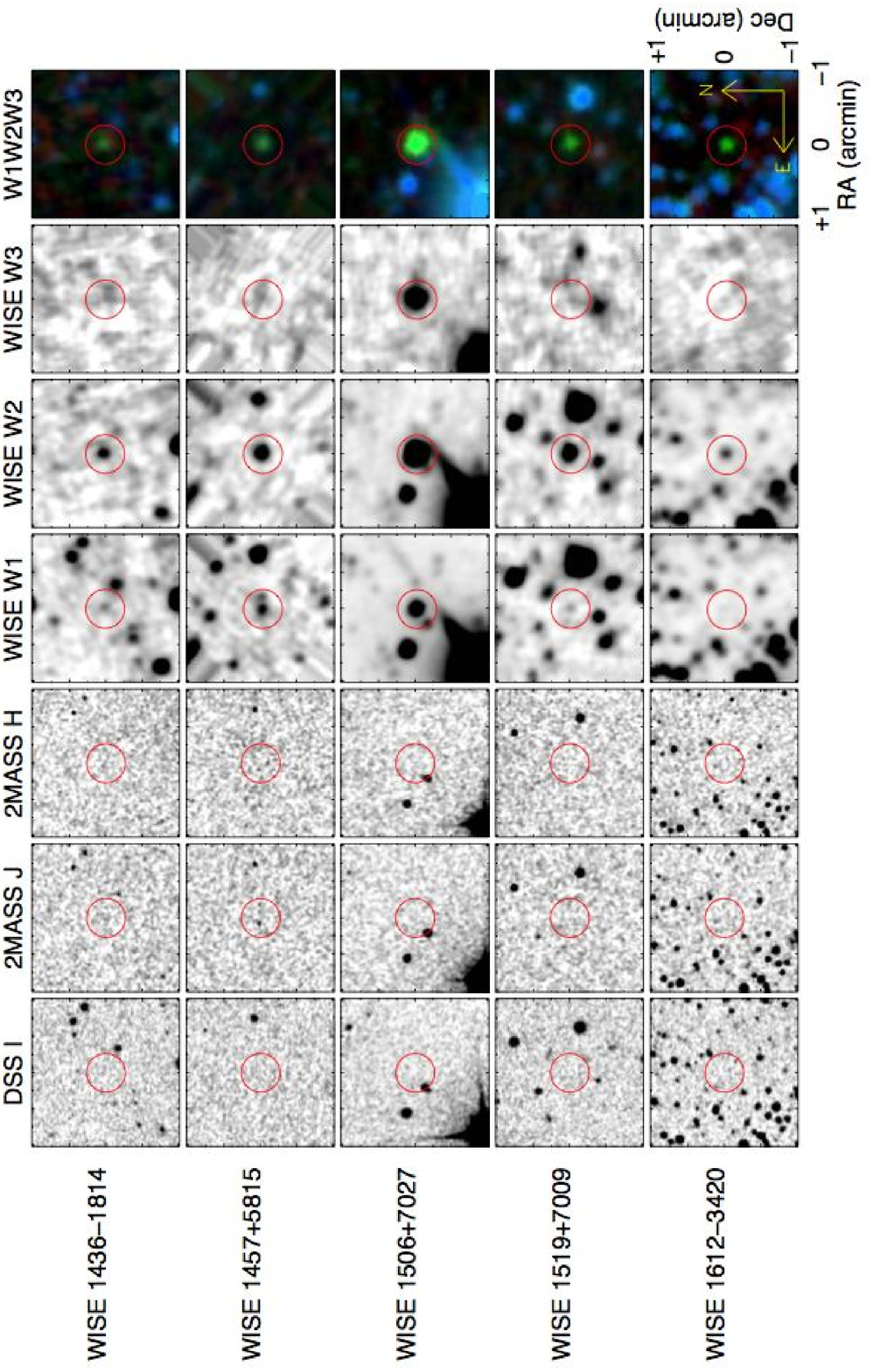}
\caption{Continued.
\label{finder_chart12}}
\end{figure}

\clearpage

\begin{figure}
\epsscale{0.85}
\figurenum{4.13}
\plotone{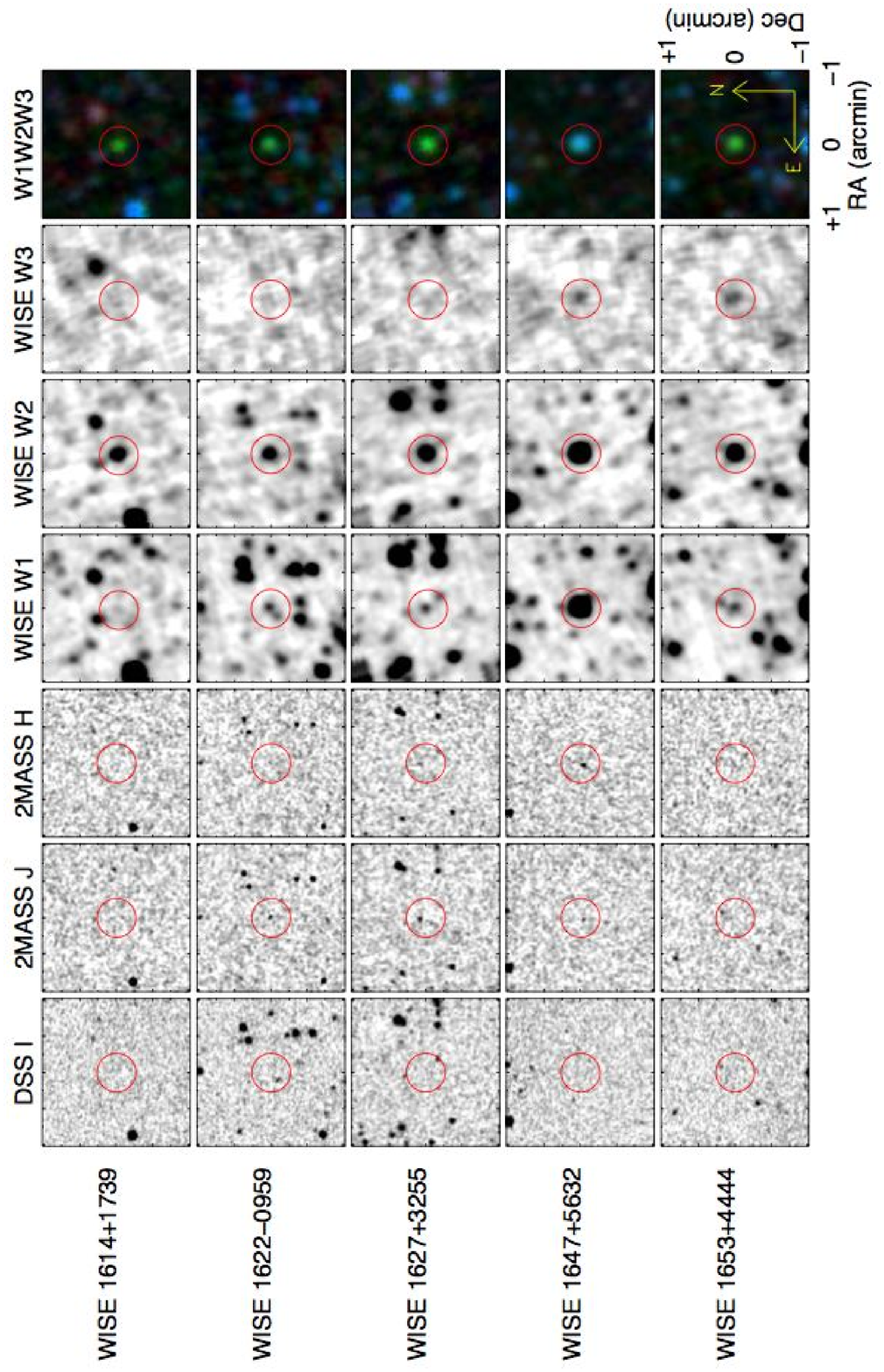}
\caption{Continued.
\label{finder_chart13}}
\end{figure}

\clearpage

\begin{figure}
\epsscale{0.85}
\figurenum{4.14}
\plotone{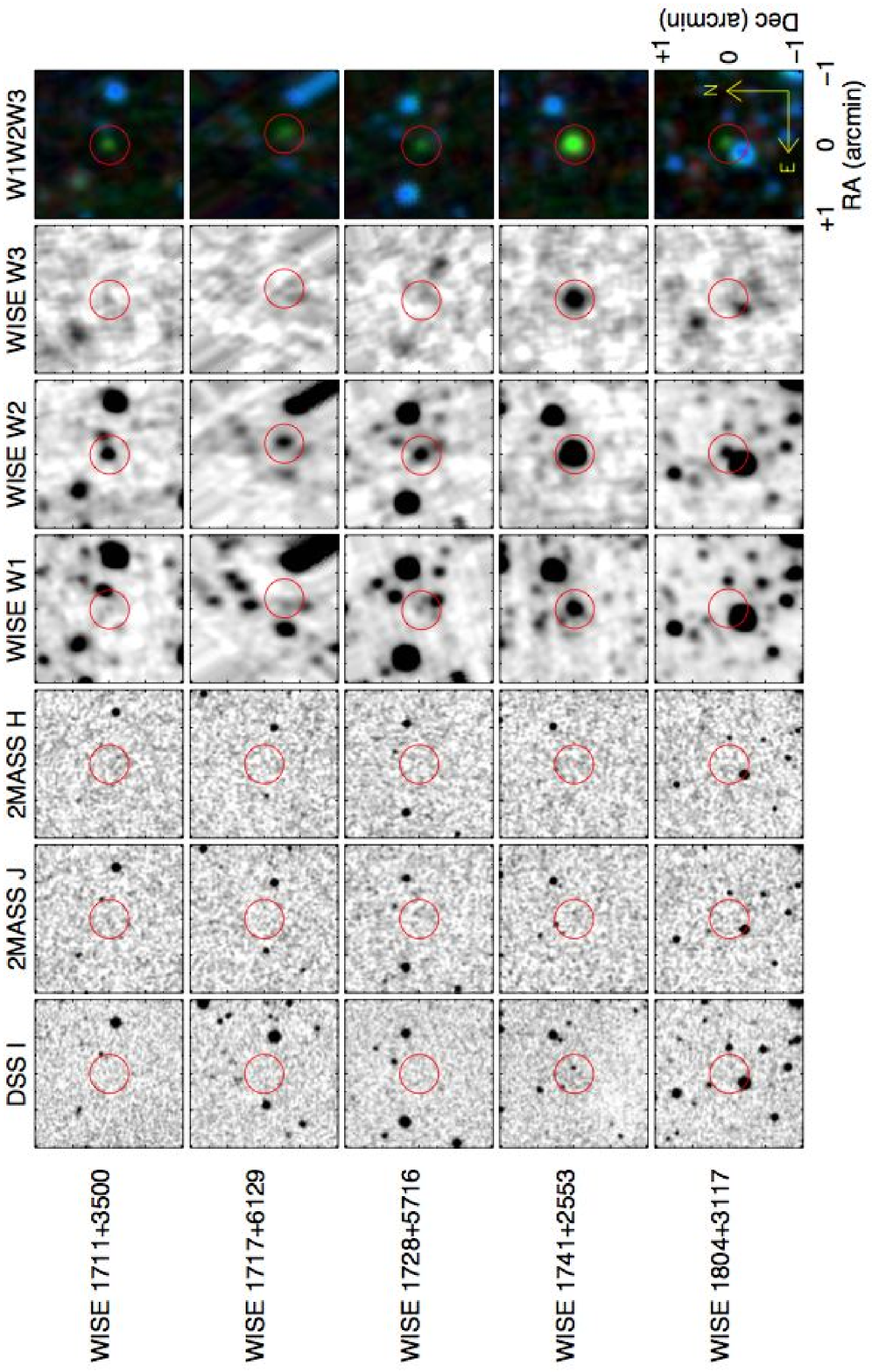}
\caption{Continued.
\label{finder_chart14}}
\end{figure}

\clearpage

\begin{figure}
\epsscale{0.85}
\figurenum{4.15}
\plotone{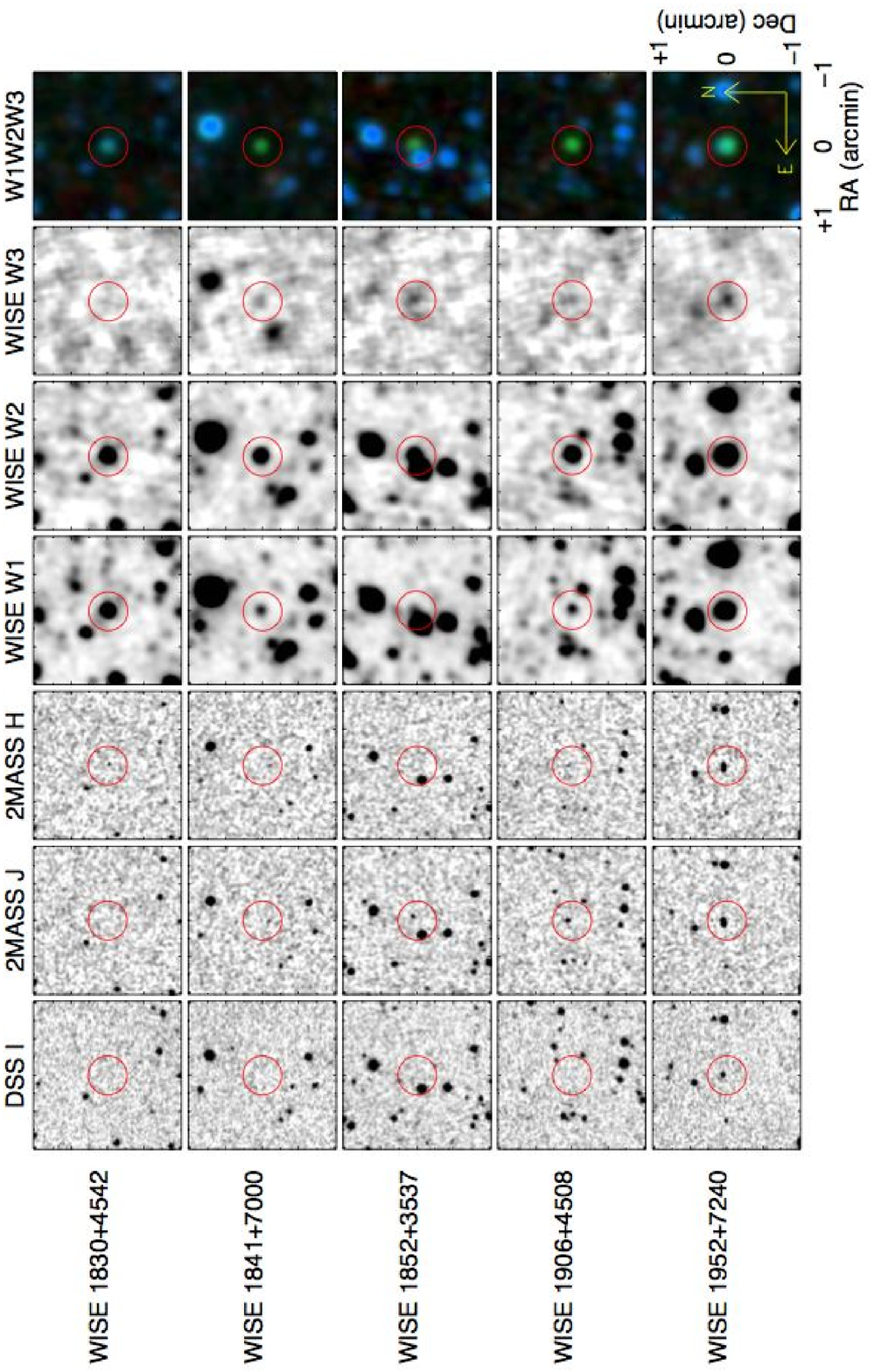}
\caption{Continued.
\label{finder_chart15}}
\end{figure}

\clearpage

\begin{figure}
\epsscale{0.85}
\figurenum{4.16}
\plotone{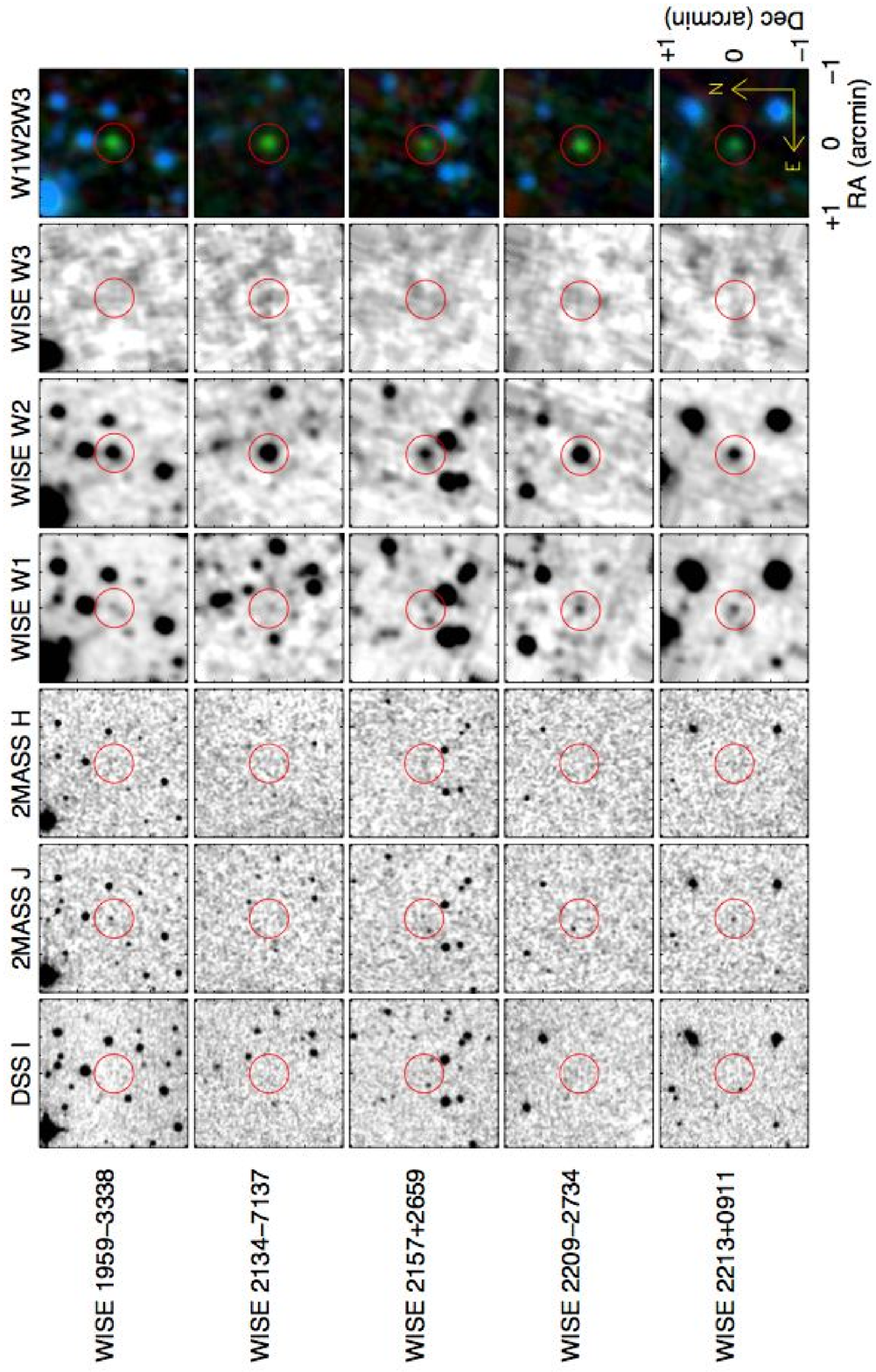}
\caption{Continued.
\label{finder_chart16}}
\end{figure}

\clearpage

\begin{figure}
\epsscale{0.85}
\figurenum{4.17}
\plotone{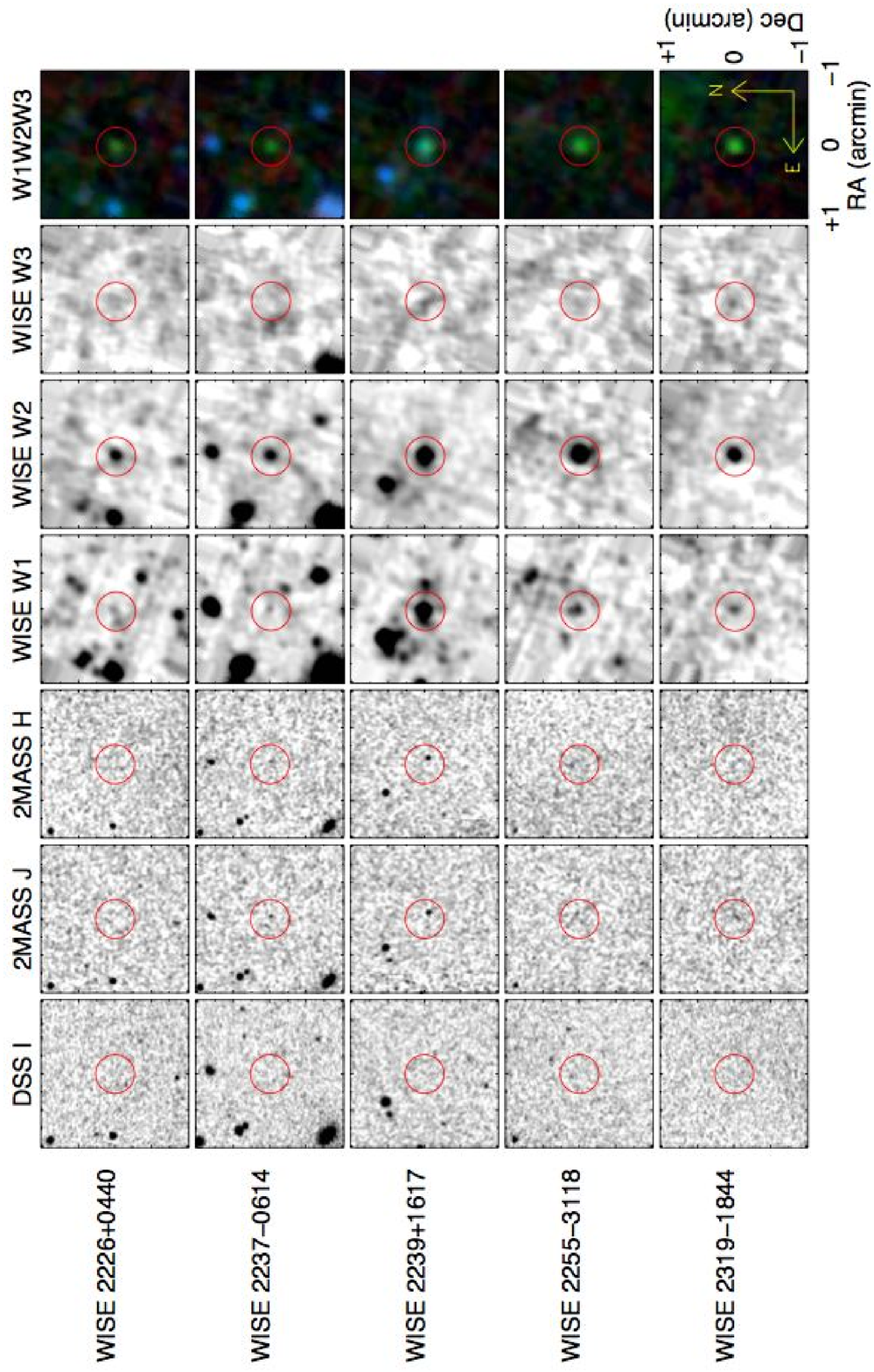}
\caption{Continued.
\label{finder_chart17}}
\end{figure}

\clearpage

\begin{figure}
\epsscale{0.85}
\figurenum{4.18}
\plotone{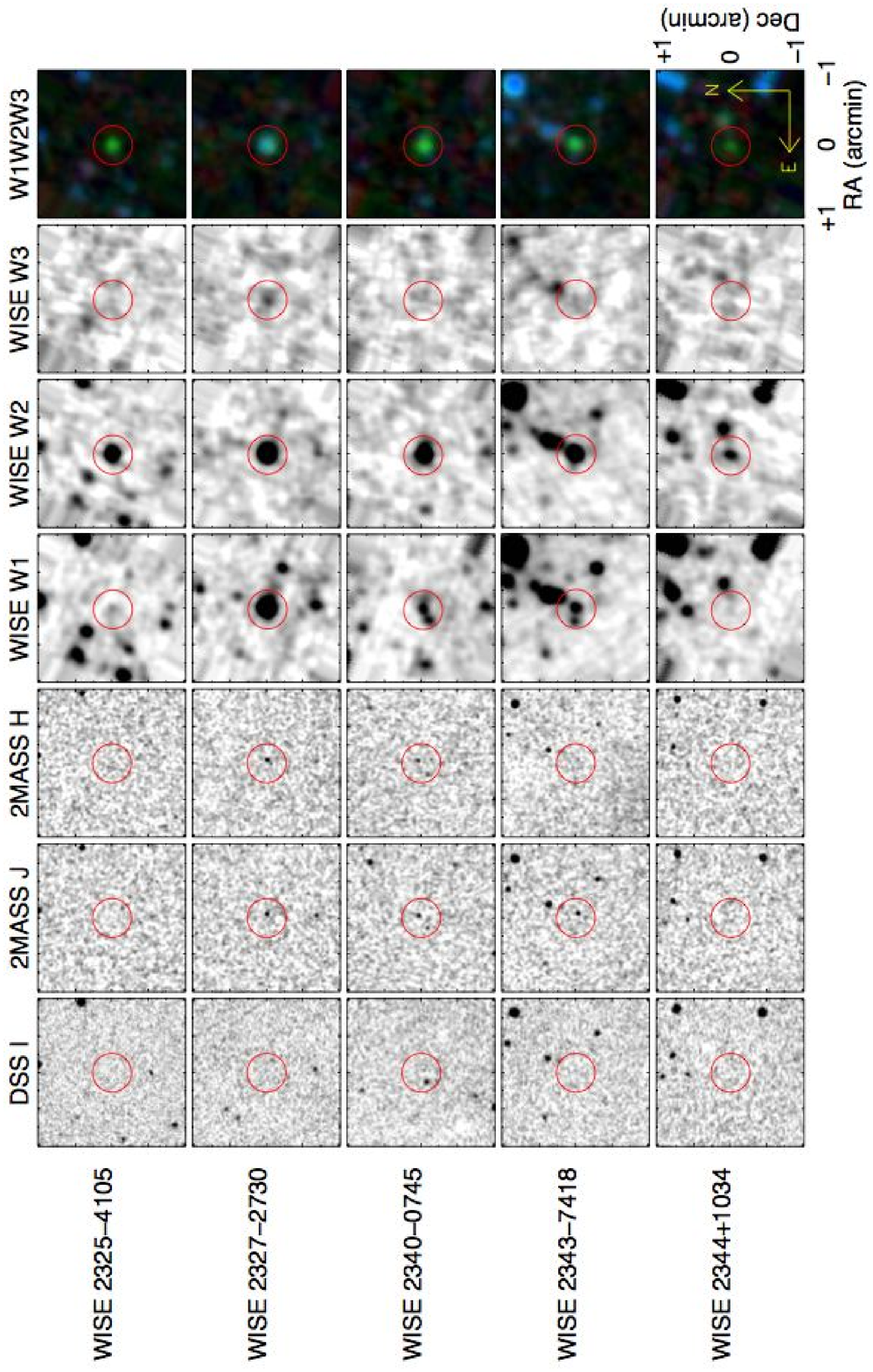}
\caption{Continued.
\label{finder_chart18}}
\end{figure}

\clearpage

\begin{figure}
\epsscale{0.25}
\figurenum{4.19}
\plotone{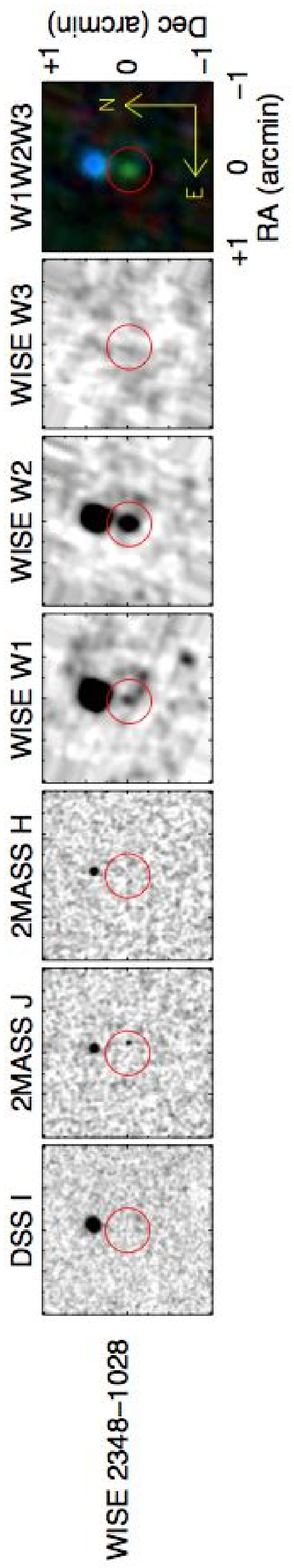}
\caption{Continued.
\label{finder_chart19}}
\end{figure}

\clearpage

\begin{figure}
\epsscale{0.85}
\figurenum{5}
\plotone{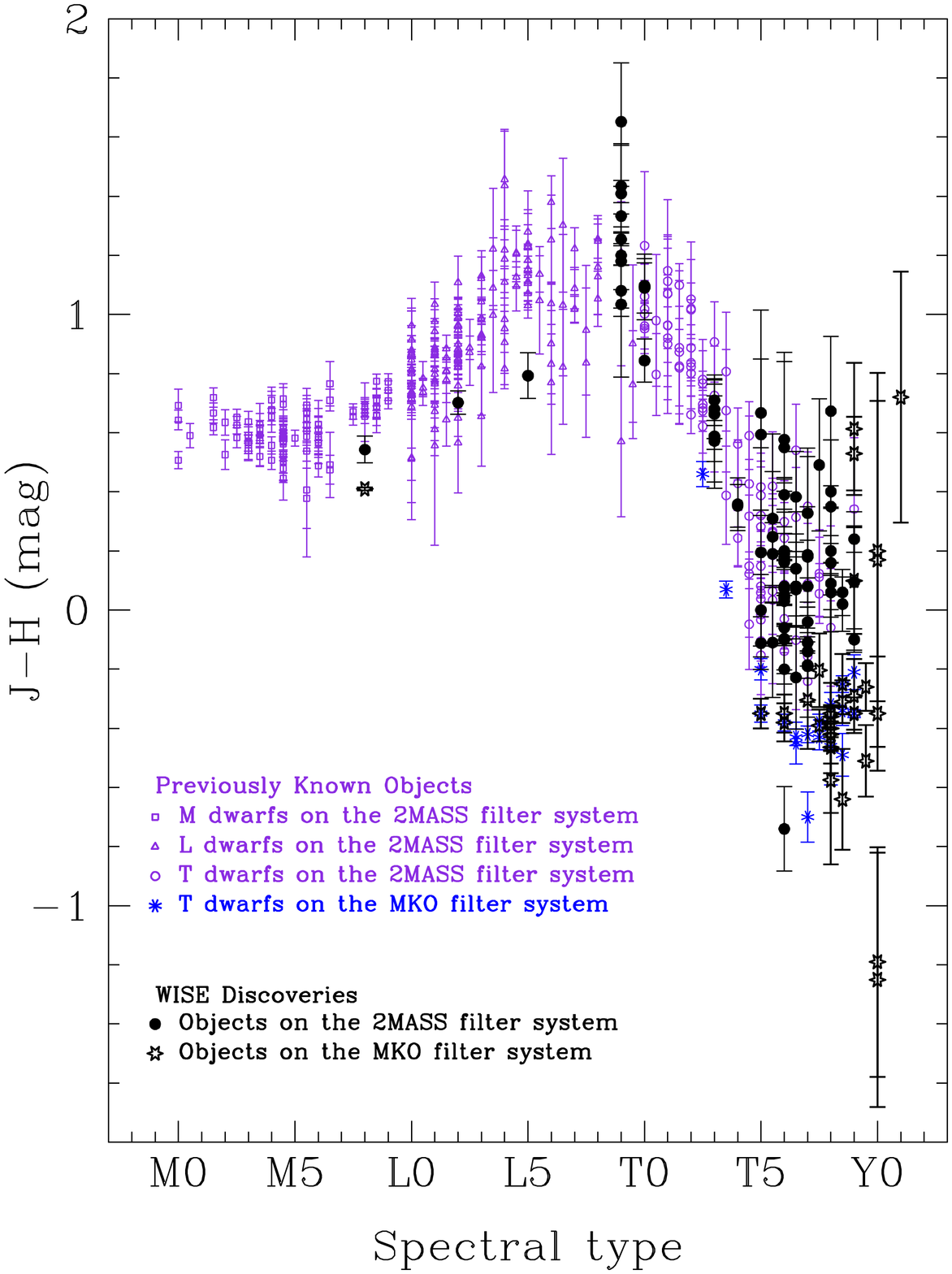}
\caption{$J-H$ color versus spectral type for objects with solid $J-H$ colors 
(not limits). Color coding and symbol selection are explained in the legend.
\label{JH_vs_type}}
\end{figure}

\clearpage

\begin{figure}
\epsscale{0.9}
\figurenum{6}
\plotone{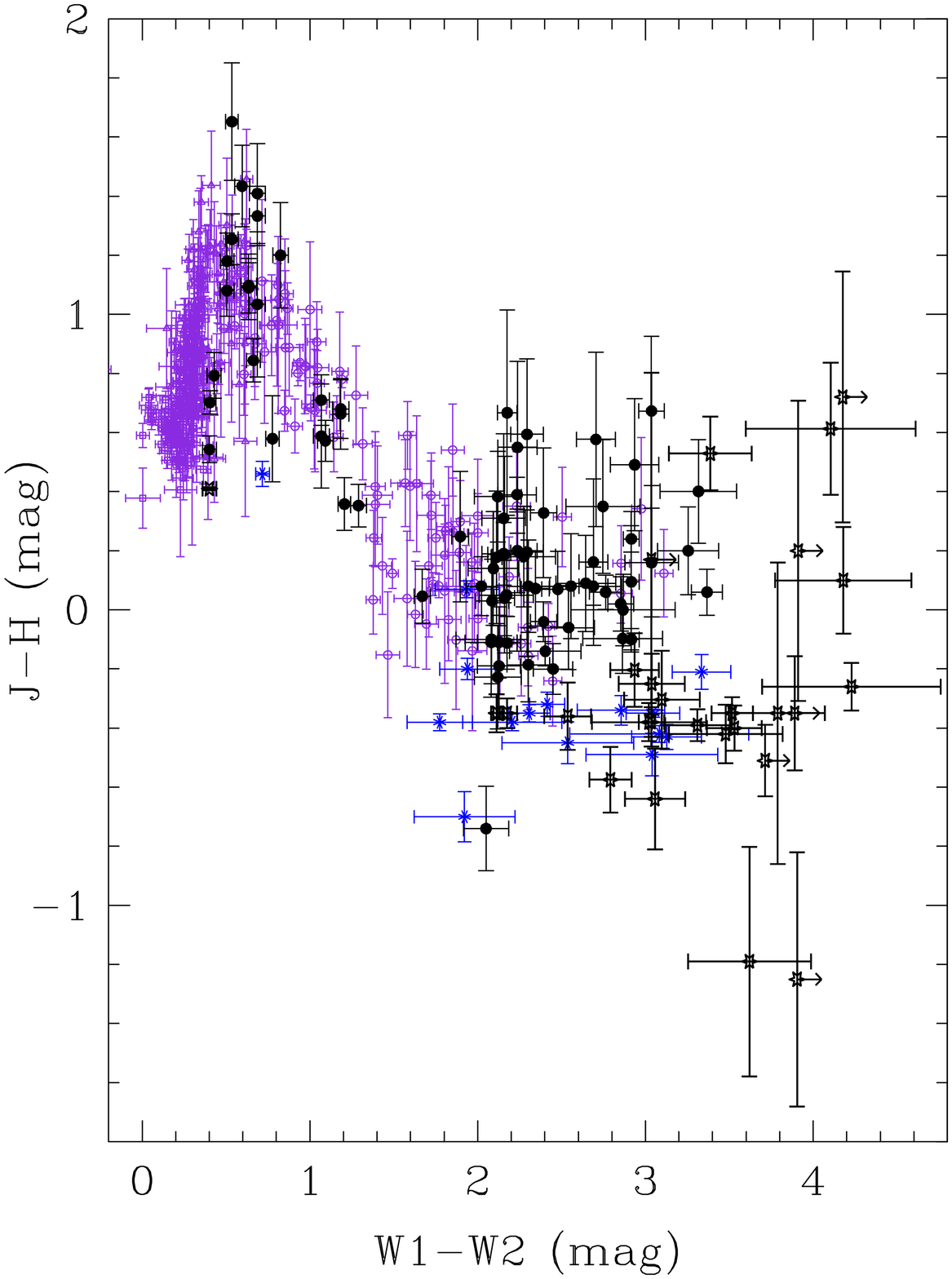}
\caption{$J-H$ color versus W1-W2 color. Symbols are the same as in Figure~\ref{JH_vs_type}.
\label{JH_vs_W1W2}}
\end{figure}

\clearpage

\begin{figure}
\epsscale{0.9}
\figurenum{7}
\plotone{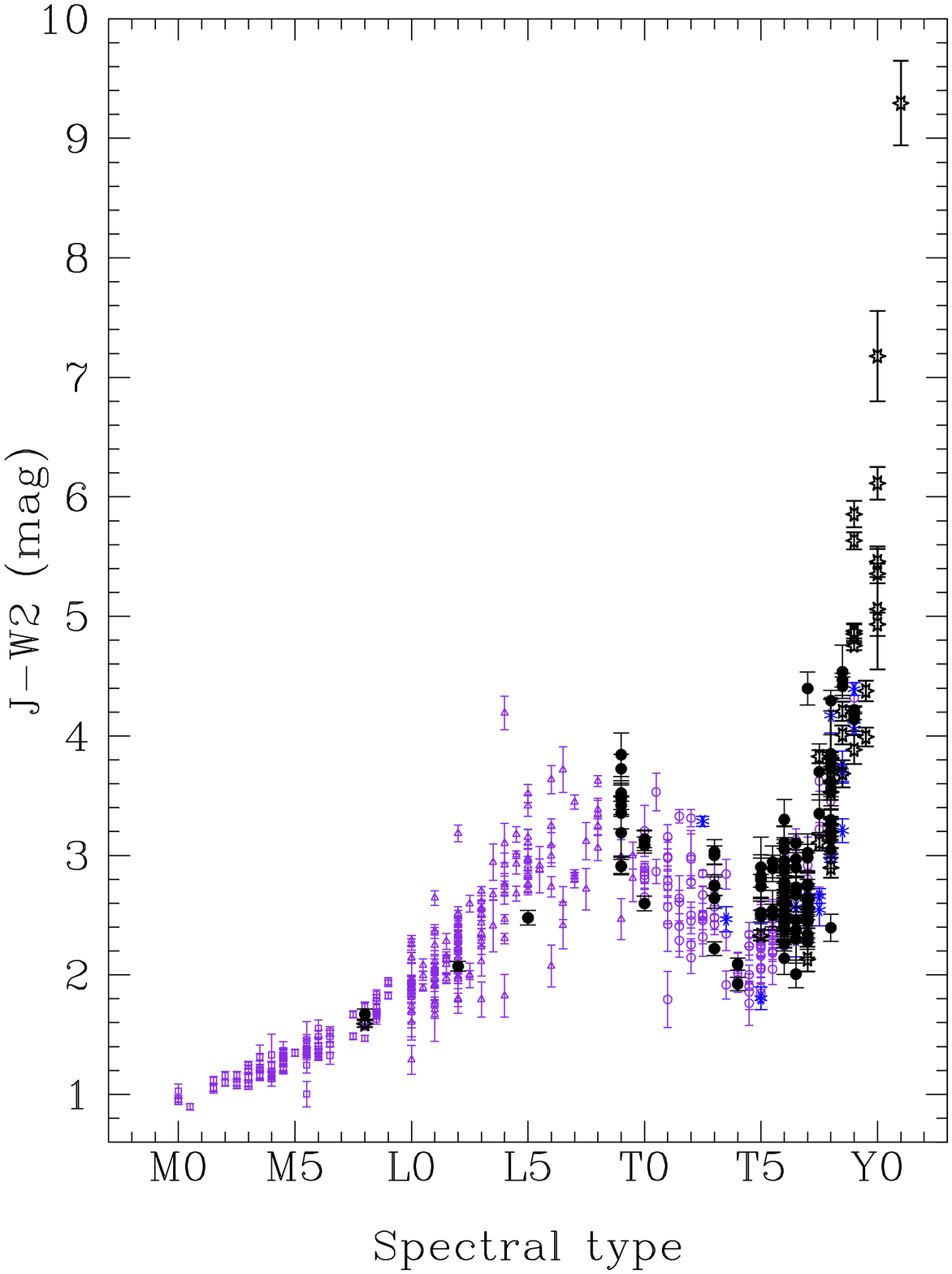}
\caption{$J$-W2 color versus spectral type. Color coding is the same as in Figure~\ref{JH_vs_type}.
\label{JW2_vs_type}}
\end{figure}

\clearpage

\begin{figure}
\epsscale{0.9}
\figurenum{8}
\plotone{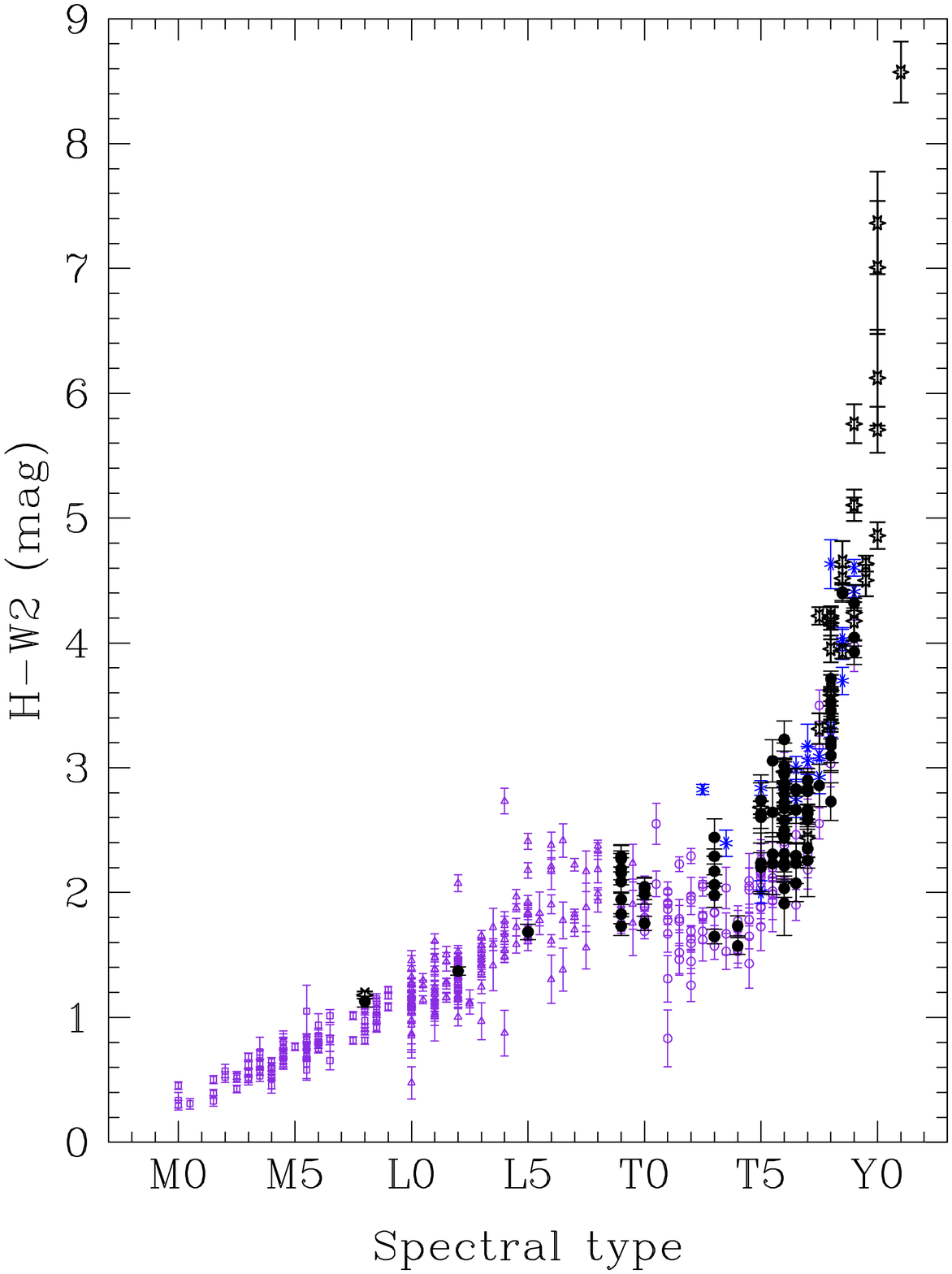}
\caption{$H$-W2 color versus spectral type. Color coding is the same as in Figure~\ref{JH_vs_type}.
\label{HW2_vs_type}}
\end{figure}

\clearpage

\begin{figure}
\epsscale{0.9}
\figurenum{9}
\plotone{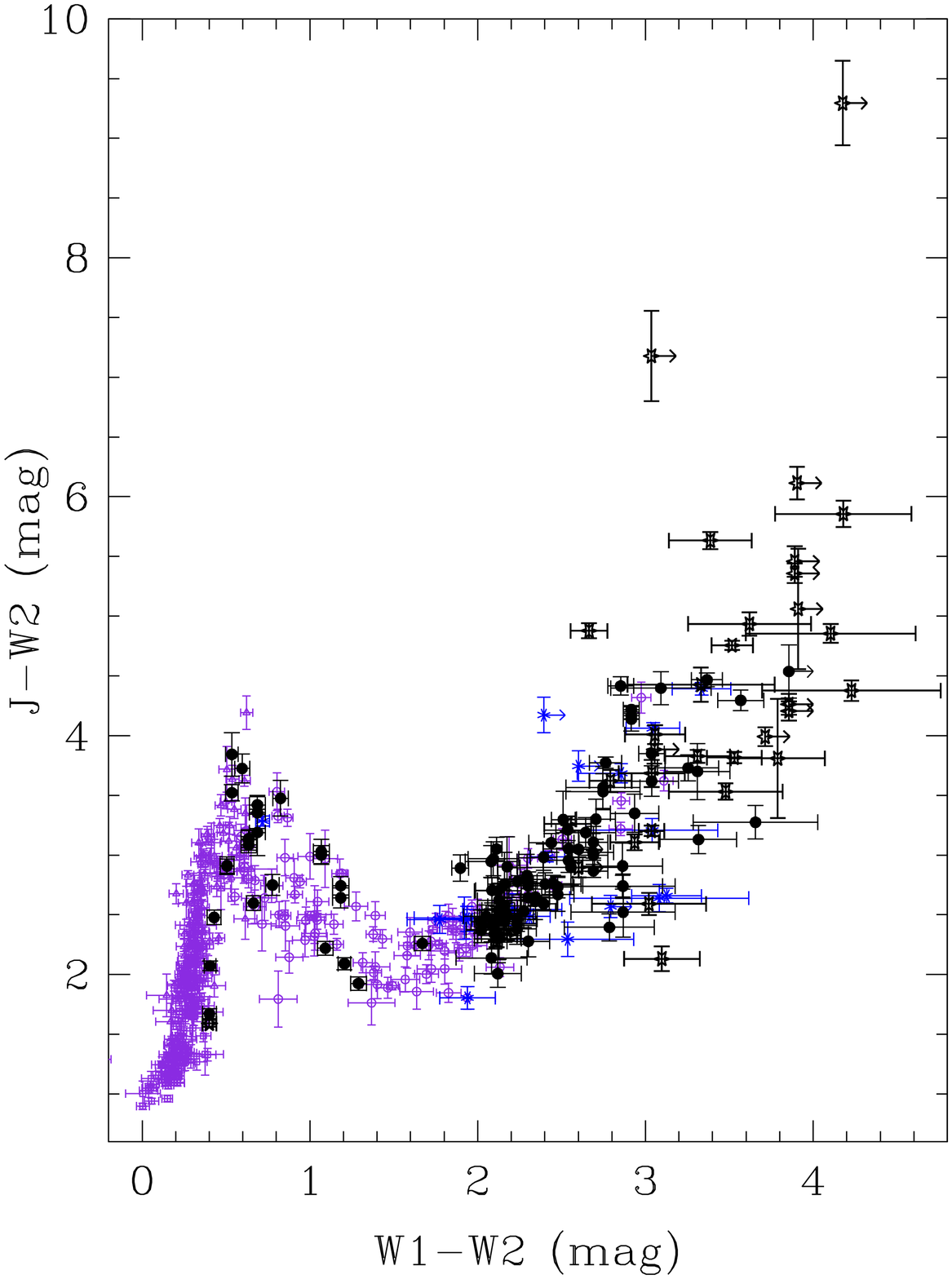}
\caption{$J$-W2 color plotted against W1-W2 color. Color coding is the same as in Figure~\ref{JH_vs_type}.
\label{JW2_vs_W1W2}}
\end{figure}

\clearpage

\begin{figure}
\epsscale{0.9}
\figurenum{10}
\plotone{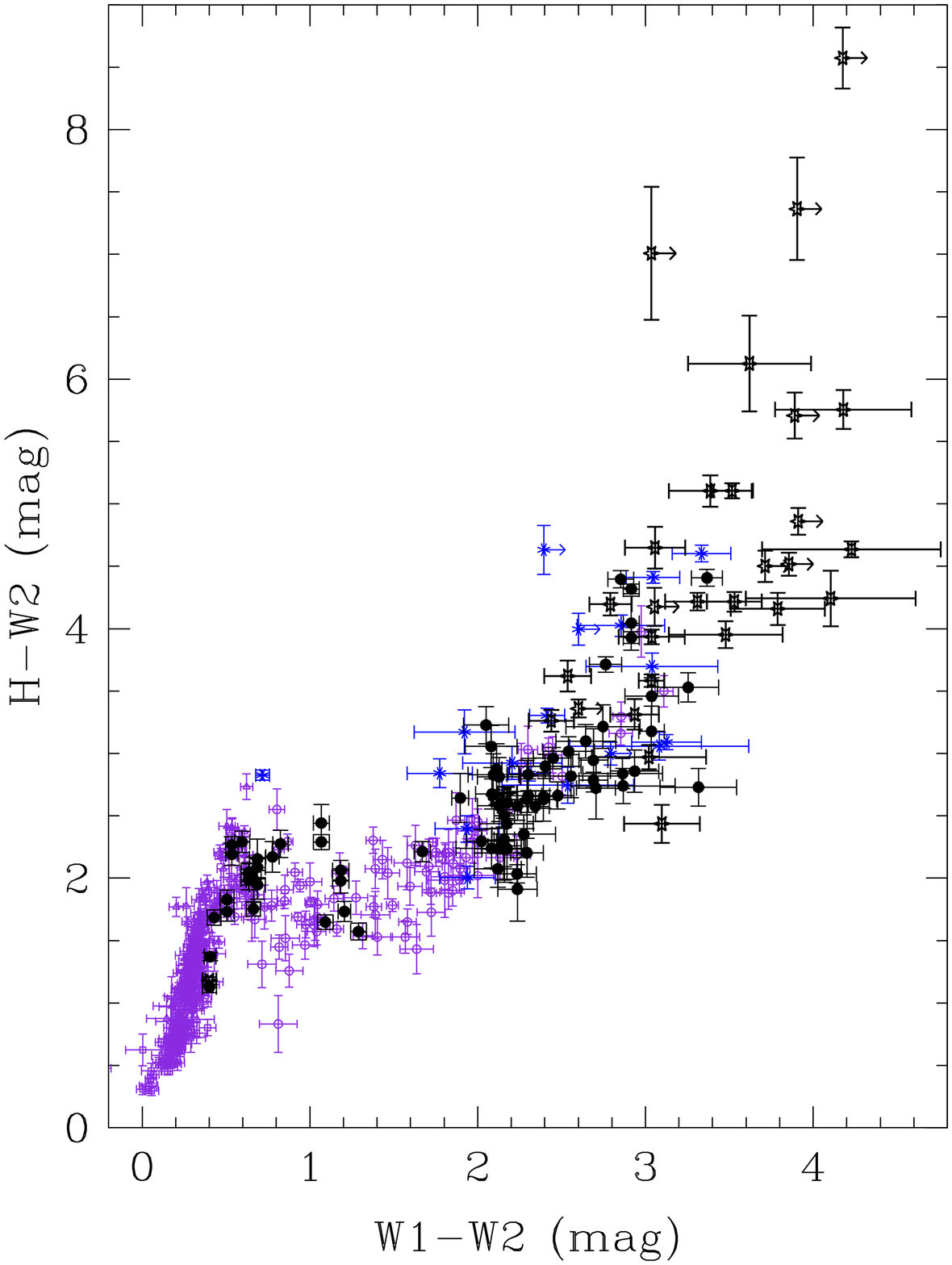}
\caption{$H$-W2 color plotted against W1-W2 color. Color coding is the same as in Figure~\ref{JH_vs_type}.
\label{HW2_vs_W1W2}}
\end{figure}

\clearpage

\begin{figure}
\epsscale{0.9}
\figurenum{11}
\plotone{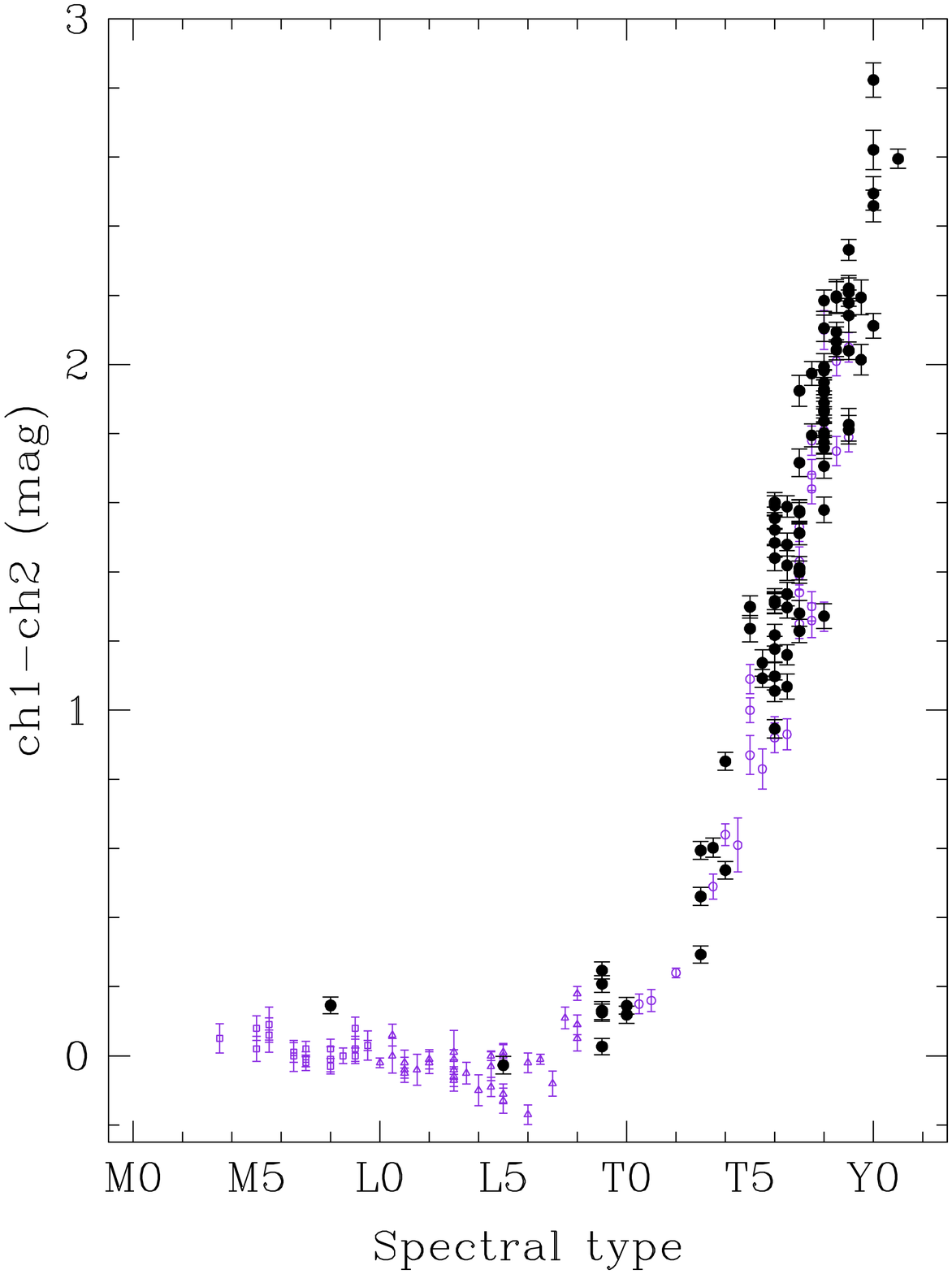}
\caption{{\it Spitzer} IRAC ch1-ch2 color as a function of spectral type. The color scheme is identical to that of 
Figure~\ref{W1W2_vs_type}.
\label{ch1ch2_vs_type}}
\end{figure}

\clearpage

\begin{figure}
\epsscale{0.9}
\figurenum{12}
\plotone{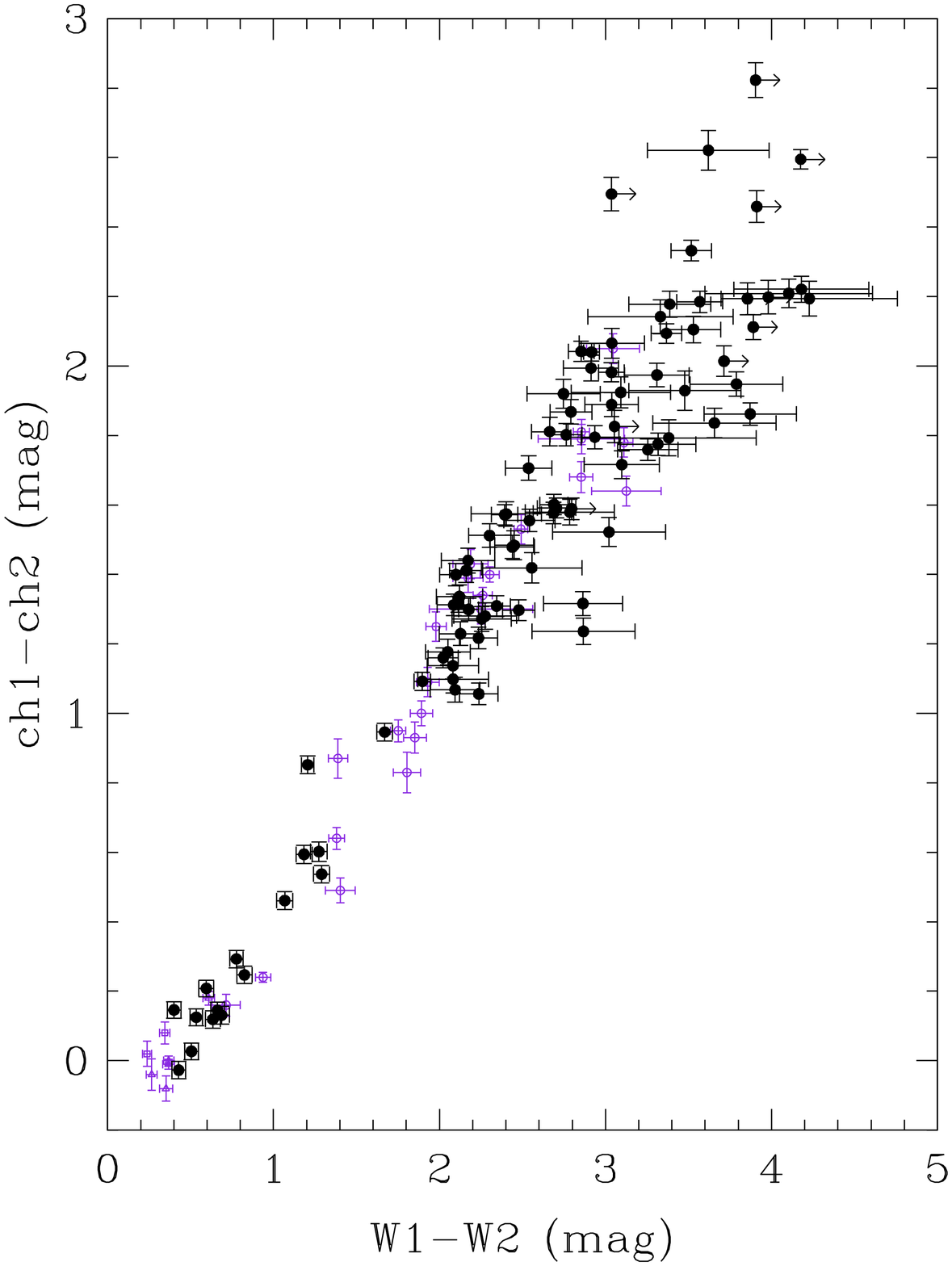}
\caption{{\it Spitzer} IRAC ch1-ch2 color as a function of W1-W2 color. The color scheme is identical to that of 
Figure~\ref{W1W2_vs_type}.
\label{ch1ch2_vs_W1W2}}
\end{figure}

\clearpage

\begin{figure}
\epsscale{0.9}
\figurenum{13}
\plotone{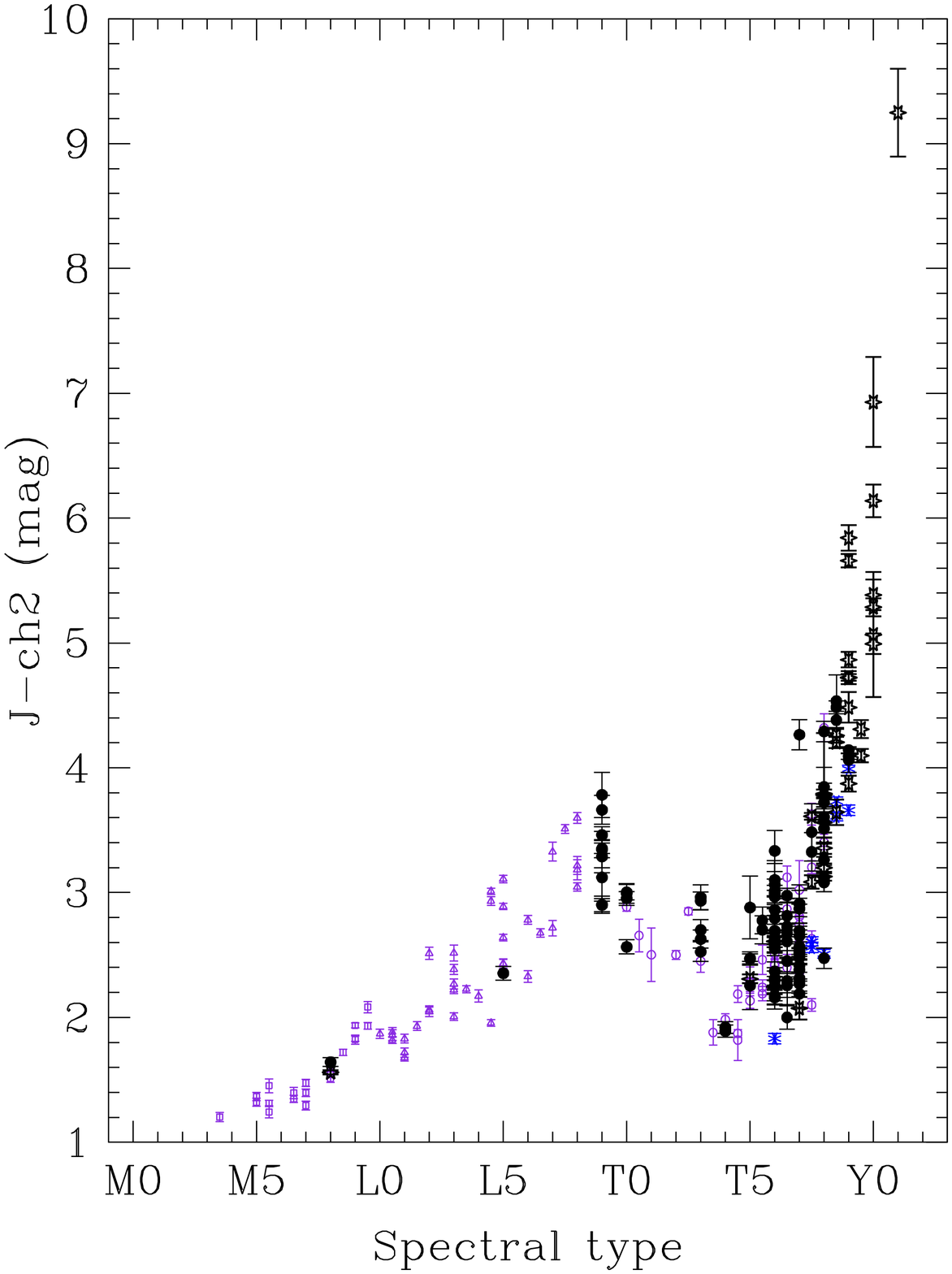}
\caption{$J$-ch2 color as a function of spectral type. Color coding is the same as in Figure~\ref{JH_vs_type}.
\label{Jch2_vs_type}}
\end{figure}

\clearpage

\begin{figure}
\epsscale{0.9}
\figurenum{14}
\plotone{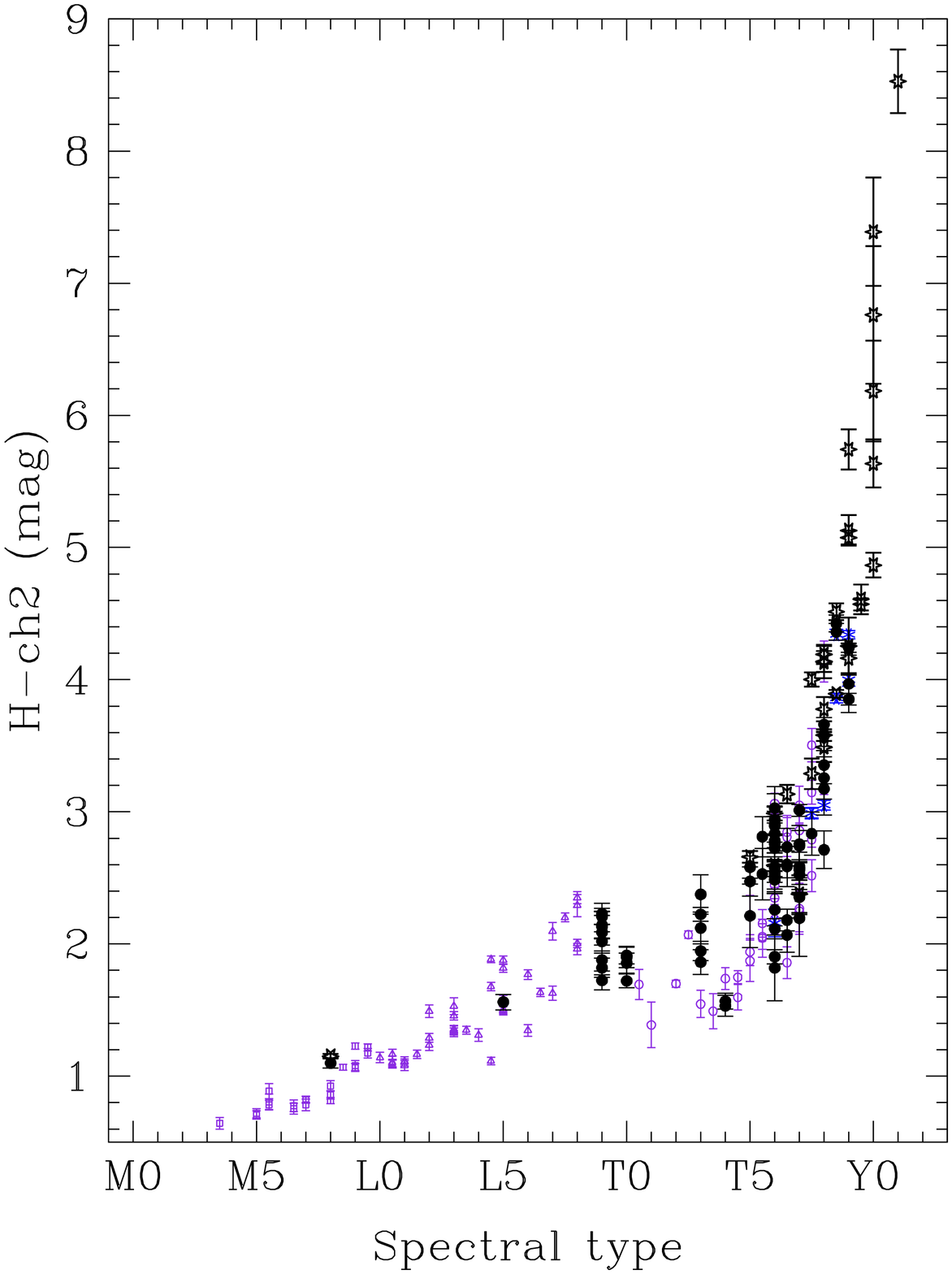}
\caption{$H$-ch2 color as a function of spectral type. Color coding is the same as in Figure~\ref{JH_vs_type}.
\label{Hch2_vs_type}}
\end{figure}

\clearpage

\begin{figure}
\epsscale{0.9}
\figurenum{15}
\plotone{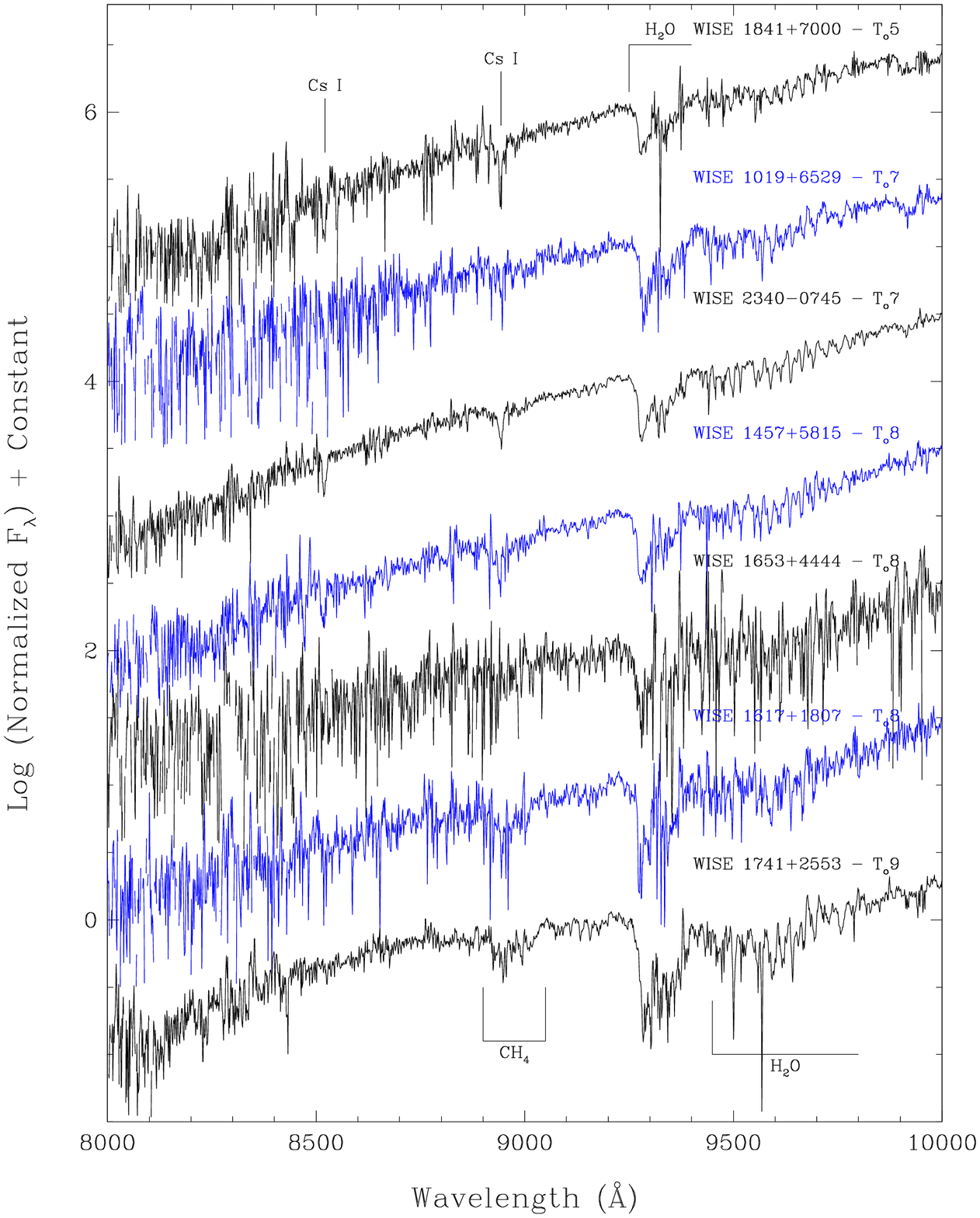}
\caption{Spectra from 8000 to 10000 \AA\ for seven WISE brown dwarf discoveries. Data have been 
corrected for telluric absorption and prominent spectral features are marked. All spectra have been
normalized at 9200 \AA\ and an integral offset added to the $y$-axis values to separate the spectra
vertically.
\label{optical_spectra}}
\end{figure}

\clearpage

\begin{figure}
\epsscale{0.9}
\figurenum{16}
\plotone{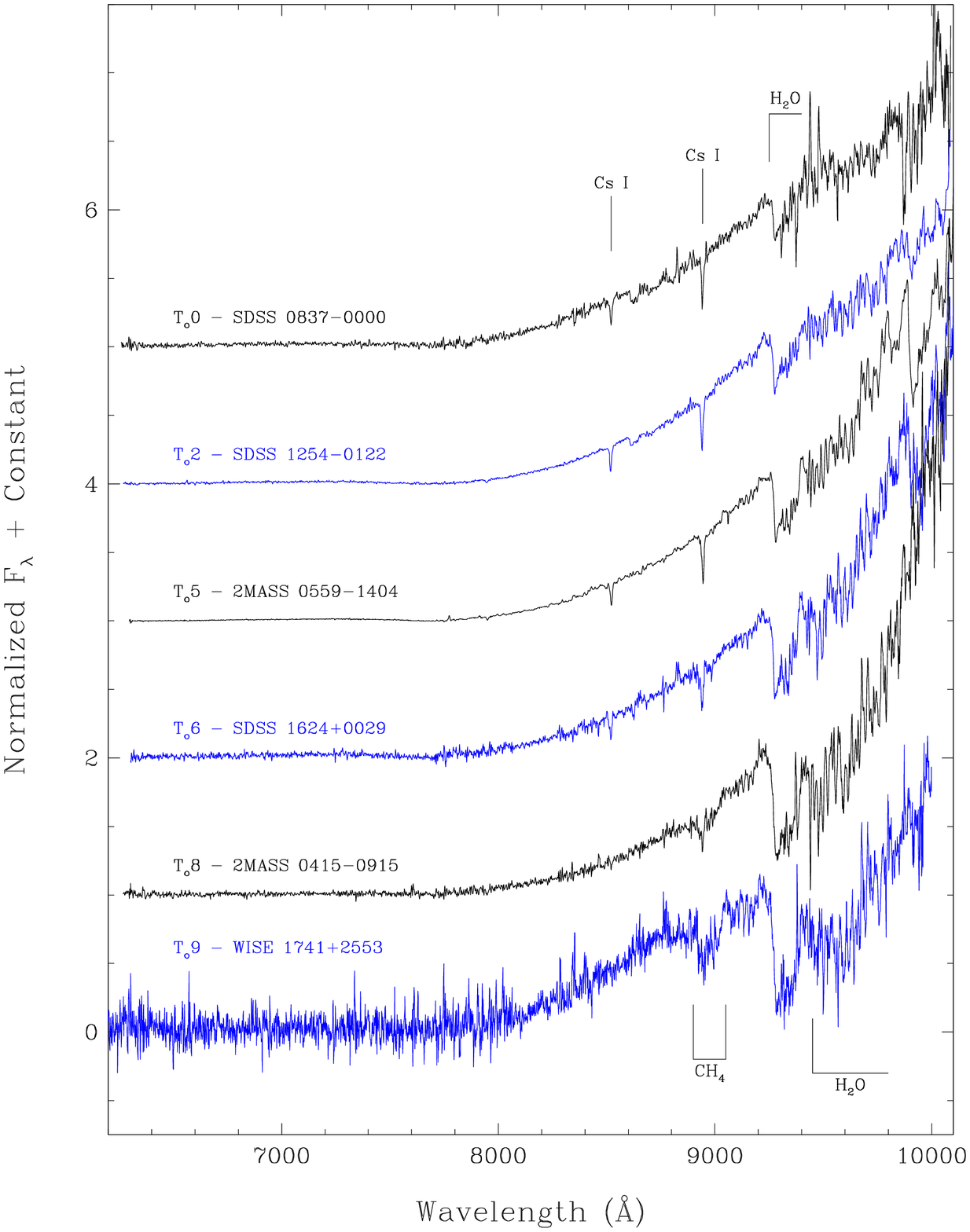}
\caption{Spectra from 6200 to 10000 \AA\ for the optical spectral standards from \cite{burgasser2006}
supplemented with the T$_o$0 standard from \cite{kirkpatrick2010} and the T$_o$9 standard proposed
here. Data have been 
corrected for telluric absorption and prominent spectral features are marked. All spectra have been
normalized at 9200 \AA\ and an integral offset added to the $y$-axis values to separate the spectra
vertically.
\label{optical_spectral_sequence}}
\end{figure}

\clearpage

\begin{figure}
\epsscale{0.9}
\figurenum{17}
\plotone{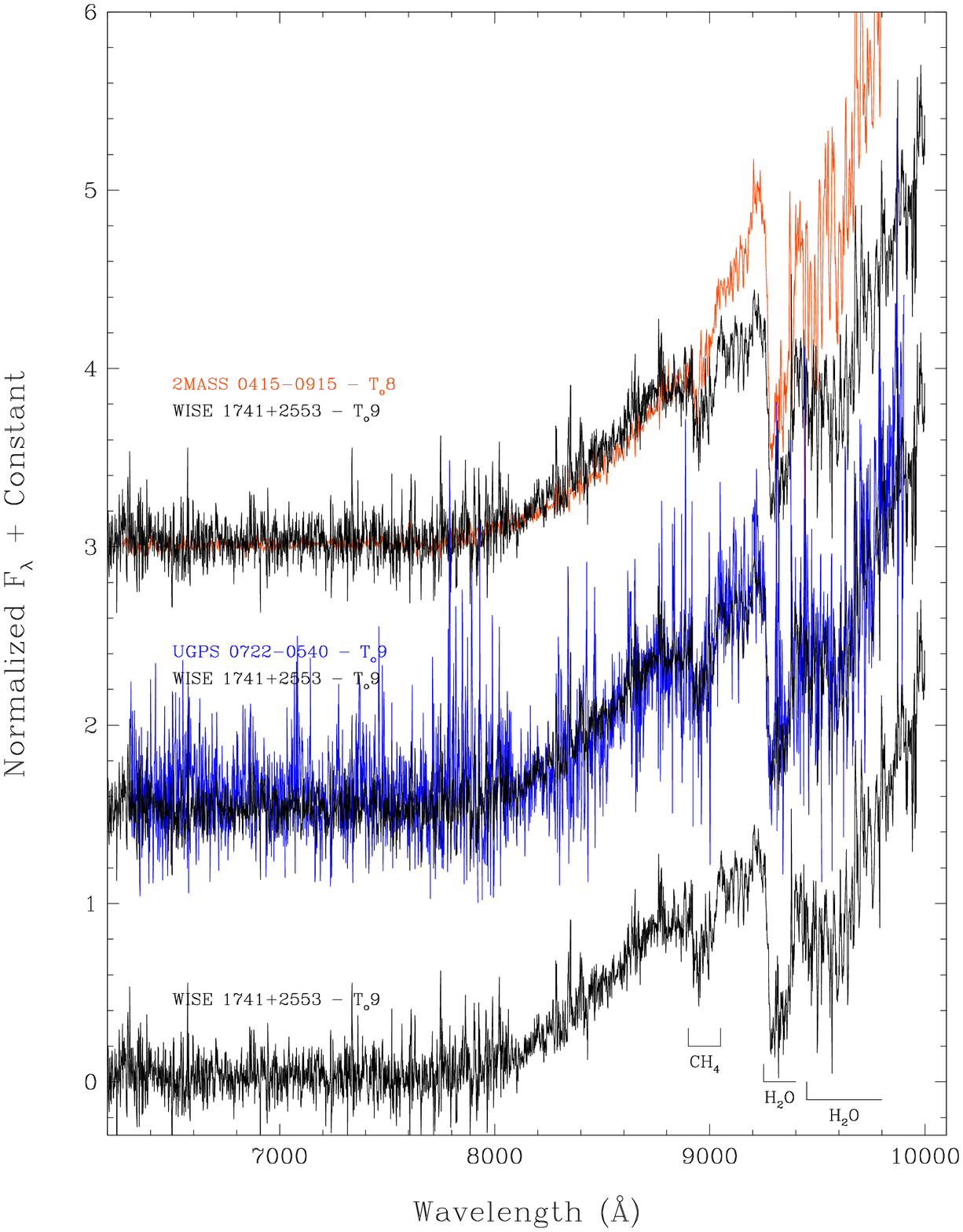}
\caption{Spectra from 6200 to 10000 \AA\ for three late-T dwarfs: the T$_o$8 standard 2MASS J04151954$-$0935066
(orange red), the T$_o$9 standard WISE 1741+2553 (black), and the T$_o$9 dwarf UGPS J072227.51$-$054031.2 (blue violet).
Data have been corrected for telluric absorption and prominent spectral features are marked. Spectra have been 
normalized at 8800 \AA\ and offsets in increments of 1.5 added to the $y$-axis values to separate the
spectra vertically except where overplotting was intended.
\label{optical_T9s}}
\end{figure}

\clearpage

\begin{figure}
\epsscale{0.9}
\figurenum{18}
\plotone{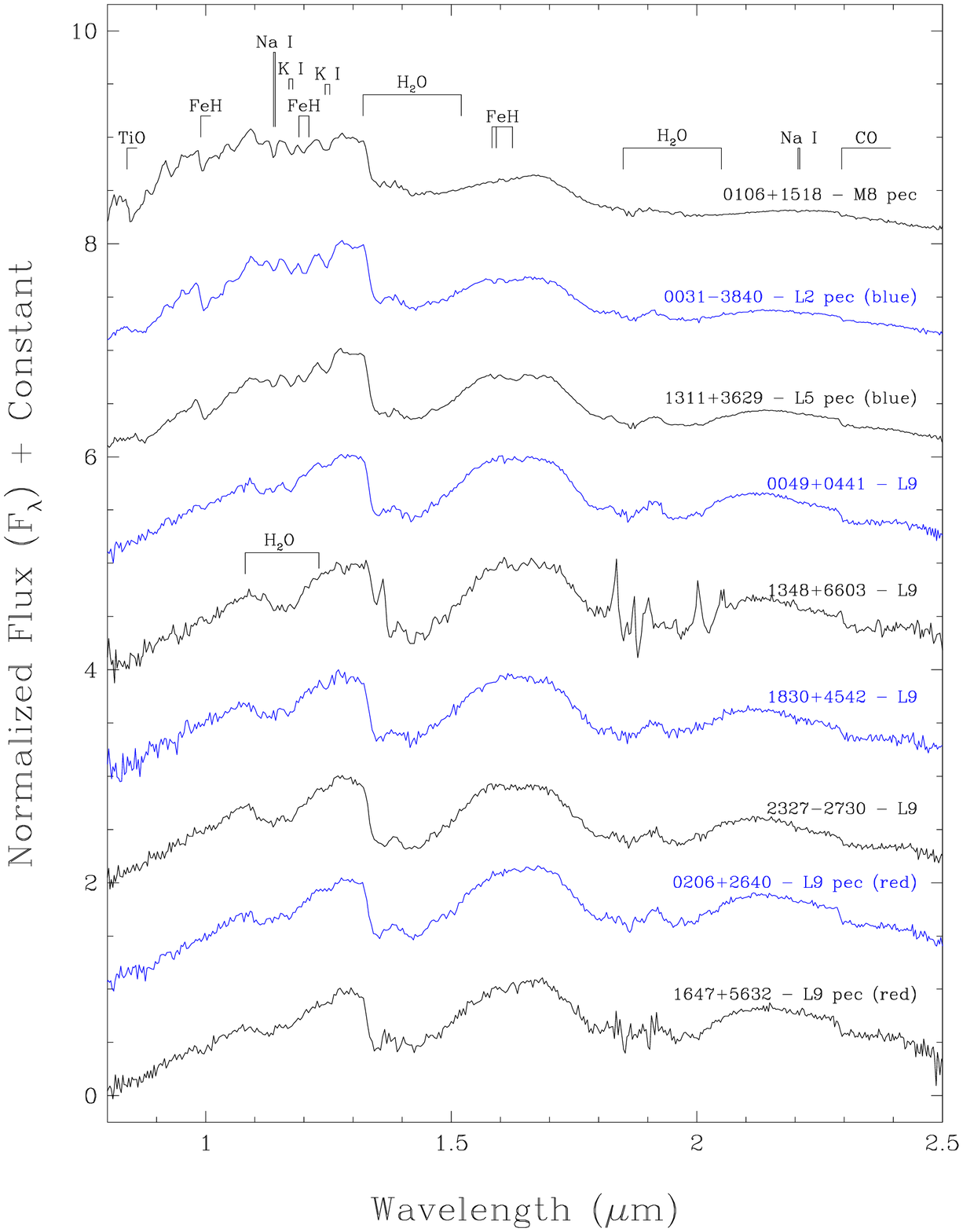}
\caption{Near-infrared spectra of confirmed WISE brown dwarfs with spectral types
earlier than T0. Spectra
have been normalized to one around 1.28 $\mu$m and integral offsets have been added to the $y$-axis values
to separate the spectra vertically.
Prominent spectral features are marked.
\label{other_spectra_lt_T0}}
\end{figure}

\clearpage

\begin{figure}
\epsscale{0.9}
\figurenum{19}
\plotone{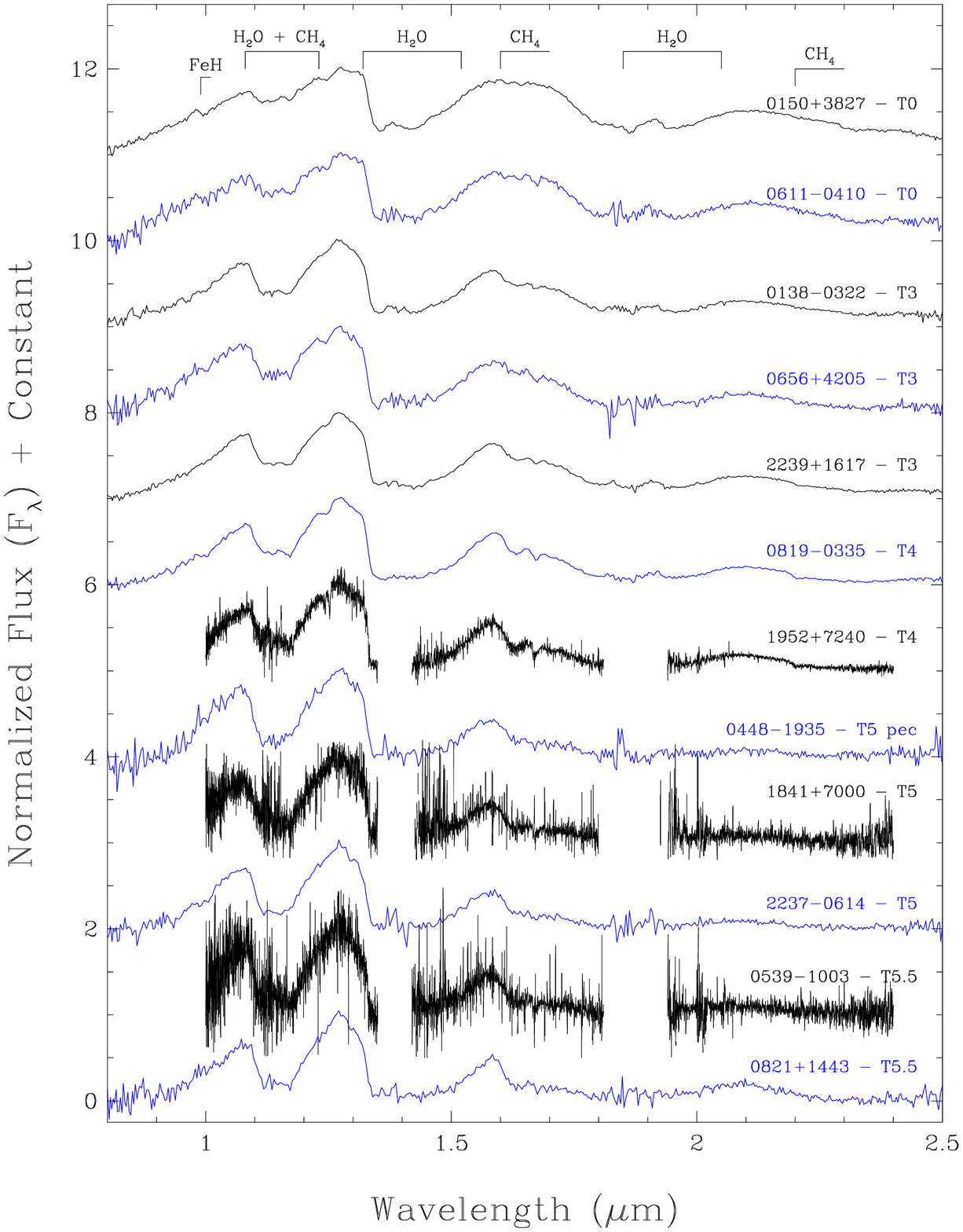}
\caption{Near-infrared spectra of confirmed WISE brown dwarfs with spectral types
from T0 to T5.5. Spectra
have been normalized to one around 1.28 $\mu$m and integral offsets have been added to the $y$-axis values
to separate the spectra vertically. For some spectra, noisy data in the depths of the telluric
water bands are not plotted. Prominent spectral features are marked.
\label{other_spectra_T0-T5.5}}
\end{figure}

\clearpage

\begin{figure}
\epsscale{0.9}
\figurenum{20}
\plotone{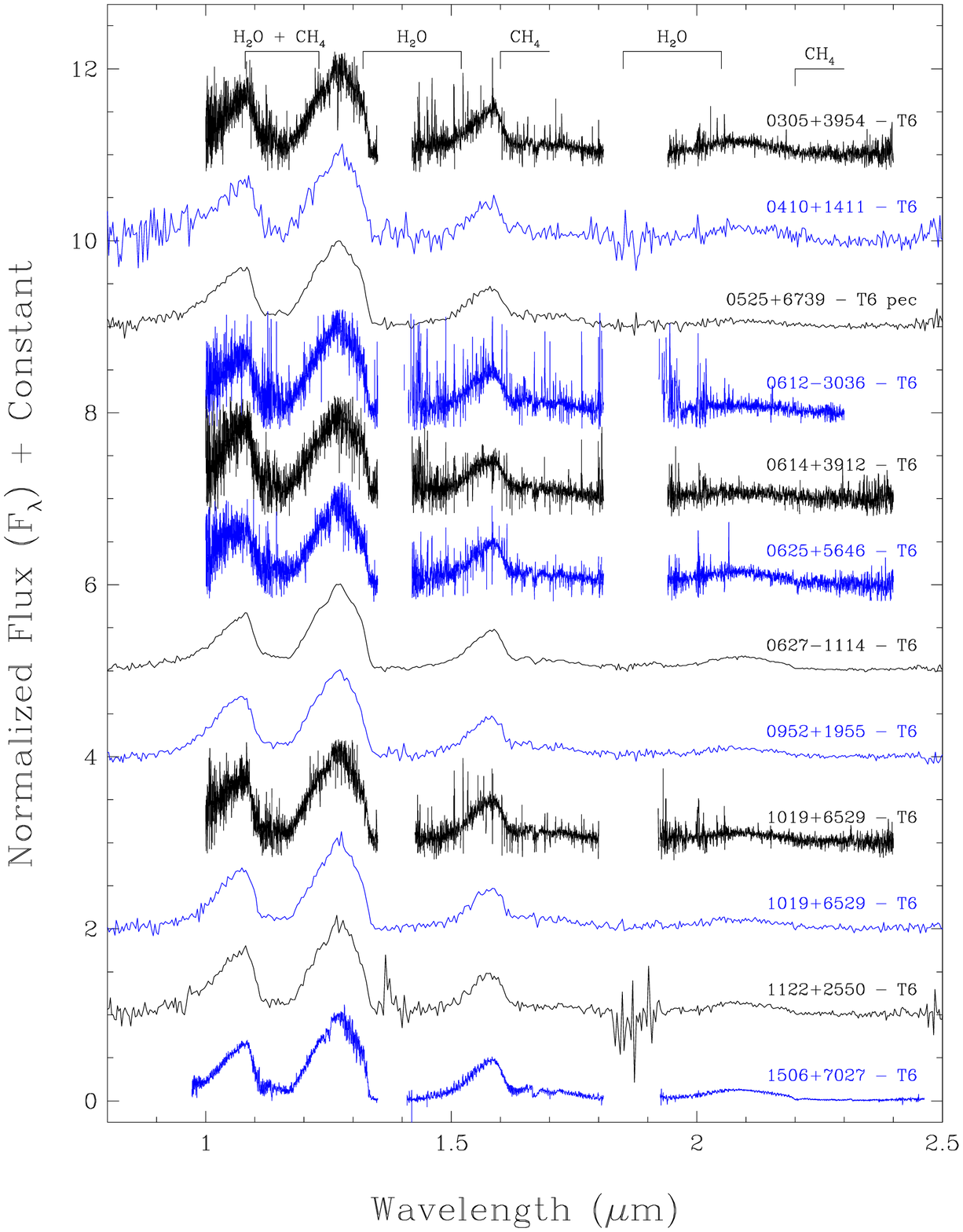}
\caption{Near-infrared spectra of confirmed WISE brown dwarfs with spectral types of T6. (Additional T6 dwarfs
are shown in Figure~\ref{other_spectra_T6-T7}.) Spectra
have been normalized to one around 1.28 $\mu$m and integral offsets have been added to the $y$-axis values
to separate the spectra vertically. For some spectra, noisy data in the depths of the telluric
water bands are not plotted. Prominent spectral features are marked.
\label{other_spectra_T6-T6}}
\end{figure}

\clearpage

\begin{figure}
\epsscale{0.9}
\figurenum{21}
\plotone{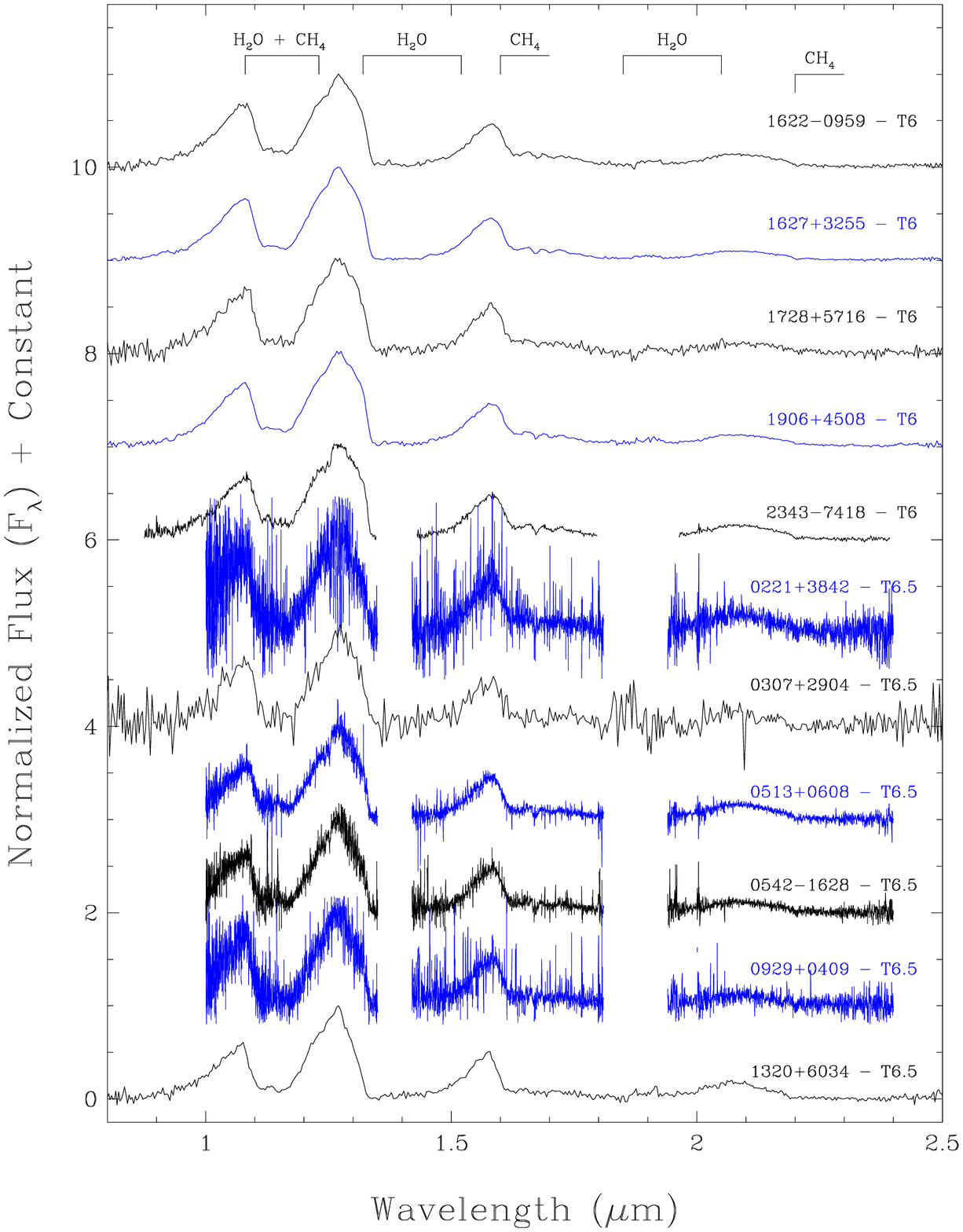}
\caption{Near-infrared spectra of confirmed WISE brown dwarfs with spectral types from T6 (continued) to T6.5. Spectra
have been normalized to one around 1.28 $\mu$m and integral offsets have been added to the $y$-axis values
to separate the spectra vertically. For some spectra, noisy data in the depths of the telluric
water bands are not plotted. Prominent spectral features are marked.
\label{other_spectra_T6-T7}}
\end{figure}

\clearpage

\begin{figure}
\epsscale{0.9}
\figurenum{22}
\plotone{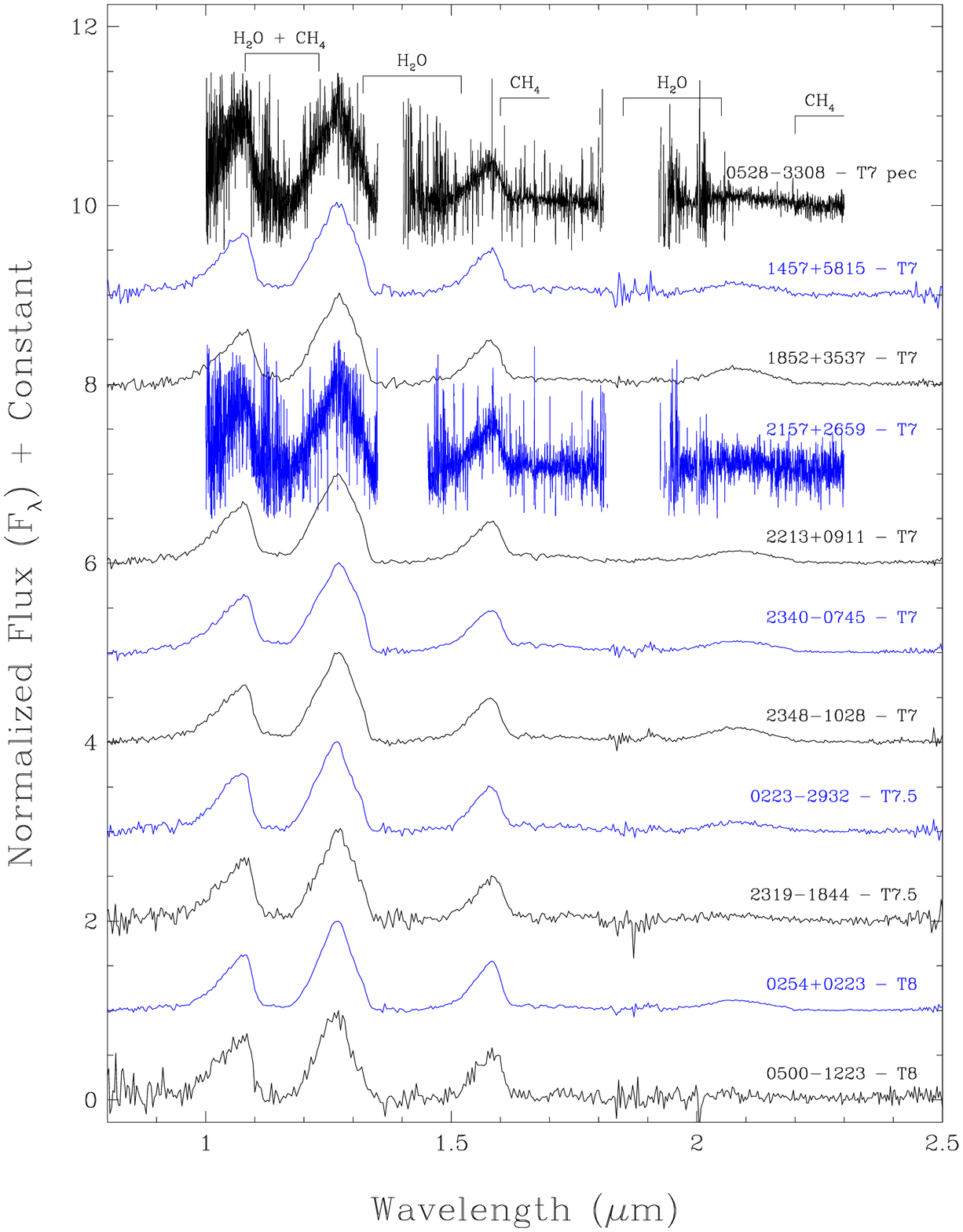}
\caption{Near-infrared spectra of confirmed WISE brown dwarfs with spectral types from T7 to T8. (Additional T8 dwarfs
are shown in Figure~\ref{other_spectra_T8-T8} and Figure~\ref{other_spectra_T8-T8.5}.) Spectra
have been normalized to one around 1.28 $\mu$m and integral offsets have been added to the $y$-axis values
to separate the spectra vertically. For some spectra, noisy data in the depths of the telluric
water bands are not plotted. Prominent spectral features are marked.
\label{other_spectra_T7-T7.5}}
\end{figure}

\clearpage

\begin{figure}
\epsscale{0.9}
\figurenum{23}
\plotone{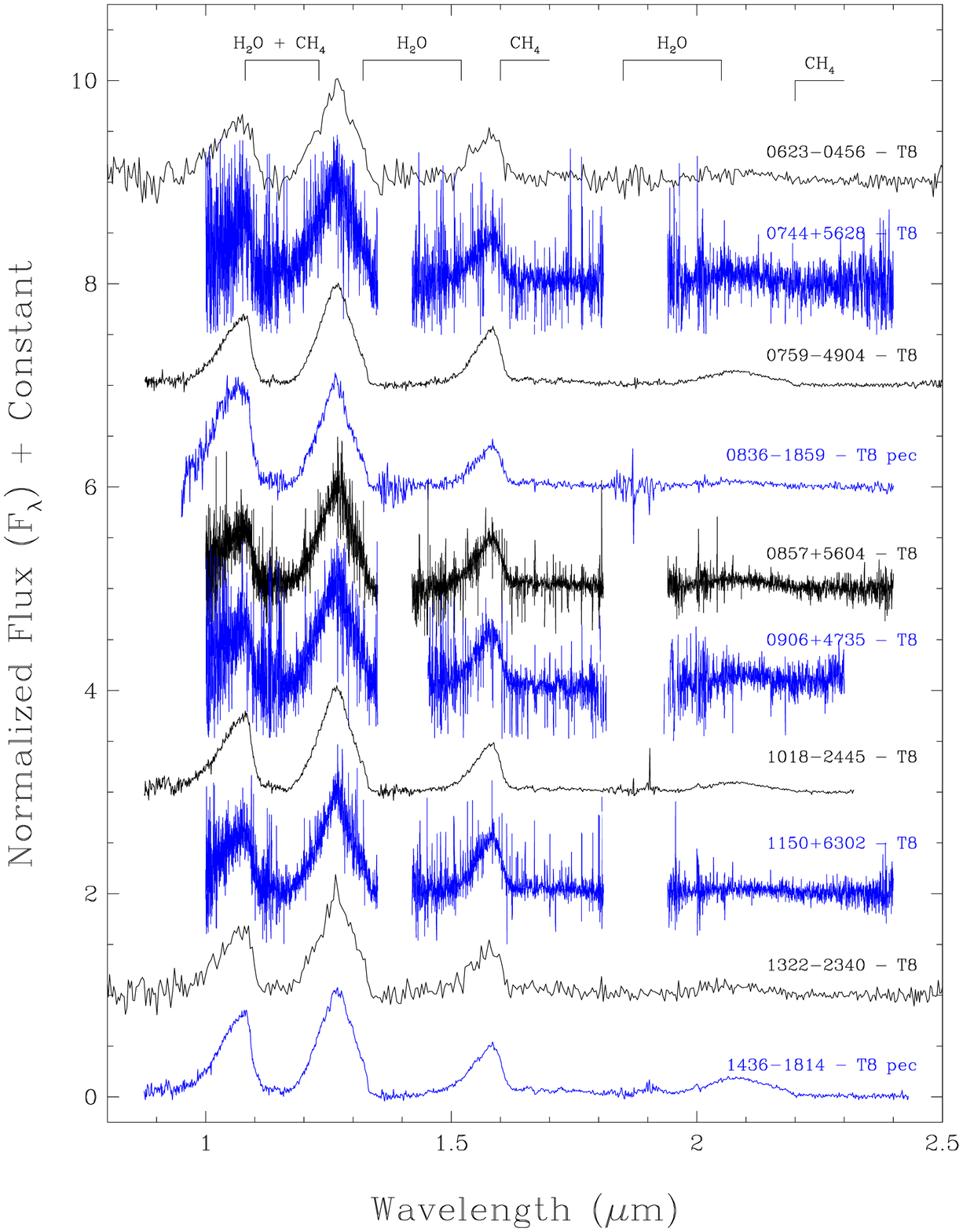}
\caption{Near-infrared spectra of confirmed WISE brown dwarfs with spectral types of T8 (continued). Spectra
have been normalized to one around 1.28 $\mu$m and integral offsets have been added to the $y$-axis values
to separate the spectra vertically. For some spectra, noisy data in the depths of the telluric
water bands are not plotted. Prominent spectral features are marked.
\label{other_spectra_T8-T8}}
\end{figure}

\clearpage

\begin{figure}
\epsscale{0.9}
\figurenum{24}
\plotone{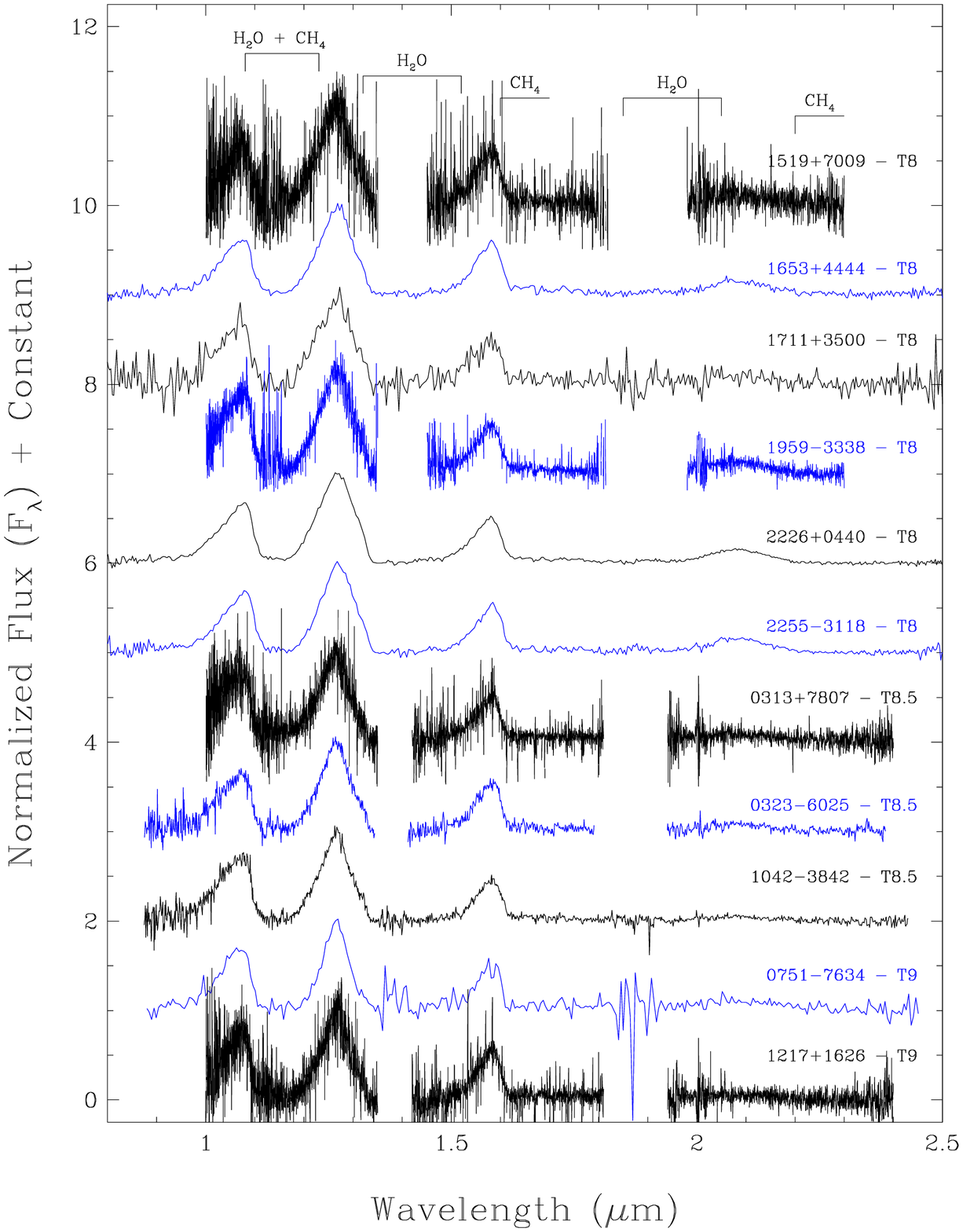}
\caption{Near-infrared spectra of confirmed WISE brown dwarfs with spectral types from T8 (continued) to T9. (Additional T9 dwarfs
are shown in Figure~\ref{other_spectra_gt_T8}.) Spectra
have been normalized to one around 1.28 $\mu$m and integral offsets have been added to the $y$-axis values
to separate the spectra vertically. For some spectra, noisy data in the depths of the telluric
water bands are not plotted. Prominent spectral features are marked.
\label{other_spectra_T8-T8.5}}
\end{figure}

\clearpage

\begin{figure}
\epsscale{0.9}
\figurenum{25}
\plotone{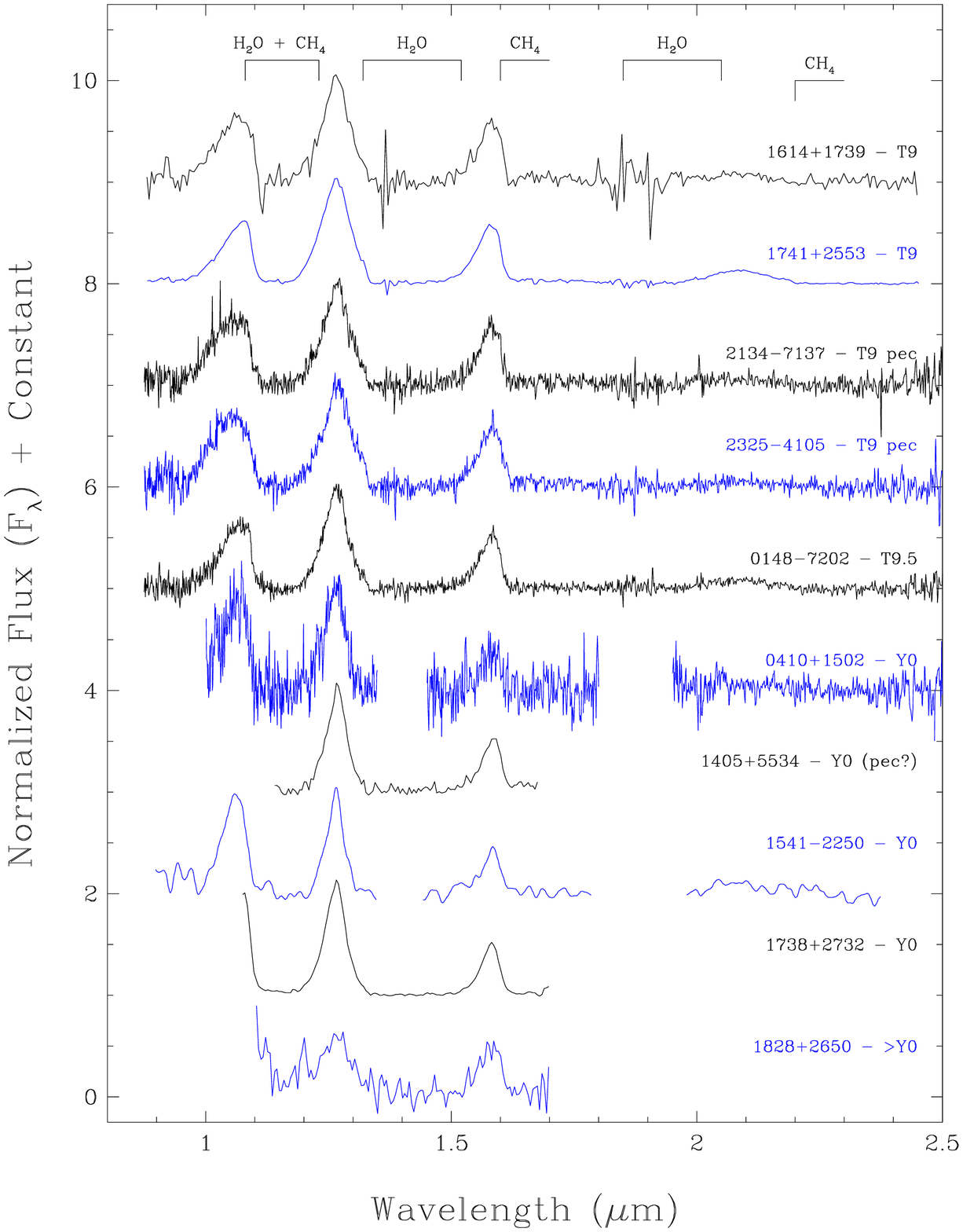}
\caption{Near-infrared spectra of confirmed WISE brown dwarfs with spectral types from T9 to early Y. Spectra
have been normalized to one around 1.28 $\mu$m and integral offsets have been added to the $y$-axis values
to separate the spectra vertically. For some spectra, noisy data in the depths of the telluric
water bands are not plotted. Prominent spectral features are marked.
\label{other_spectra_gt_T8}}
\end{figure}

\clearpage

\begin{figure}
\epsscale{0.9}
\figurenum{26}
\plotone{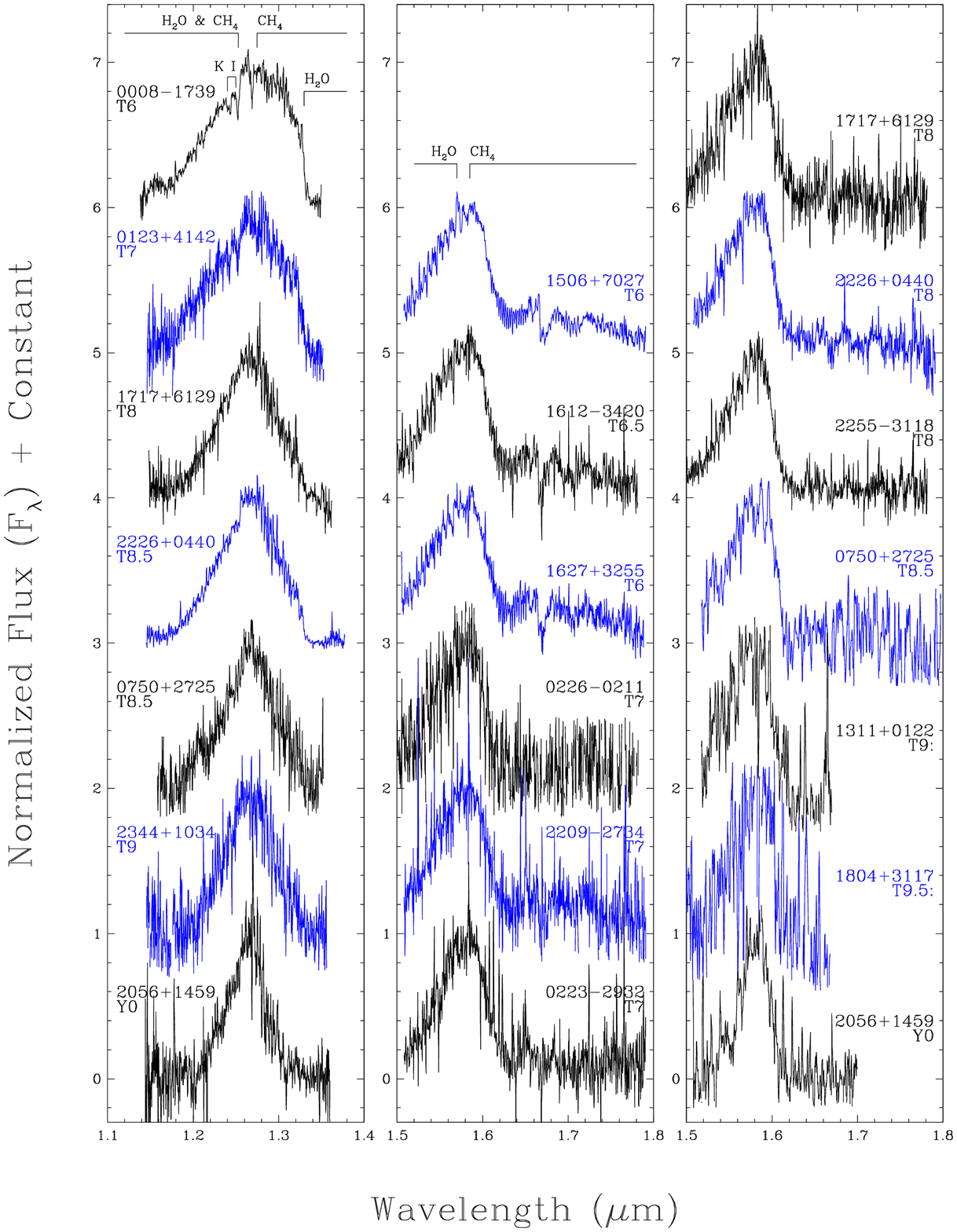}
\caption{Keck/NIRSPEC spectra of confirmed WISE brown dwarfs. $J$-band spectra are shown
in the left panel and $H$-band spectra are shown in the middle and right panels. Spectra
have been normalized to one at peak flux and integral offsets have been added to the $y$-axis values
to separate the spectra vertically. The bottom three spectra in the rightmost plot -- those
of WISE 1311+0122, WISE 1804+3117, and WISE 2056+1459 -- have been smoothed with a five-pixel boxcar.
Prominent spectral features are marked.
\label{NIRSPEC_spectra}}
\end{figure}

\clearpage

\begin{figure}
\epsscale{0.9}
\figurenum{27}
\plotone{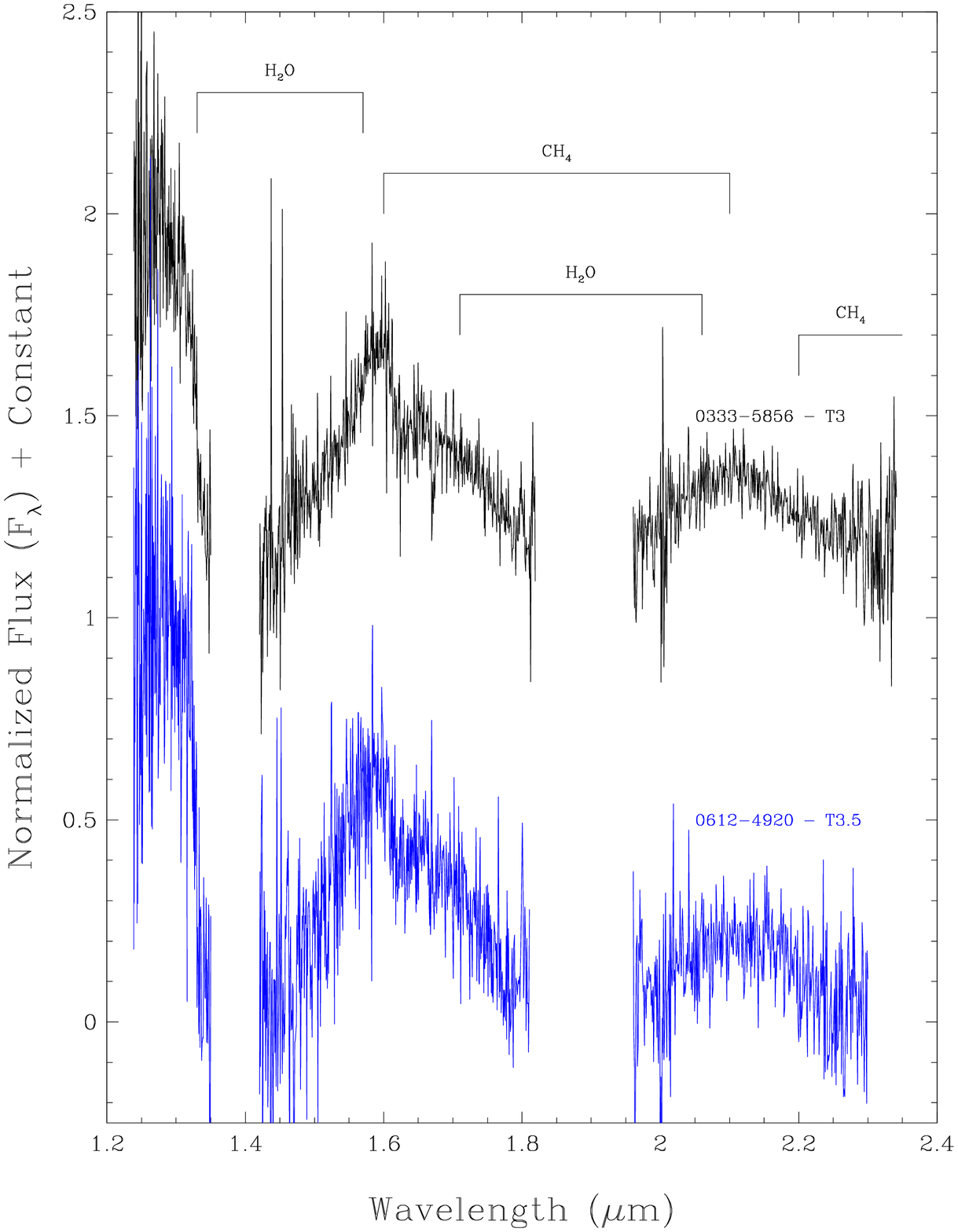}
\caption{SOAR/OSIRIS spectra of two confirmed WISE brown dwarfs. Noisy data in the depths of the
telluric water bands near 1.4 and 1.9 $\mu$m are not plotted. Spectra
have been normalized to one at peak flux and an offset of 1 has been added to the $y$-axis 
value of WISE 0333$-$5856 to separate it vertically from WISE 0612$-$4920.
Prominent spectral features are marked.
\label{OSIRIS_spectra}}
\end{figure}

\clearpage

\begin{figure}
\epsscale{0.85}
\figurenum{28}
\plotone{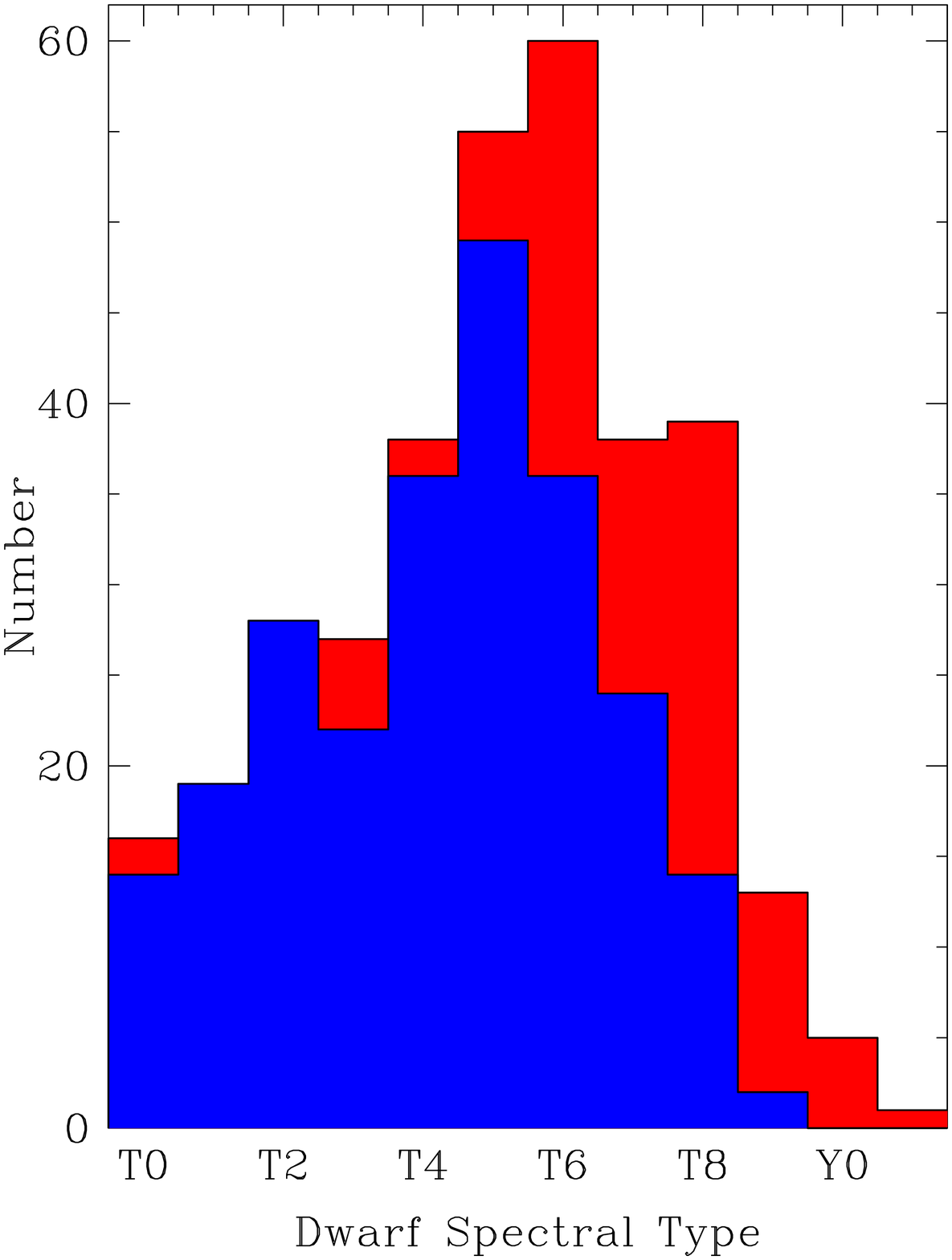}
\caption{A histogram of spectral types for published objects with near-infrared types of T0 or later.
The distribution of types of previously published brown dwarfs is shown in blue
and the distribution of new types for discoveries in this paper is shown in red. Objects are counted
into bins of integral subtypes (e.g., objects of type T7 and T7.5 are shown in the T7 bin). The
previously published objects were taken from DwarfArchives.org on 2011 May 15 and are supplemented
with new discoveries by \cite{albert2011} and \cite{burningham2011}.
\label{spec_type_histogram}}
\end{figure}

\clearpage

\begin{figure}
\epsscale{0.8}
\figurenum{29}
\plotone{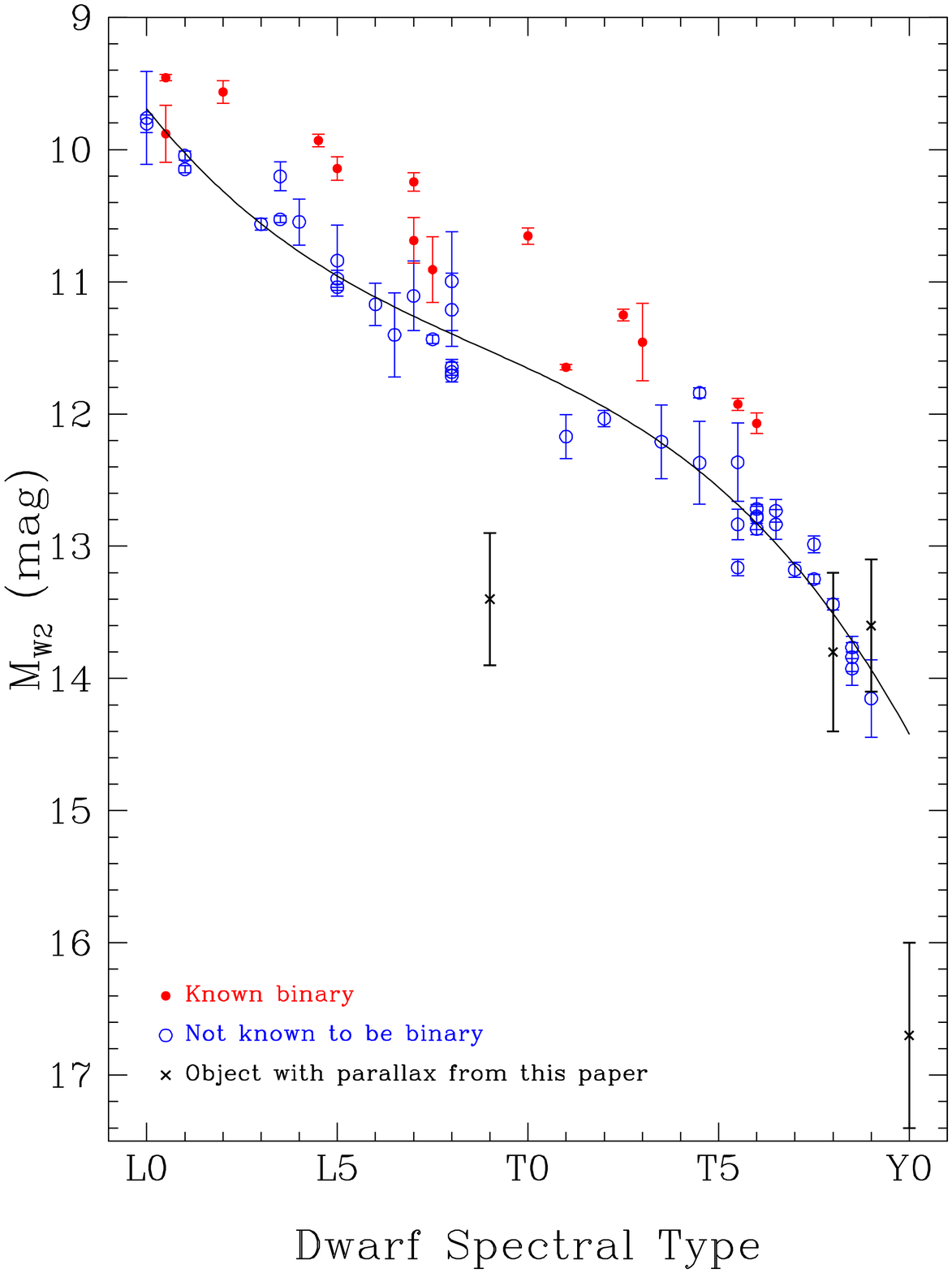}
\caption{Absolute W2 magnitude plotted against spectral type for objects with measured trigonometric
parallaxes. Red points are those objects known to be binary through high-resolution imaging. All others
are shown as blue points. The solid line shows a third-order relation fit through the blue points, as described
in the text. WISE discoveries with trigonometric parallaxes from Tabe~\ref{prelim_parallaxes} are shown with
black x's. The point for WISE 1541$-$2250 at lower right, if confirmed via continued astrometric monitoring,
suggests that an extrapolation of the fitted relation cooler than T9 may result in Y dwarf spectrophotometric distance
estimates that are too large.
\label{MW2_vs_type}}
\end{figure}

\clearpage

\begin{figure}
\epsscale{0.8}
\figurenum{30}
\plotone{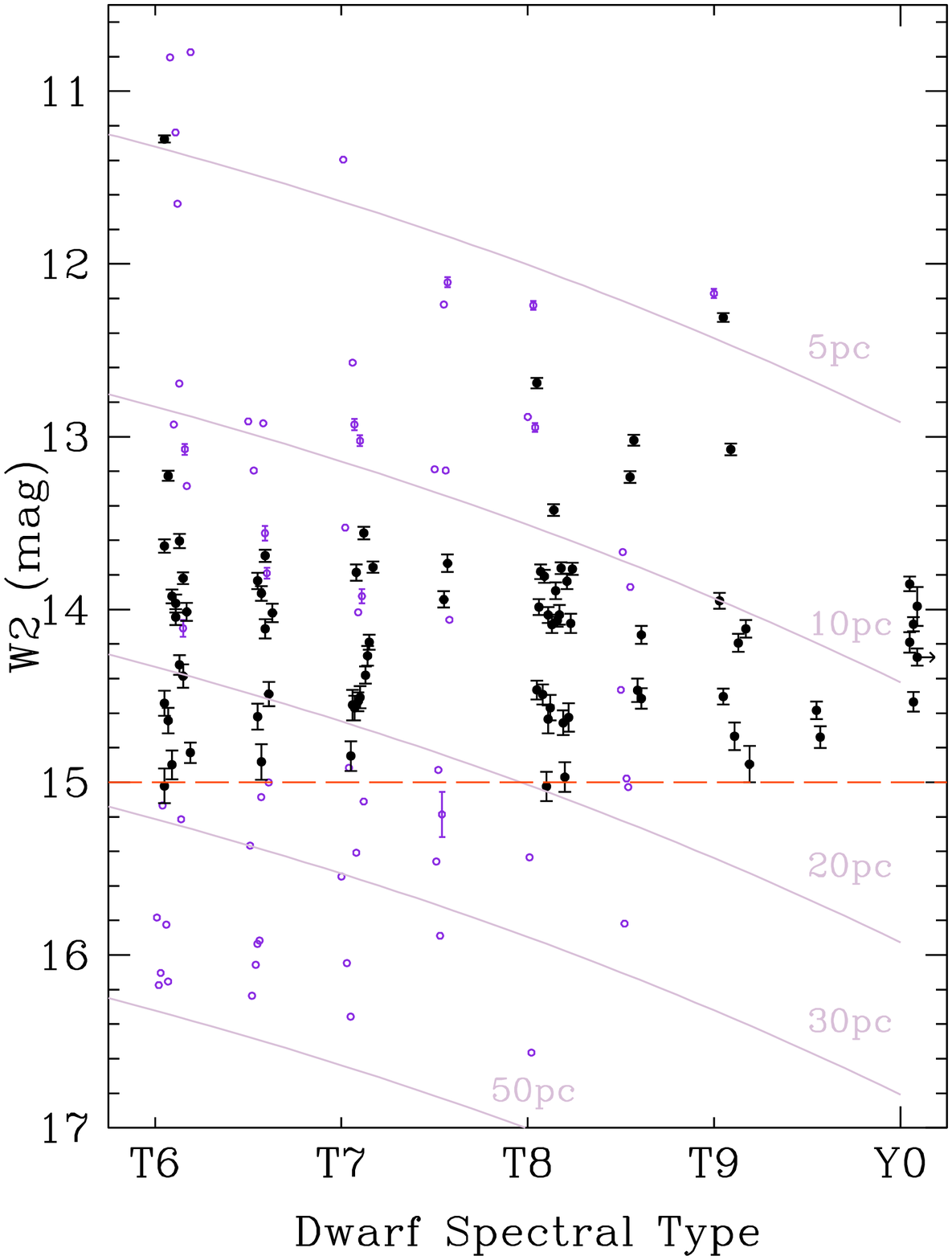}
\caption{W2 vs.\ spectral type for previously known objects (open, blue violet points) and WISE 
discoveries (solid, black points). Slight offsets have been added to the spectral subclass of each
object so that points suffer from less overlap along the $x$-axis.
The distance relation from Figure~\ref{MW2_vs_type} is plotted at various distances from 5 to 50 pc 
(grey lines) to aid
the viewer in estimating distances to plotted objects.
The dashed line in orange red shows the approximate W2 magnitude limit of our current WISE search.
(Note: Error bars are shown on the open points when W2 photometry is known. Otherwise, open points 
are plotted at the W2 magnitudes estimated from the objects' spectral types and near-infrared
magnitudes and are plotted without error bars.)
\label{distances}}
\end{figure}

\clearpage

\begin{figure}
\epsscale{0.8}
\figurenum{31}
\plotone{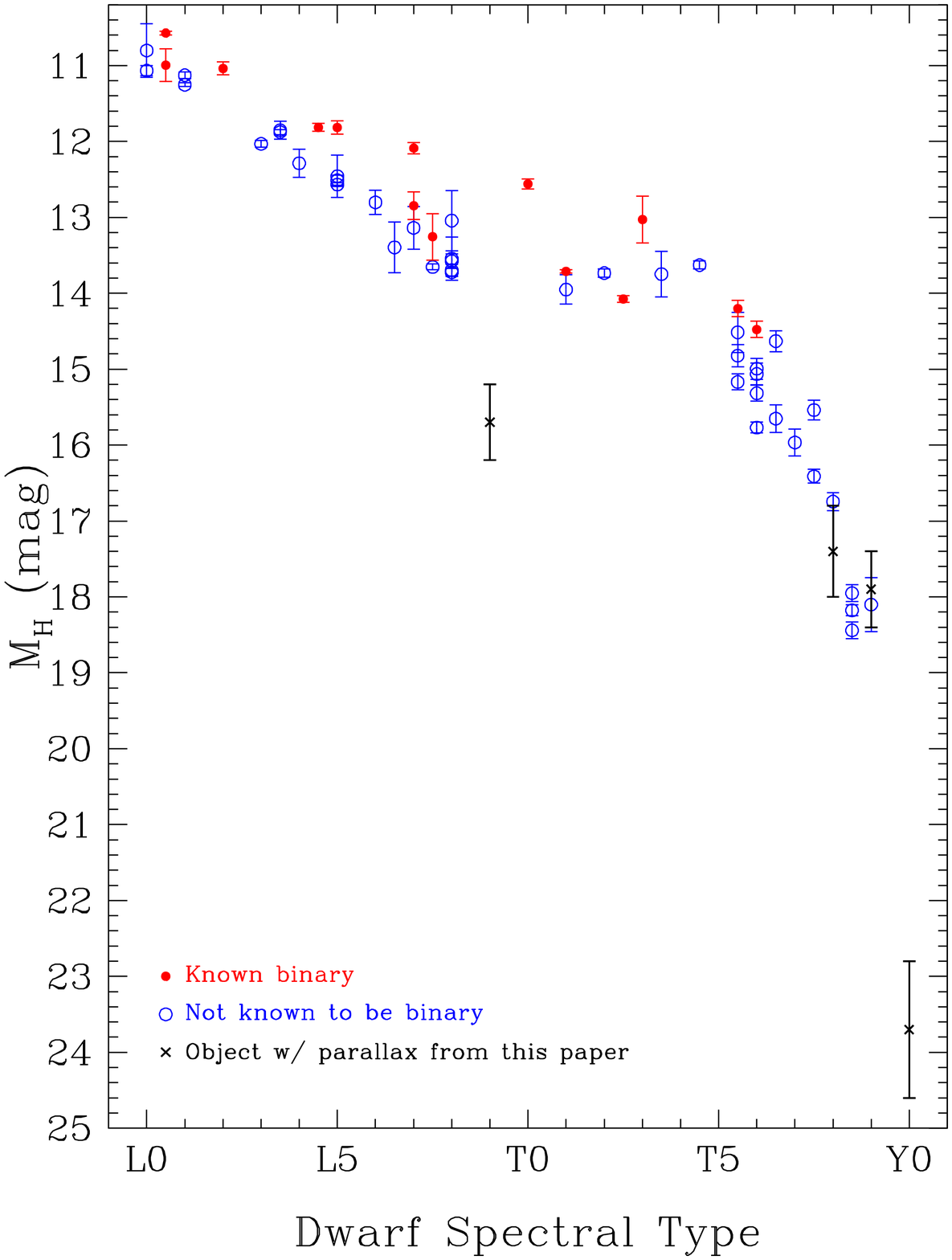}
\caption{Absolute $H$ magnitude plotted against spectral type for objects with measured trigonometric
parallaxes in Figure~\ref{MW2_vs_type}. Note the rapid dimming of the $H$-band magnitude at the latest T types.
The point for WISE 1541$-$2250 at lower right, if confirmed via continued astrometric monitoring,
suggests that this $H$-band dimming accelerates as objects cool to the Y dwarf class.
\label{MH_vs_type}}
\end{figure}

\clearpage

\begin{figure}
\epsscale{0.8}
\figurenum{32}
\plotone{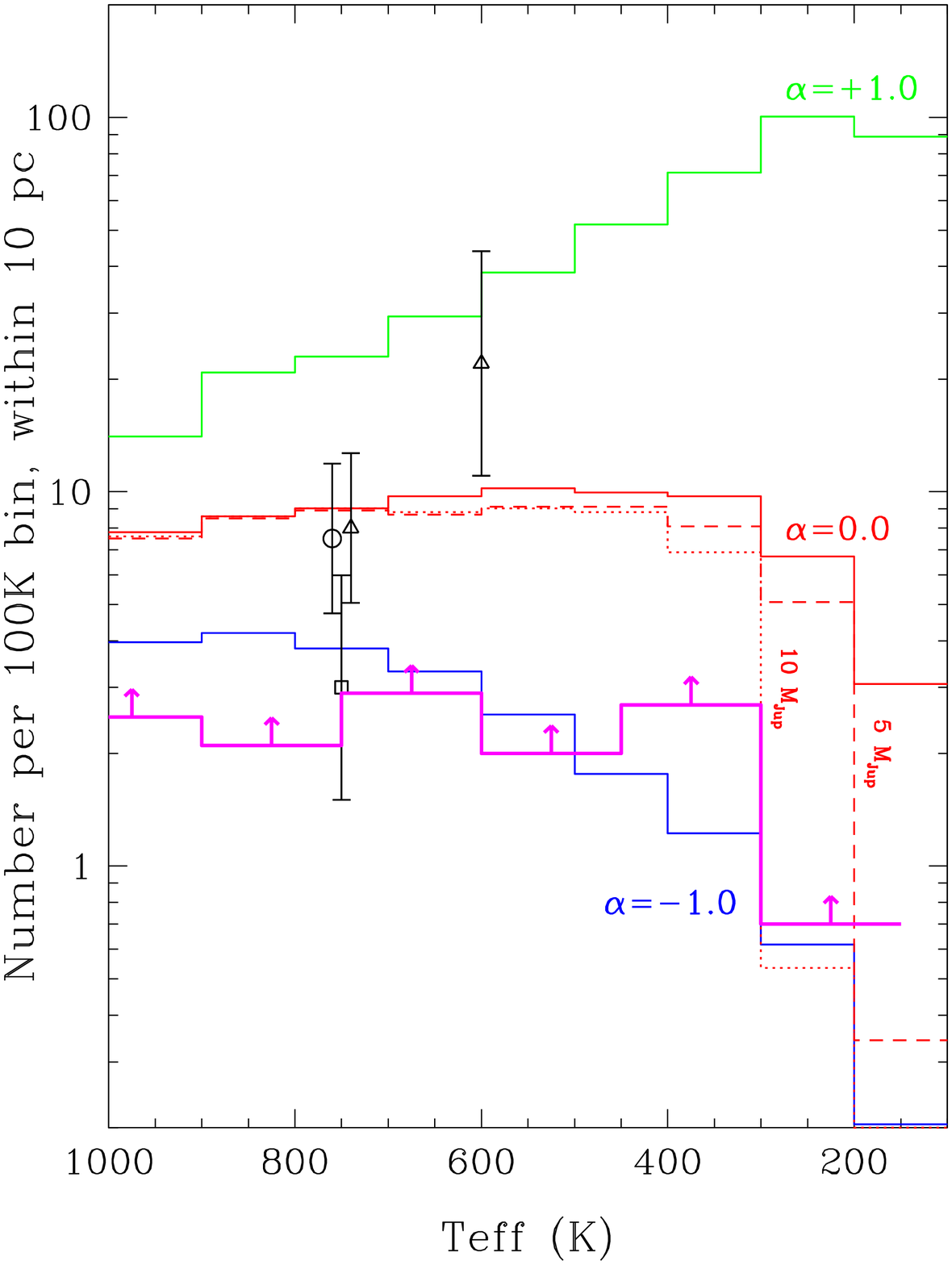}
\caption{The predicted number of brown dwarfs within 10 pc for three different mass
functions ($dN/dM \propto M^{-\alpha}$ with $\alpha = -1, 0, 1$ shown in green, red, and blue, respectively) having a minimum formation
mass of 1 $M_{Jup}$ (\citealt{burgasser2004}). Also shown for the $\alpha$ = 0 model
(dashed and dotted red lines) is the change in the expected number of brown dwarfs
when the minimum formation mass is varied. Recent measurements of the observed
space densities of T dwarfs are shown as open symbols --  \cite{metchev2008} (circle), \cite{burningham2010mf} (square), and \cite{reyle2010} (triangles). Lower limits to the 
space densities using a full accounting of objects in the Solar Neighborhood and based largely on
early WISE results (Table~\ref{space_density_numbers}) is shown in magenta.
\label{space_density}}
\end{figure}

\clearpage

\begin{figure}
\epsscale{0.9}
\figurenum{A1}
\plotone{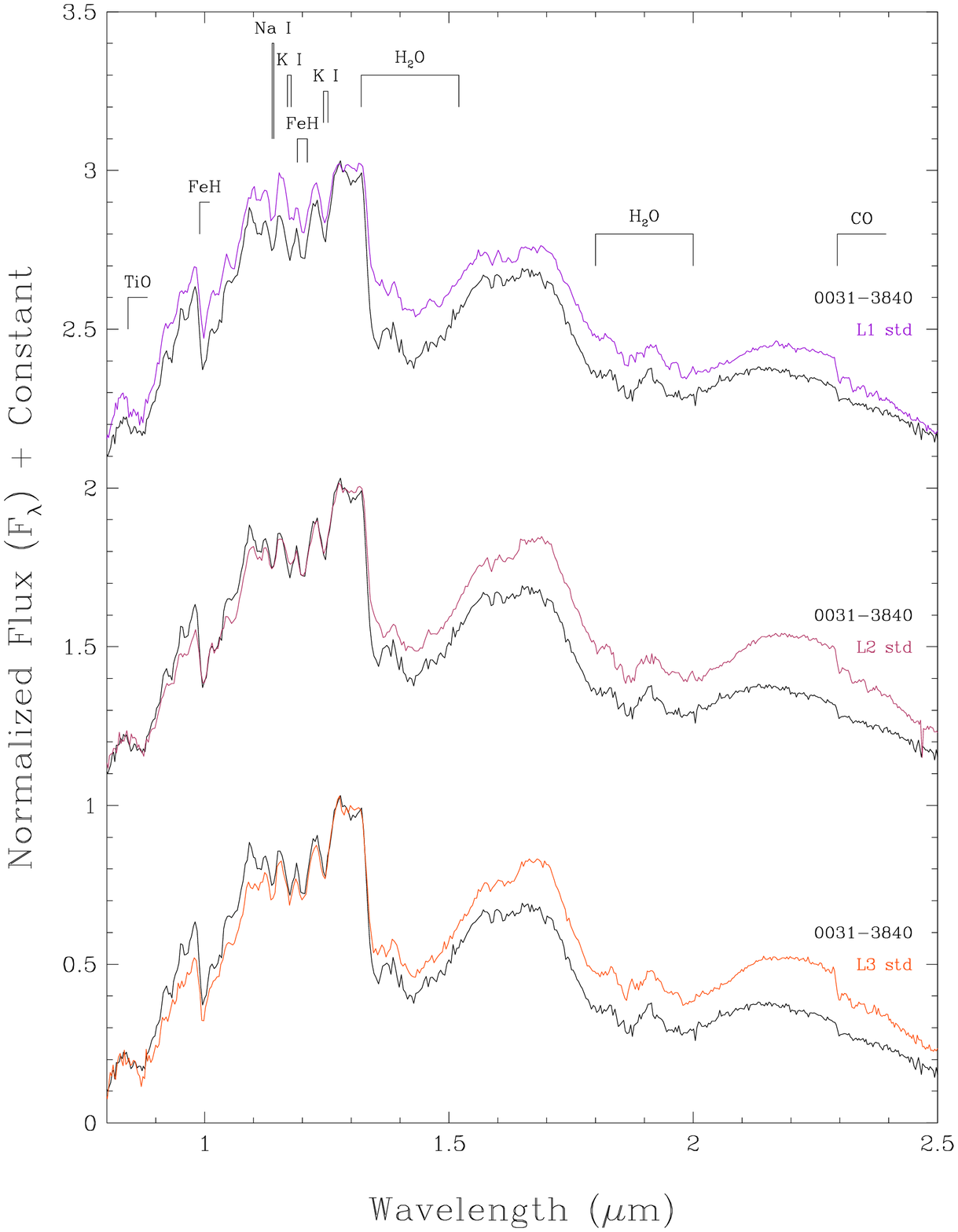}
\caption{The near-infrared spectrum of WISE 0031$-$3840 (black) compared to the
L1 (dark violet), L2 (maroon) and L3 (orange red) spectral standards from
\cite{kirkpatrick2010}. Spectra have been normalized to one at 1.28 $\mu$m
and integral offsets have been added to the $y$-axis values to separate the 
spectra vertically except where overplotting was intended. Prominent spectral
features are marked.
\label{0031m3840}}
\end{figure}

\clearpage

\begin{figure}
\epsscale{0.9}
\figurenum{A2}
\plotone{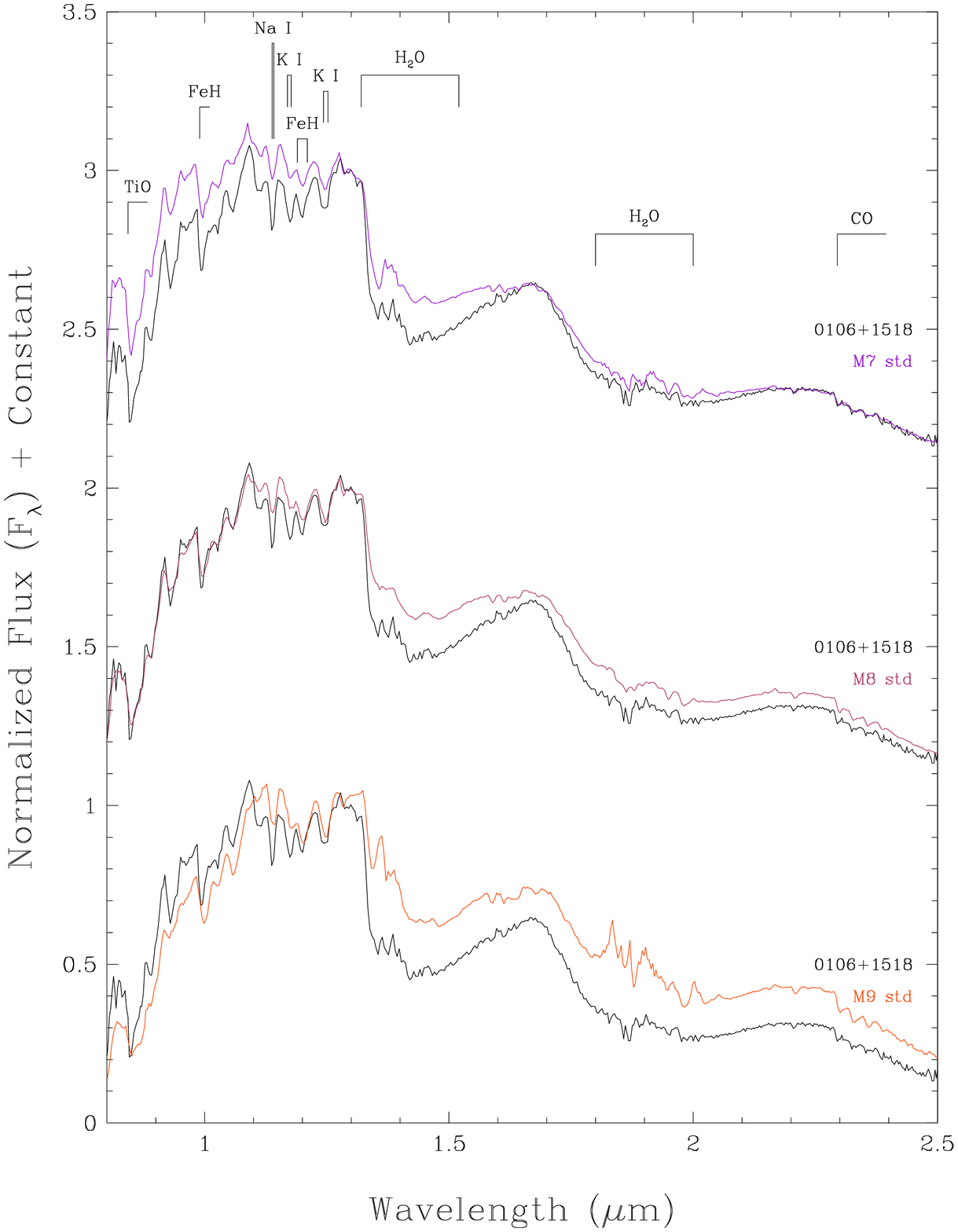}
\caption{The near-infrared spectrum of WISE 0106+1518 (black) compared to the
M7 (dark violet), M8 (maroon) and M9 (orange red) spectral standards from
\cite{kirkpatrick2010}. Spectra have been normalized to one at 1.28 $\mu$m
and integral offsets have been added to the $y$-axis values to separate the 
spectra vertically except where overplotting was intended. Prominent spectral
features are marked.
\label{0106p1518}}
\end{figure}

\clearpage

\begin{figure}
\epsscale{0.9}
\figurenum{A3}
\plotone{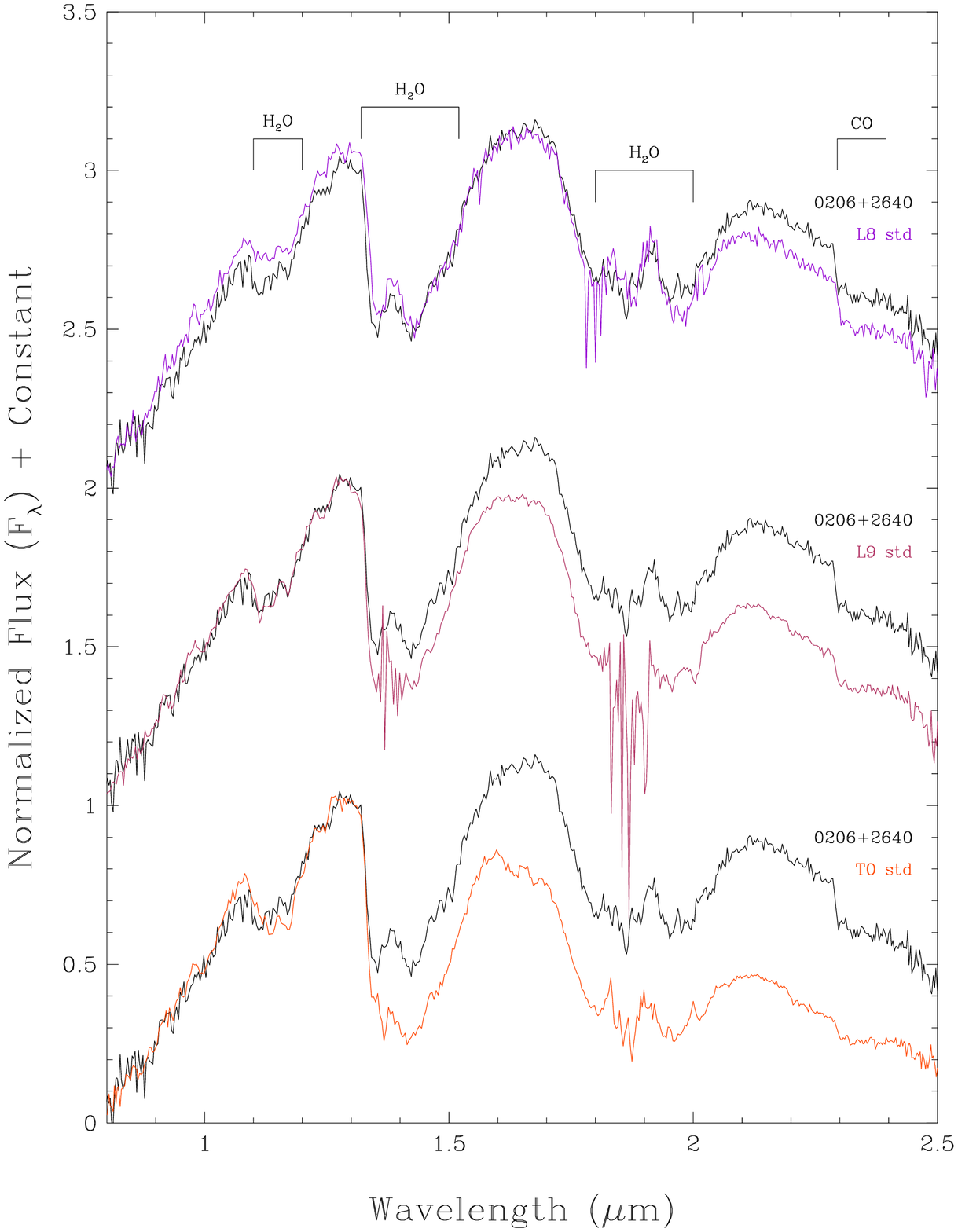}
\caption{The near-infrared spectrum of WISE 0206+2640 (black) compared to the
L8 (dark violet), L9 (maroon) and T0 (orange red) spectral standards from
\cite{kirkpatrick2010} and \cite{burgasser2006}. Spectra have been normalized to one at 1.28 $\mu$m
and integral offsets have been added to the $y$-axis values to separate the 
spectra vertically except where overplotting was intended. Prominent spectral
features are marked.
\label{0206p2640}}
\end{figure}

\clearpage

\begin{figure}
\epsscale{0.9}
\figurenum{A4}
\plotone{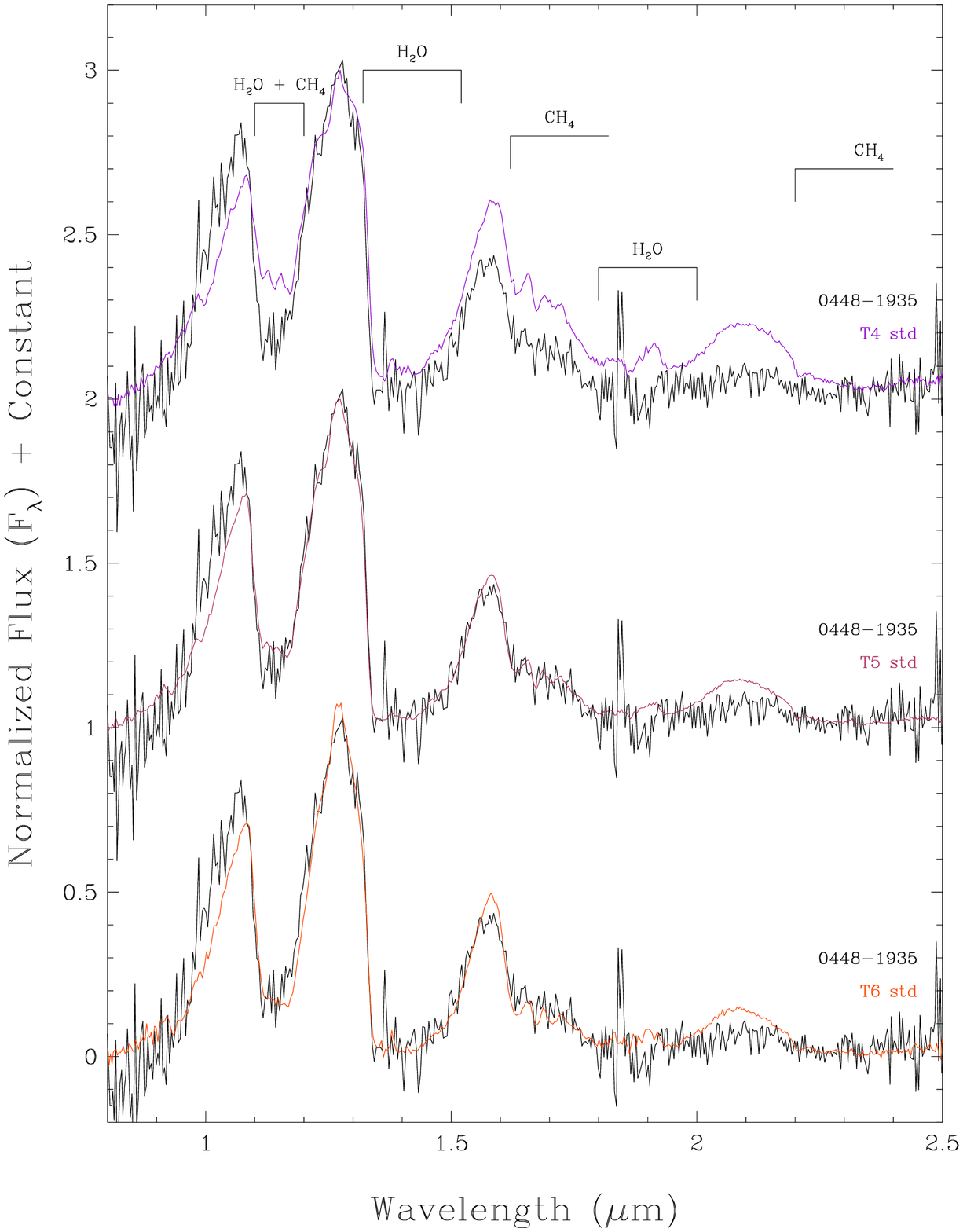}
\caption{The near-infrared spectrum of WISE 0448$-$1935 (black) compared to the
T4 (dark violet), T5 (maroon) and T6 (orange red) spectral standards from
\cite{burgasser2006}. Spectra have been normalized to one at 1.28 $\mu$m
and integral offsets have been added to the $y$-axis values to separate the 
spectra vertically except where overplotting was intended. Prominent spectral
features are marked.
\label{0448m1935}}
\end{figure}

\clearpage

\begin{figure}
\epsscale{0.9}
\figurenum{A5}
\plotone{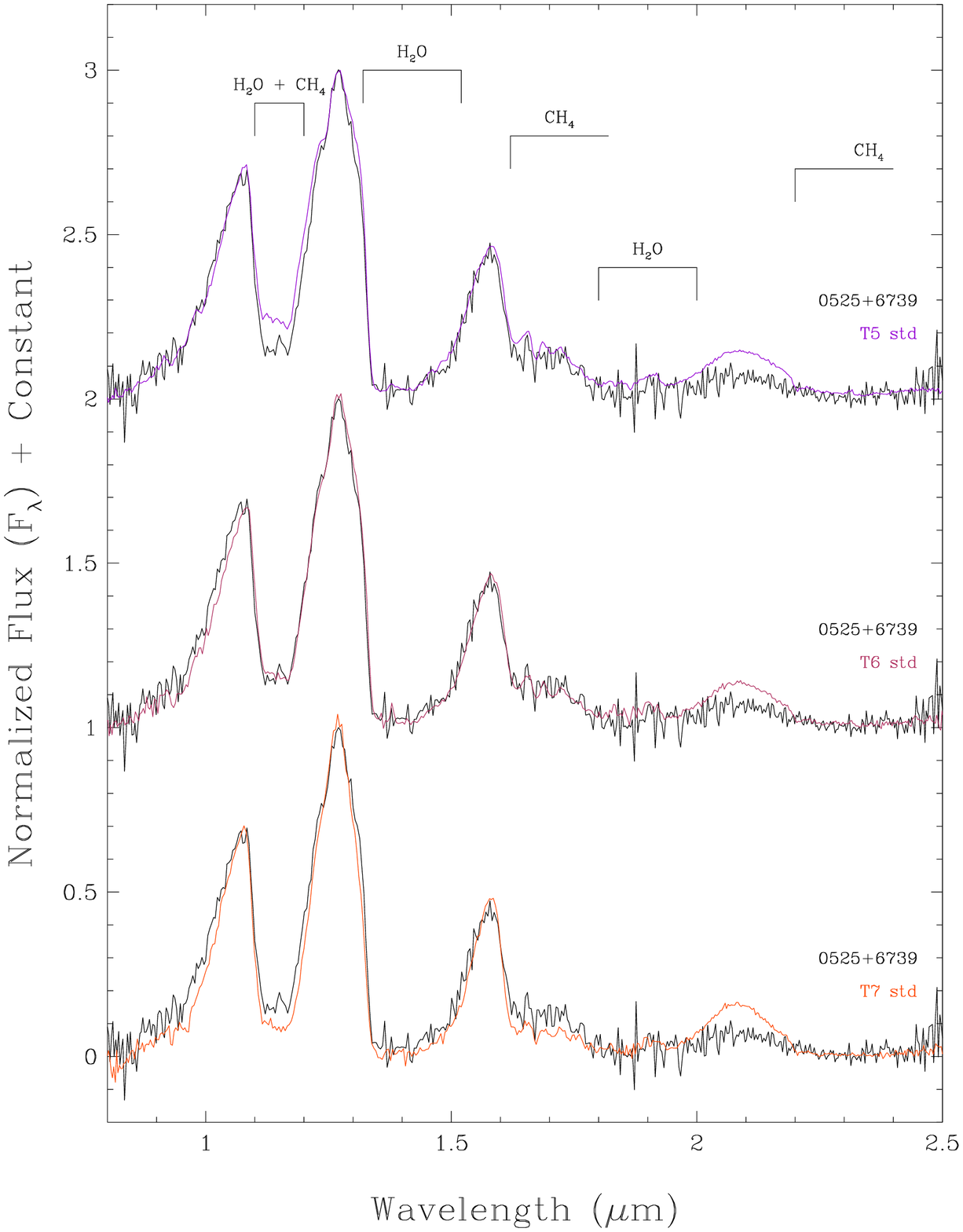}
\caption{The near-infrared spectrum of WISE 0525+6739 (black) compared to the
T5 (dark violet), T6 (maroon) and T7 (orange red) spectral standards from
\cite{burgasser2006}. Spectra have been normalized to one at 1.28 $\mu$m
and integral offsets have been added to the $y$-axis values to separate the 
spectra vertically except where overplotting was intended. Prominent spectral
features are marked.
\label{0525p6739}}
\end{figure}

\clearpage

\begin{figure}
\epsscale{0.9}
\figurenum{A6}
\plotone{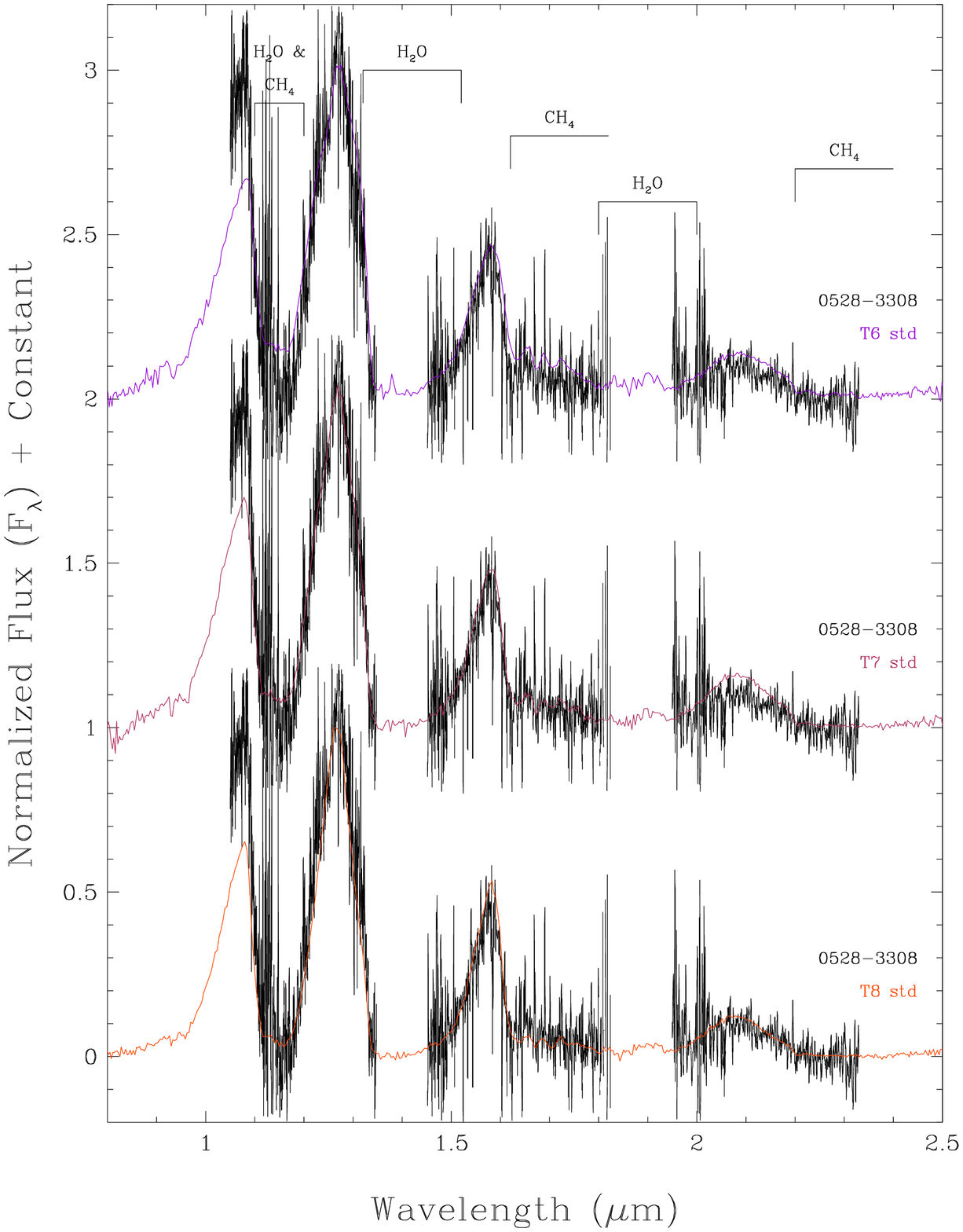}
\caption{The near-infrared spectrum of WISE 0528$-$3308 (black) compared to the
T6 (dark violet), T7 (maroon) and T8 (orange red) spectral standards from
\cite{burgasser2006}. Spectra have been normalized to one at 1.28 $\mu$m
and integral offsets have been added to the $y$-axis values to separate the 
spectra vertically except where overplotting was intended. Prominent spectral
features are marked. The spectrum of WISE 0528$-$3308 has been smoothed with
a 5-pixel boxcar.
\label{0528m3308}}
\end{figure}

\clearpage

\begin{figure}
\epsscale{0.9}
\figurenum{A7}
\plotone{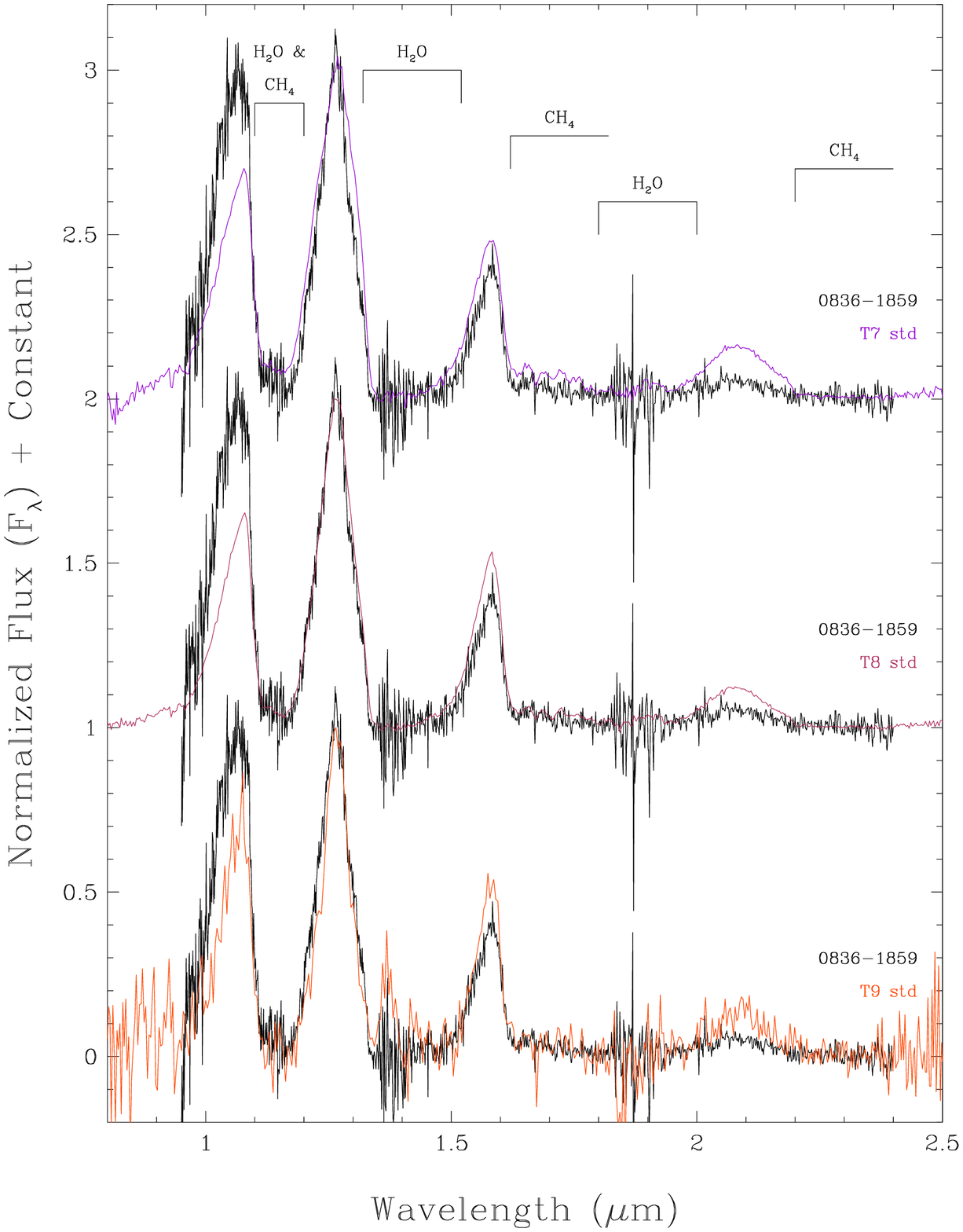}
\caption{The near-infrared spectrum of WISE 0836$-$1859 (black) compared to the
T7 (dark violet), T8 (maroon) and T9 (orange red) spectral standards from
\cite{burgasser2006} and Cushing et al.\ (accepted). Spectra have been normalized to one at 1.28 $\mu$m
and integral offsets have been added to the $y$-axis values to separate the 
spectra vertically except where overplotting was intended. Prominent spectral
features are marked.
\label{0836m1859}}
\end{figure}

\clearpage

\begin{figure}
\epsscale{0.9}
\figurenum{A8}
\plotone{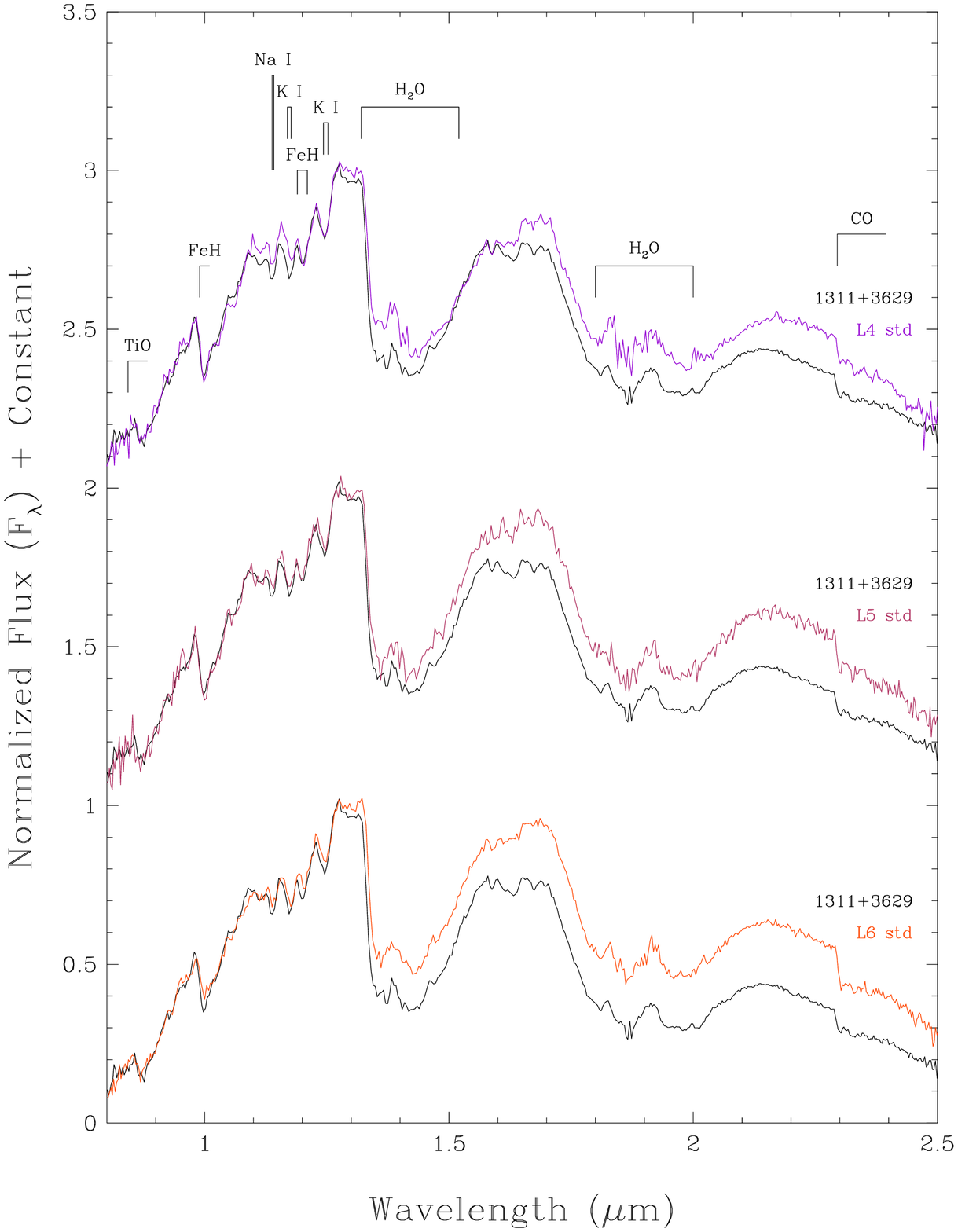}
\caption{The near-infrared spectrum of WISE 1311+3629 (black) compared to the
L4 (dark violet), L5 (maroon) and L6 (orange red) spectral standards from
\cite{kirkpatrick2010}. Spectra have been normalized to one at 1.28 $\mu$m
and integral offsets have been added to the $y$-axis values to separate the 
spectra vertically except where overplotting was intended. Prominent spectral
features are marked.
\label{1311p3629}}
\end{figure}

\clearpage

\begin{figure}
\epsscale{0.9}
\figurenum{A9}
\plotone{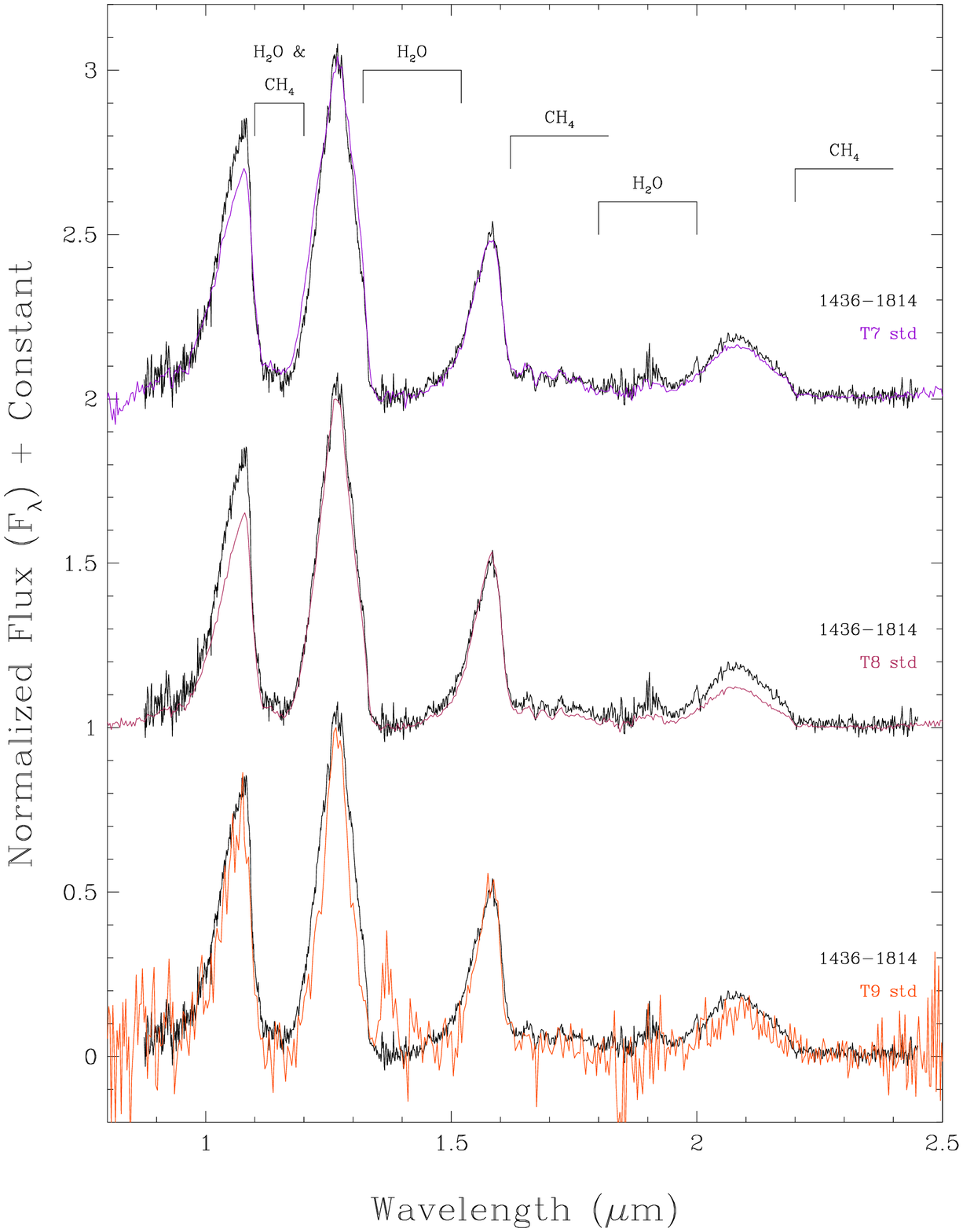}
\caption{The near-infrared spectrum of WISE 1436$-$1814 (black) compared to the
T7 (dark violet), T8 (maroon) and T9 (orange red) spectral standards from
\cite{burgasser2006} and Cushing et al.\ (accepted). Spectra have been normalized to one at 1.28 $\mu$m
and integral offsets have been added to the $y$-axis values to separate the 
spectra vertically except where overplotting was intended. Prominent spectral
features are marked.
\label{1436m1814}}
\end{figure}

\clearpage

\begin{figure}
\epsscale{0.9}
\figurenum{A10}
\plotone{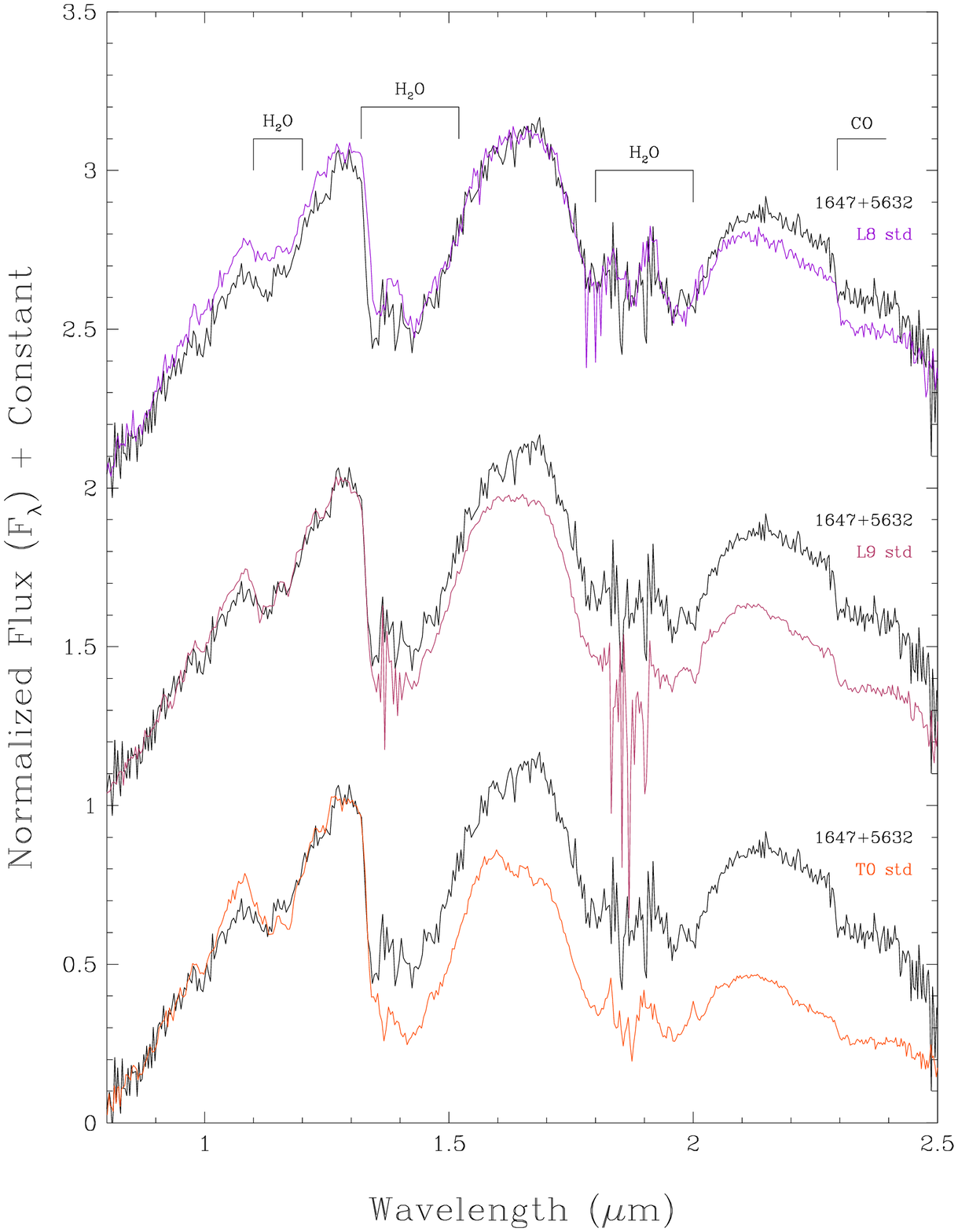}
\caption{The near-infrared spectrum of WISE 1647+5632 (black) compared to the
L8 (dark violet), L9 (maroon) and T0 (orange red) spectral standards from
\cite{kirkpatrick2010} and \cite{burgasser2006}. Spectra have been normalized to one at 1.28 $\mu$m
and integral offsets have been added to the $y$-axis values to separate the 
spectra vertically except where overplotting was intended. Prominent spectral
features are marked.
\label{1647p5632}}
\end{figure}

\clearpage

\begin{figure}
\epsscale{0.9}
\figurenum{A11}
\plotone{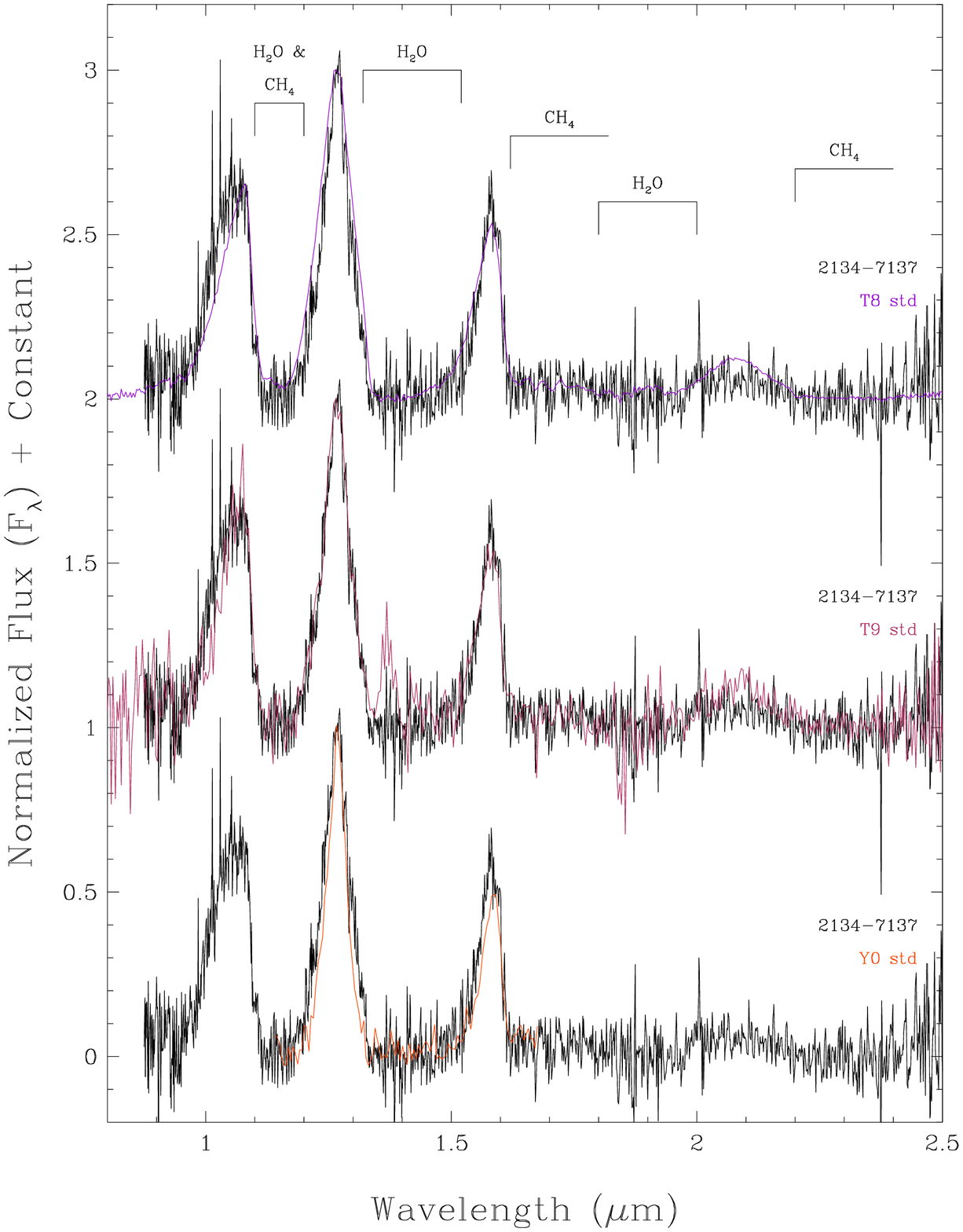}
\caption{The near-infrared spectrum of WISE 2134$-$7137 (black) compared to the
T8 (dark violet), T9 (maroon) and Y0 (orange red) spectral standards from
\cite{burgasser2006} and Cushing et al.\ (accepted). Spectra have been normalized to one at 1.28 $\mu$m
and integral offsets have been added to the $y$-axis values to separate the 
spectra vertically except where overplotting was intended. Prominent spectral
features are marked. 
\label{2134m7137}}
\end{figure}

\clearpage

\begin{figure}
\epsscale{0.9}
\figurenum{A12}
\plotone{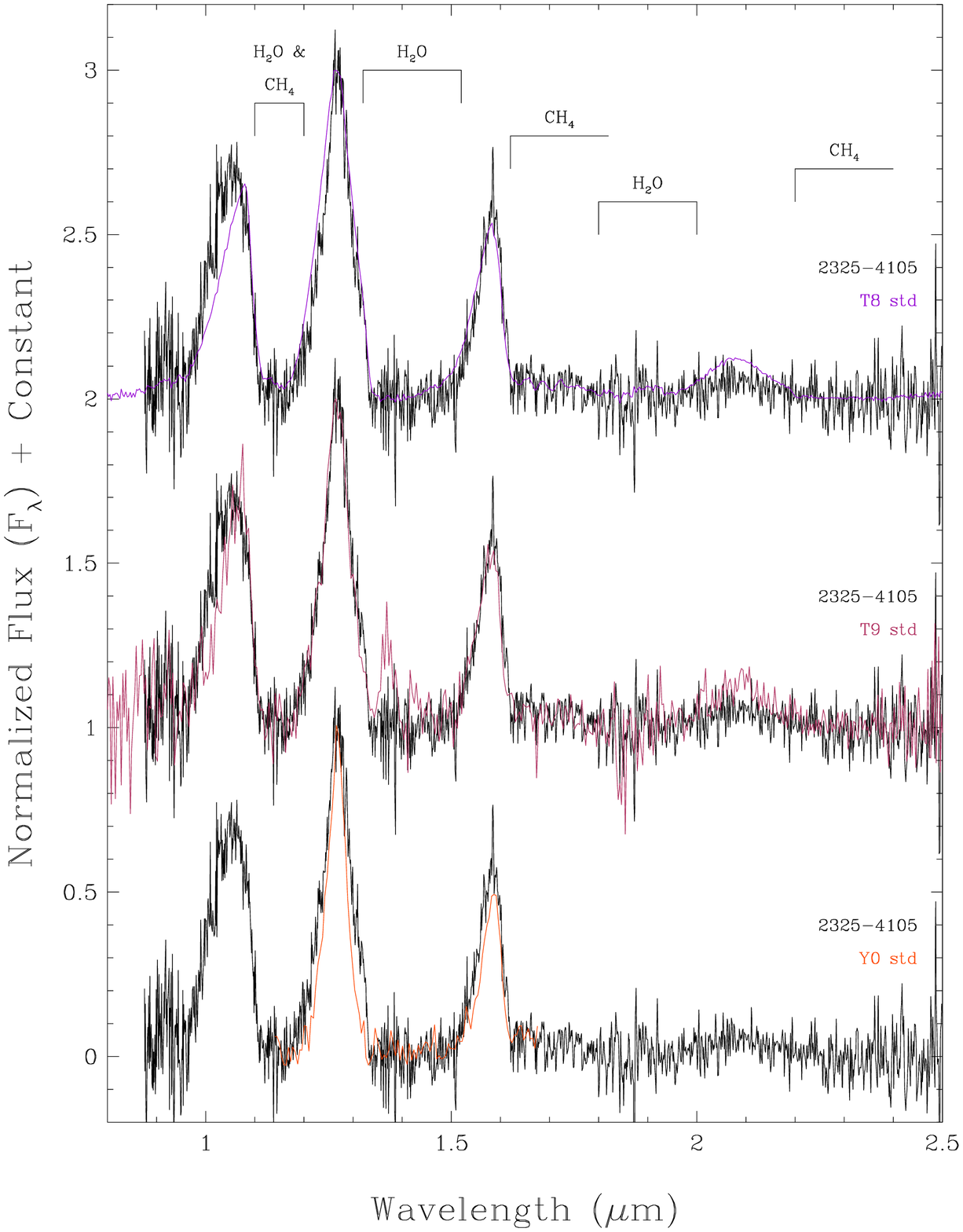}
\caption{The near-infrared spectrum of WISE 2325$-$4105 (black) compared to the
T8 (dark violet), T9 (maroon) and Y0 (orange red) spectral standards from
\cite{burgasser2006} and Cushing et al.\ (accepted). Spectra have been normalized to one at 1.28 $\mu$m
and integral offsets have been added to the $y$-axis values to separate the 
spectra vertically except where overplotting was intended. Prominent spectral
features are marked. 
\label{2325m4105}}
\end{figure}


\begin{thebibliography}{}
\bibitem[Albert et al.(2011)]{albert2011} Albert, L., Artigau, 
   {\'E}., Delorme, P., Reyl{\'e}, C., Forveille, T., Delfosse, X., 
   \& Willott, C.~J.\ 2011, \aj, 141, 203
\bibitem[Artigau et al.(2010)]{artigau2010} Artigau, {\'E}., 
   Radigan, J., Folkes, S., Jayawardhana, R., Kurtev, R., Lafreni{\`e}re, D., 
   Doyon, R., \& Borissova, J.\ 2010, \apjl, 718, L38
\bibitem[Artigau et al.(2006)]{artigau2006} Artigau, {\'E}., Doyon, 
   R., Lafreni{\`e}re, D., Nadeau, D., Robert, J., 
   \& Albert, L.\ 2006, \apjl, 651, L57
\bibitem[Baraffe et al.(2003)]{baraffe2003} Baraffe, I., Chabrier, 
   G., Allard, F., \& Hauschildt, P.\ 2003, Brown Dwarfs, 211, 41
\bibitem[Barman et al.(2005)]{barman2005} Barman, T.~S., 
   Hauschildt, P.~H., \& Allard, F.\ 2005, \apj, 632, 1132
\bibitem[Bate \& Bonnell(2005)]{bate2005} Bate, 
   M.~R., \& Bonnell, I.~A.\ 2005, \mnras, 356, 1201 
\bibitem[Becklin \& Zuckerman(1988)]{becklin1988} Becklin, E.~E., 
   \& Zuckerman, B.\ 1988, \nat, 336, 656
\bibitem[Bertin \& Arnouts(1996)]{bertin1996} Bertin, E., \& Arnouts, 
   S.\ 1996, \aaps, 117, 393
\bibitem[Bessell(2005)]{bessell2005} Bessell, M.~S.\ 2005, \araa, 43, 293
\bibitem[Bessell \& Brett(1988)]{bessell1988} Bessell, M.~S., 
   \& Brett, J.~M.\ 1988, \pasp, 100, 1134
\bibitem[Bloom et al.(2006)]{bloom2006} Bloom, J.~S., Starr, 
   D.~L., Blake, C.~H., Skrutskie, M.~F., 
\& Falco, E.~E.\ 2006, Astronomical Data Analysis Software and Systems XV, 351, 751
\bibitem[Boss(2004)]{boss2004} Boss, A.~P.\ 2004, \mnras, 350, L57
\bibitem[Bouvier et al.(2009)]{bouvier2009} Bouvier, J., Kendall, T., 
   \& Meeus, G.\ 2009, American Institute of Physics Conference Series, 
   1094, 497
\bibitem[Bouy et al.(2003)]{bouy2003} Bouy, H., Brandner, W., 
   Mart{\'{\i}}n, E.~L., Delfosse, X., Allard, F., 
   \& Basri, G.\ 2003, \aj, 126, 1526
\bibitem[Bowler et al.(2010)]{bowler2010} Bowler, B.~P., Liu, 
   M.~C., \& Dupuy, T.~J.\ 2010, \apj, 710, 45
\bibitem[Boyd \& Whitworth(2005)]{boyd2005} Boyd, D.~F.~A., \& 
   Whitworth, A.~P.\ 2005, \aap, 430, 1059 
\bibitem[Burgasser et al.(2011)]{burgasser2011b} Burgasser, A.~J., 
   Sitarski, B.~N., Gelino, C.~R., Logsdon, S.~E., 
   \& Perrin, M.~D.\ 2011, arXiv:1107.1484
\bibitem[Burgasser et al.(2011)]{burgasser2011} Burgasser, A.~J., et 
   al.\ 2011, arXiv:1104.2537
\bibitem[Burgasser et al.(2010b)]{burgasser2010-ross458c} Burgasser, A.~J., et 
   al.\ 2010, \apj, 725, 1405
\bibitem[Burgasser et al.(2010)]{burgasser2010} Burgasser, A.~J., 
   Looper, D., \& Rayner, J.~T.\ 2010, \aj, 139, 2448
\bibitem[Burgasser(2008)]{burgasser2008} Burgasser, A.~J.\ 2008, 14th 
   Cambridge Workshop on Cool Stars, Stellar Systems, and the Sun, 384, 126 
\bibitem[Burgasser et al.(2008)]{burgasser2008-2M1126} Burgasser, A.~J., 
   Looper, D.~L., Kirkpatrick, J.~D., Cruz, K.~L., 
   \& Swift, B.~J.\ 2008, \apj, 674, 451 
\bibitem[Burgasser(2007)]{burgasser2007} Burgasser, A.~J.\ 2007, \aj, 
   134, 1330
\bibitem[Burgasser et al.(2006)]{burgasser2006} Burgasser, A.~J., 
   Geballe, T.~R., Leggett, S.~K., Kirkpatrick, J.~D., 
   \& Golimowski, D.~A.\ 2006, \apj, 637, 1067
\bibitem[Burgasser et al.(2006)]{burgasser2006BBK} Burgasser, A.~J., 
   Burrows, A., \& Kirkpatrick, J.~D.\ 2006, \apj, 639, 1095
\bibitem[Burgasser(2004)]{burgasser2004} Burgasser, A.~J.\ 2004, 
   \apjs, 155, 191
\bibitem[Burgasser et al.(2004)]{burgasser2004b} Burgasser, A.~J., 
   McElwain, M.~W., Kirkpatrick, J.~D., Cruz, K.~L., Tinney, C.~G., 
   \& Reid, I.~N.\ 2004, \aj, 127, 2856
\bibitem[Burgasser et al.(2003)]{burgasser2003} Burgasser, A.~J., 
   Kirkpatrick, J.~D., Liebert, J., \& Burrows, A.\ 2003, \apj, 594, 510
\bibitem[Burgasser et al.(2003)]{burgasser2003-2mass1503} Burgasser, A.~J., 
   Kirkpatrick, J.~D., McElwain, M.~W., Cutri, R.~M., Burgasser, A.~J., 
   \& Skrutskie, M.~F.\ 2003, \aj, 125, 850
\bibitem[Burgasser et al.(2003)]{burgasser2003c} Burgasser, A.~J., 
   McElwain, M.~W., \& Kirkpatrick, J.~D.\ 2003, \aj, 126, 2487
\bibitem[Burgasser et al.(2002)]{burgasser2002} Burgasser, A.~J., et 
   al.\ 2002, \apj, 564, 421
\bibitem[Burgasser et al.(2000)]{burgasser2000} Burgasser, A.~J., et 
   al.\ 2000, \aj, 120, 1100
\bibitem[Burgasser et al.(2000)]{burgasser2000-gl570d} Burgasser, A.~J., et 
   al.\ 2000, \apjl, 531, L57
\bibitem[Burgasser et al.(1999)]{burgasser1999} Burgasser, A.~J., et 
   al.\ 1999, \apjl, 522, L65
\bibitem[Burningham et al.(2011b)]{burningham2011ugps0521} Burningham, B., et 
   al.\ 2011, \mnras, 414, L90
\bibitem[Burningham et al.(2011)]{burningham2011} Burningham, B., et 
   al.\ 2011, \mnras, L259 
\bibitem[Burningham et al.(2010b)]{burningham2010mf} Burningham, B., et 
   al.\ 2010, \mnras, 406, 1885
\bibitem[Burningham et al.(2010)]{burningham2010} Burningham, B., et 
   al.\ 2010, \mnras, 404, 1952
\bibitem[Burningham et al.(2009)]{burningham2009} Burningham, B., et 
   al.\ 2009, \mnras, 395, 1237
\bibitem[Burningham et al.(2008)]{burningham2008} Burningham, B., et 
   al.\ 2008, \mnras, 391, 320
\bibitem[Burrows et al.(2003)]{burrows2003} Burrows, A., Sudarsky, 
   D., \& Lunine, J.~I.\ 2003, \apj, 596, 587
\bibitem[Chiu et al.(2006)]{chiu2006} Chiu, K., Fan, X., 
   Leggett, S.~K., Golimowski, D.~A., Zheng, W., Geballe, T.~R., Schneider, 
   D.~P., \& Brinkmann, J.\ 2006, \aj, 131, 2722
\bibitem[Costa et al.(2006)]{costa2006} Costa, E., M{\'e}ndez, 
   R.~A., Jao, W.-C., Henry, T.~J., Subasavage, J.~P., 
   \& Ianna, P.~A.\ 2006, \aj, 132, 1234
\bibitem[Cruz et al.(2007)]{cruz2007} Cruz, K.~L., et al.\ 2007, 
   \aj, 133, 439
\bibitem[Cruz et al.(2004)]{cruz2004} Cruz, K.~L., Burgasser, 
   A.~J., Reid, I.~N., \& Liebert, J.\ 2004, \apjl, 604, L61
\bibitem[Cruz et al.(2003)]{cruz2003} Cruz, K.~L., Reid, I.~N., 
   Liebert, J., Kirkpatrick, J.~D., \& Lowrance, P.~J.\ 2003, \aj, 126, 2421
\bibitem[Cushing et al.(2004)]{cushing2004} Cushing, M.~C., Vacca, 
   W.~D., \& Rayner, J.~T.\ 2004, \pasp, 116, 362
\bibitem[Dahn et al.(2002)]{dahn2002} Dahn, C.~C., et al.\ 2002, 
   \aj, 124, 117
\bibitem[Dahn et al.(1988)]{dahn1988} Dahn, C.~C., et al.\ 1988, 
   \aj, 95, 237
\bibitem[Deacon et al.(2009)]{deacon2009} Deacon, N.~R., Hambly, 
  N.~C., King, R.~R., \& McCaughrean, M.~J.\ 2009, \mnras, 394, 857
\bibitem[Deacon et al.(2005)]{deacon2005} Deacon, N.~R., Hambly, 
   N.~C., \& Cooke, J.~A.\ 2005, \aap, 435, 363
\bibitem[Delfosse et al.(1999)]{delfosse1999} Delfosse, X., Tinney, 
   C.~G., Forveille, T., Epchtein, N., Borsenberger, J., Fouqu{\'e}, P., 
   Kimeswenger, S., \& Tiph{\`e}ne, D.\ 1999, \aaps, 135, 41
\bibitem[Delfosse et al.(1997)]{delfosse1997} Delfosse, X., et 
   al.\ 1997, \aap, 327, L25
\bibitem[Delorme et al.(2010)]{delorme2010} Delorme, P., et al.\ 2010, \aap, 518, A39
\bibitem[Delorme et al.(2008)]{delorme2008} Delorme, P., et al.\ 2008, \aap, 482, 961
\bibitem[Eddington(1940)]{eddington1940} Eddington, A.~S., Sir 1940, 
   \mnras, 100, 354
\bibitem[Eddington(1913)]{eddington1913} Eddington, A.~S.\ 1913, 
   \mnras, 73, 359
\bibitem[Eisenhardt et al.(2010)]{eisenhardt2010} Eisenhardt, 
   P.~R.~M., et al.\ 2010, \aj, 139, 2455 
\bibitem[Ellis et al.(2005)]{ellis2005} Ellis, S.~C., Tinney, 
   C.~G., Burgasser, A.~J., Kirkpatrick, J.~D., 
   \& McElwain, M.~W.\ 2005, \aj, 130, 2347
\bibitem[Faherty et al.(2009)]{faherty2009} Faherty, J.~K., 
   Burgasser, A.~J., Cruz, K.~L., Shara, M.~M., Walter, F.~M., 
   \& Gelino, C.~R.\ 2009, \aj, 137, 1
\bibitem[Fan et al.(2000)]{fan2000} Fan, X., et al.\ 2000, \aj, 
   119, 928
\bibitem[Fazio et al.(2004)]{fazio2004} Fazio, G.~G., et al.\ 
   2004, \apjs, 154, 10 
\bibitem[Fortney et al.(2005)]{fortney2005} Fortney, J.~J., Marley, 
   M.~S., Lodders, K., Saumon, D., \& Freedman, R.\ 2005, \apjl, 627, L69 
\bibitem[Geballe et al.(2002)]{geballe2002} Geballe, T.~R., et al.\ 
   2002, \apj, 564, 466
\bibitem[Gelino et al.(2011)]{gelino2011} Gelino, C.~R., et al.\ 
   2011, \aj, 142, 57
\bibitem[Giclas et al.(1971)]{giclas1971} Giclas, H.~L., Burnham, 
   R., \& Thomas, N.~G.\ 1971, Flagstaff, Arizona: Lowell Observatory, 1971,
\bibitem[Gizis(2002)]{gizis2002} Gizis, J.~E.\ 2002, \apj, 575, 
   484
\bibitem[Gizis et al.(2000)]{gizis2000} Gizis, J.~E., Monet, 
   D.~G., Reid, I.~N., Kirkpatrick, J.~D., Liebert, J., 
   \& Williams, R.~J.\ 2000, \aj, 120, 1085
\bibitem[Goldman et al.(2010)]{goldman2010} Goldman, B., Marsat, 
   S., Henning, T., Clemens, C., \& Greiner, J.\ 2010, \mnras, 405, 1140
\bibitem[Hall(2002)]{hall2002} Hall, P.~B.\ 2002, \apjl, 564, 
   L89
\bibitem[Hamuy et al.(1994)]{hamuy1994} Hamuy, M., Suntzeff, 
   N.~B., Heathcote, S.~R., Walker, A.~R., Gigoux, P., 
   \& Phillips, M.~M.\ 1994, \pasp, 106, 566
\bibitem[Hanisch et al.(2001)]{hanisch2001} Hanisch, R.~J., Farris, A., 
   Greisen, E.~W., Pence, W.~D., Schlesinger, B.~M., Teuben, P.~J., 
   Thompson, R.~W., \& Warnock, A., III 2001, \aap, 376, 359
\bibitem[Hawkins \& Bessell(1988)]{hawkins1988} Hawkins, M.~R.~S., 
   \& Bessell, M.~S.\ 1988, \mnras, 234, 177
\bibitem[Hawley et al.(2002)]{hawley2002} Hawley, S.~L., et al.\ 
   2002, \aj, 123, 3409
\bibitem[Hayashi \& Nakano(1963)]{hayashi1963} Hayashi, C., \& Nakano, 
   T.\ 1963, Progress of Theoretical Physics, 30, 460
\bibitem[Herter et al.(2008)]{herter2008} Herter, T.~L., et al.\ 
   2008, \procspie, 7014, 
\bibitem[Horne(1986)]{horne1986} Horne, K.\ 1986, \pasp, 98, 609
\bibitem[Kanneganti et al.(2009)]{kanneganti2009} Kanneganti, S., 
   Park, C., Skrutskie, M.~F., Wilson, J.~C., Nelson, M.~J., Smith, A.~W., 
   \& Lam, C.~R.\ 2009, \pasp, 121, 885
\bibitem[Kendall et al.(2007)]{kendall2007} Kendall, T.~R., Jones, 
   H.~R.~A., Pinfield, D.~J., Pokorny, R.~S., Folkes, S., Weights, D., 
   Jenkins, J.~S., \& Mauron, N.\ 2007, \mnras, 374, 445
\bibitem[Kirkpatrick et al.(2010)]{kirkpatrick2010} Kirkpatrick, J.~D., 
   et al.\ 2010, \apjs, 190, 100
\bibitem[Kirkpatrick(2008)]{kirkpatrick2008} Kirkpatrick, J.~D.\ 2008, 
   14th Cambridge Workshop on Cool Stars, Stellar Systems, and the Sun, 384, 
   85
\bibitem[Kirkpatrick et al.(2008)]{kirkpatrick2008b} Kirkpatrick, J.~D., 
   et al.\ 2008, \apj, 689, 1295
\bibitem[Kirkpatrick et al.(2006)]{kirkpatrick2006} Kirkpatrick, J.~D., 
   Barman, T.~S., Burgasser, A.~J., McGovern, M.~R., McLean, I.~S., Tinney, 
   C.~G., \& Lowrance, P.~J.\ 2006, \apj, 639, 1120
\bibitem[Kirkpatrick(2005)]{kirkpatrick2005} Kirkpatrick, J.~D.\ 2005, 
   \araa, 43, 195 
\bibitem[Kirkpatrick et al.(2001)]{kirkpatrick2001} Kirkpatrick, J.~D., 
   Dahn, C.~C., Monet, D.~G., Reid, I.~N., Gizis, J.~E., Liebert, J., 
   \& Burgasser, A.~J.\ 2001, \aj, 121, 3235
\bibitem[Kirkpatrick(2000)]{kirkpatrick2000} Kirkpatrick, J.~D.\ 2000, 
   From Giant Planets to Cool Stars, 212, 20
\bibitem[Kirkpatrick et al.(2000)]{kirkpatrick2000b} Kirkpatrick, J.~D., 
   et al.\ 2000, \aj, 120, 447
\bibitem[Kirkpatrick et al.(1999)]{kirkpatrick1999} Kirkpatrick, J.~D., 
   et al.\ 1999, \apj, 519, 802 
\bibitem[Kirkpatrick et al.(1997)]{kirkpatrick1997} Kirkpatrick, J.~D., 
   Beichman, C.~A., \& Skrutskie, M.~F.\ 1997, \apj, 476, 311 
\bibitem[Kirkpatrick \& McCarthy(1994)]{kirkpatrick1994} Kirkpatrick, 
   J.~D., \& McCarthy, D.~W., Jr.\ 1994, \aj, 107, 333 
\bibitem[Kirkpatrick et al.(1994)]{kirkpatrick1994b} Kirkpatrick, J.~D., 
   McGraw, J.~T., Hess, T.~R., Liebert, J., 
   \& McCarthy, D.~W., Jr.\ 1994, \apjs, 94, 749
\bibitem[Knapp et al.(2004)]{knapp2004} Knapp, G.~R., et al.\ 
   2004, \aj, 127, 3553
\bibitem[Koekemoer et al.(2002)]{koekemoer2002} Koekemoer, A.~M., 
   Fruchter, A.~S., Hook, R.~N., \& Hack, W.\ 2002, The 2002 HST Calibration
   Workshop : Hubble after the Installation of the ACS and the NICMOS Cooling System, 337
\bibitem[Kouzuma \& Yamaoka(2010)]{kouzuma2010} Kouzuma, S., \& 
   Yamaoka, H.\ 2010, \aap, 509, A64 
\bibitem[Kumar(1963)]{kumar1963} Kumar, S.~S.\ 1963, \apj, 137, 
   1121
\bibitem[K{\"u}mmel et al.(2009)]{kummel2009} K{\"u}mmel, M., 
   Walsh, J.~R., Pirzkal, N., Kuntschner, H., 
   \& Pasquali, A.\ 2009, \pasp, 121, 59
\bibitem[Lawrence et al.(2007)]{lawrence2007} Lawrence, A., et al.\ 
   2007, \mnras, 379, 1599 
\bibitem[Leggett et al.(2010)]{leggett2010} Leggett, S.~K., et al.\ 
   2010, \apj, 710, 1627
\bibitem[Leggett et al.(2009)]{leggett2009} Leggett, S.~K., et al.\ 
   2009, \apj, 695, 1517
\bibitem[Leggett et al.(2000)]{leggett2000} Leggett, S.~K., et al.\ 
   2000, \apjl, 536, L35
\bibitem[L{\'e}pine \& Shara(2005)]{lepine2005} L{\'e}pine, S., 
   \& Shara, M.~M.\ 2005, \aj, 129, 1483
\bibitem[Liebert et al.(2003)]{liebert2003} Liebert, J., 
   Kirkpatrick, J.~D., Cruz, K.~L., Reid, I.~N., Burgasser, A., Tinney, C.~G., 
   \& Gizis, J.~E.\ 2003, \aj, 125, 343
\bibitem[Liu et al.(2011)]{liu2011} Liu, M.~C., et al.\ 2011, 
   arXiv:1103.0014
\bibitem[Lodieu et al.(2002)]{lodieu2002} Lodieu, N., Scholz, R.-D., 
   \& McCaughrean, M.~J.\ 2002, \aap, 389, L20
\bibitem[Loh et al.(2004)]{loh2004} Loh, E.~D., Biel, J.~D., 
   Chen, J.-J., Davis, M., Laporte, R., 
   \& Loh, O.~Y.\ 2004, \procspie, 5492, 1644
\bibitem[Looper et al.(2008)]{looper2008} Looper, D.~L., et al.\ 
   2008, \apj, 686, 528
\bibitem[Looper et al.(2007)]{looper2007} Looper, D.~L., 
   Kirkpatrick, J.~D., \& Burgasser, A.~J.\ 2007, \aj, 134, 1162
\bibitem[L{\'o}pez-Morales(2007)]{lopez-morales2007} L{\'o}pez-Morales, 
   M.\ 2007, \apj, 660, 732
\bibitem[Lucas et al.(2010)]{lucas2010} Lucas, P.~W., et al.\ 
   2010, \mnras, 408, L56
\bibitem[Lucas et al.(2006)]{lucas2006} Lucas, P.~W., Weights, D.~J., 
   Roche, P.~F., \& Riddick, F.~C.\ 2006, \mnras, 373, L60
\bibitem[Luhman et al.(2011)]{luhman2011} Luhman, K.~L., 
   Burgasser, A.~J., \& Bochanski, J.~J.\ 2011, \apjl, 730, L9
\bibitem[Luhman et al.(2007)]{luhman2007} Luhman, K.~L., et al.\ 
   2007, \apj, 654, 570
\bibitem[Luhman(2007)]{luhman2007-cham} Luhman, K.~L.\ 2007, \apjs, 173, 104
\bibitem[Luhman et al.(2000)]{luhman2000} Luhman, K.~L., Rieke,
  G.~H., Young, E.~T., Cotera, A.~S., Chen, H., Rieke, M.~J., Schneider, G.,
  \& Thompson, R.~I.\ 2000, \apj, 540, 1016
\bibitem[Lutz \& Kelker(1973)]{lutz1973} Lutz, T.~E., \& Kelker, 
   D.~H.\ 1973, \pasp, 85, 573
\bibitem[Luyten(1980)]{luyten1980} Luyten, W.~J.\ 1980, NLTT 
   Catalogue.~Volume\_III.~0\_\_to -30\_., by Luyten, W.~J..~ Published by 
   University of Minnesota, Mineapolis, Minnesota, USA, 283 p.,
\bibitem[Luyten(1979)]{luyten1979-lhs} Luyten, W.~J.\ 1979, 
   Minneapolis: University of Minnesota, 1979, 2nd ed.,
\bibitem[Luyten(1979)]{luyten1979b} Luyten, W.~J.\ 1979, New Luyten 
   Catalogue of stars with proper motions larger than two tenths of an 
   arcsecond, 1, 0 (1979), 0
\bibitem[Luyten(1979)]{luyten1979c} Luyten, W.~J.\ 1979, New Luyten 
   Catalogue of stars with proper motions larger than two tenths of an 
   arcsecond, 2, 0 (1979), 0
\bibitem[Luyten \& Kowal(1975)]{luyten1975} Luyten, W.~J., \& 
   Kowal, C.~T.\ 1975, Proper motion survey with the forty-eight inch 
   Schmidt telescope.~XLIII.~One hundred and six faint stars with large 
   proper motions., by Luyten, W.~J.; Kowal, C.~T..~ Separate print 
   Univ.~Minnesota, Minneapolis, Minnesota, 2 p., 43, 1
\bibitem[Luyten(1974)]{luyten1974b} Luyten, W.~J.\ 1974, Proper 
   Motion Survey with the forty-eight inch Schmidt telescope.\_XXXVII.~Proper 
   motions for 4483 faint stars., by Luyten, W.~J..~ Separate print 
   Univ.~Minnesota, Minneapolis, Minnesota, 44 p., 37, 1
\bibitem[Luyten(1974)]{luyten1974a} Luyten, W.~J.\ 1974, Proper 
   Motion Survey with the forty-eight inch Schmidt telescope.~XXXVI.~Proper 
   motions for 6955 faint stars., by Luyten, W.~J..~ Separate print 
   Univ.~Minnesota, Minneapolis, Minnesota, 64 p., 36, 1
\bibitem[Luyten(1972)]{luyten1972} Luyten, W.~J.\ 1972, Proper 
   Motion Survey with the forty-eight inch Schmidt telescope.~XXXI.~Proper 
   motions for 2520 faint stars., by Luyten, W.~J..~ Separate print 
   Univ.~Minnesota, Minneapolis, Minnesota, 24 p., 31, 1
\bibitem[Mainzer et al.(2011)]{mainzer2011} Mainzer, A., et al.\ 
   2011, \apj, 726, 30
\bibitem[Malmquist (1920)]{malmquist1920} Malmquist, K. G. 1920, Medd. Lund Astron. Obs. Ser., 2., No. 22
\bibitem[Marocco et al.(2010)]{marocco2010} Marocco, F., et al.\ 2010, \aap, 524, A38
\bibitem[Marois et al.(2008)]{marois2008} Marois, C., Macintosh, 
   B., Barman, T., Zuckerman, B., Song, I., Patience, J., Lafreni{\`e}re, D., 
   \& Doyon, R.\ 2008, Science, 322, 1348
\bibitem[Mart{\'{\i}}n et 
   al.(2010)]{martin2010} Mart{\'{\i}}n, E.~L., et al.\ 2010, \aap, 517, A53
\bibitem[Martini et al.(2004)]{martini2004} Martini, P., Persson, 
   S.~E., Murphy, D.~C., Birk, C., Shectman, S.~A., Gunnels, S.~M., 
   \& Koch, E.\ 2004, \procspie, 5492, 1653
\bibitem[Metchev et al.(2008)]{metchev2008} Metchev, S.~A., 
   Kirkpatrick, J.~D., Berriman, G.~B., \& Looper, D.\ 2008, \apj, 676, 1281
\bibitem[McLean et al.(2003)]{mclean2003} McLean, I.~S., McGovern, 
   M.~R., Burgasser, A.~J., Kirkpatrick, J.~D., Prato, L., 
   \& Kim, S.~S.\ 2003, \apj, 596, 561
\bibitem[McLean et al.(2000)]{mclean2000} McLean, I.~S., Graham, 
   J.~R., Becklin, E.~E., Figer, D.~F., Larkin, J.~E., Levenson, N.~A., 
   \& Teplitz, H.~I.\ 2000, \procspie, 4008, 1048
\bibitem[McLean et al.(1998)]{mclean1998} McLean, I.~S., et al.\ 
   1998, \procspie, 3354, 566
\bibitem[McLean et al.(1994)]{mclean1994} McLean, I.~S., et al.\ 
   1994, \procspie, 2198, 457
\bibitem[Milligan et al.(1996)]{milligan1996} Milligan, S., Cranton, 
   B.~W., \& Skrutskie, M.~F.\ 1996, \procspie, 2863, 2
\bibitem[Monet et al.(2003)]{monet2003} Monet, D.~G., et al.\ 
   2003, \aj, 125, 984
\bibitem[Muench et al.(2002)]{muench2002} Muench, A.~A., Lada, E.~A., 
   Lada, C.~J., \& Alves, J.\ 2002, \apj, 573, 366
\bibitem[Noll et al.(2000)]{noll2000} Noll, K.~S., Geballe, 
   T.~R., Leggett, S.~K., \& Marley, M.~S.\ 2000, \apjl, 541, L75
\bibitem[Oke et al.(1995)]{oke1995} Oke, J. B., et al.\ 1995, PASP, 107, 
   375
\bibitem[Padoan et al.(2005)]{padoan2005} Padoan, P., Kritsuk, A., 
   Michael, Norman, L., \& Nordlund, {\AA}.\ 2005, \memsai, 76, 187
\bibitem[Patten et al.(2006)]{patten2006} Patten, B.~M., et al.\ 
   2006, \apj, 651, 502
\bibitem[Perryman et al.(1997)]{perryman1997} Perryman, M.~A.~C., et al.\ 1997, \aap, 323, L49
\bibitem[Phan-Bao et al.(2008)]{phan-bao2008} Phan-Bao, N., et al.\ 
   2008, \mnras, 383, 831
\bibitem[Pinfield et al.(2008)]{pinfield2008} Pinfield, D.~J., et 
   al.\ 2008, \mnras, 390, 304
\bibitem[Rayner et al.(2003)]{rayner2003} Rayner, J.~T., Toomey, 
   D.~W., Onaka, P.~M., Denault, A.~J., Stahlberger, W.~E., Vacca, W.~D., 
   Cushing, M.~C., \& Wang, S.\ 2003, \pasp, 115, 362
\bibitem[Reid et al.(2008)]{reid2008} Reid, I.~N., Cruz, K.~L., 
   Kirkpatrick, J.~D., Allen, P.~R., Mungall, F., Liebert, J., Lowrance, P., 
   \& Sweet, A.\ 2008, \aj, 136, 1290
\bibitem[Reid et al.(2002)]{reid2002} Reid, I.~N., Kirkpatrick, 
   J.~D., Liebert, J., Gizis, J.~E., Dahn, C.~C., 
   \& Monet, D.~G.\ 2002, \aj, 124, 519 
\bibitem[Reid et al.(2001)]{reid2001} Reid, I.~N., Gizis, J.~E., 
   Kirkpatrick, J.~D., \& Koerner, D.~W.\ 2001, \aj, 121, 489
\bibitem[Reid et al.(2000)]{reid2000} Reid, I.~N., Kirkpatrick, 
   J.~D., Gizis, J.~E., Dahn, C.~C., Monet, D.~G., Williams, R.~J., Liebert, 
   J., \& Burgasser, A.~J.\ 2000, \aj, 119, 369
\bibitem[Reipurth \& Clarke(2001)]{reipurth2001} Reipurth, B., \& Clarke, 
   C.\ 2001, \aj, 122, 432
\bibitem[Reyl{\'e} et al.(2010)]{reyle2010} Reyl{\'e}, C., et al.\ 2010, \aap, 522, A112
\bibitem[Rodriguez et al.(2011)]{rodriguez2011} Rodriguez, D.~R., 
   Zuckerman, B., Melis, C., \& Song, I.\ 2011, \apjl, 732, L29
\bibitem[Ross(1928)]{ross1928} Ross, F.~E.\ 1928, \aj, 38, 117
\bibitem[Ruiz et al.(1997)]{ruiz1997} Ruiz, M.~T., Leggett, 
   S.~K., \& Allard, F.\ 1997, \apjl, 491, L107
\bibitem[Ruiz et al.(1991)]{ruiz1991} Ruiz, M.~T., Takamiya, 
   M.~Y., \& Roth, M.\ 1991, \apjl, 367, L59
\bibitem[Schilbach et al.(2009)]{schilbach2009} Schilbach, E., 
   R{\"o}ser, S., \& Scholz, R.-D.\ 2009, \aap, 493, L27
\bibitem[Schmidt(1968)]{schmidt1968} Schmidt, M.\ 1968, \apj, 151, 
   393
\bibitem[Schmidt et al.(2010)]{schmidt2010} Schmidt, S.~J., West, 
   A.~A., Burgasser, A.~J., Bochanski, J.~J., 
   \& Hawley, S.~L.\ 2010, \aj, 139, 1045
\bibitem[Schneider et al.(2002)]{schneider2002} Schneider, D.~P., et 
   al.\ 2002, \aj, 123, 458
\bibitem[Scholz(2010)]{scholz2010} Scholz, R.-D.\ 2010, \aap, 510, L8
\bibitem[Scholz(2010)]{scholz2010b} Scholz, R.-D.\ 2010, \aap, 515, 
   A92
\bibitem[Scholz et al.(2003)]{scholz2003} Scholz, R.-D., McCaughrean, 
   M.~J., Lodieu, N., \& Kuhlbrodt, B.\ 2003, \aap, 398, L29
\bibitem[Sheppard \& Cushing(2009)]{sheppard2009} Sheppard, S.~S., 
   \& Cushing, M.~C.\ 2009, \aj, 137, 304 
\bibitem[Simcoe et al.(2010)]{simcoe2010} Simcoe, R.~A., et al.\ 
   2010, \procspie, 7735,
\bibitem[Simcoe et al.(2008)]{simcoe2008} Simcoe, R.~A., et al.\ 
   2008, \procspie, 7014,
\bibitem[Skrutskie et al.(2006)]{skrutskie2006} Skrutskie, M.~F., et 
   al.\ 2006, \aj, 131, 1163
\bibitem[Smith et al.(2007)]{smith2007} Smith, J.~D.~T., et al.\ 
   2007, \pasp, 119, 1133
\bibitem[Stephens \& Leggett(2004)]{stephens2004} Stephens, D.~C., 
   \& Leggett, S.~K.\ 2004, \pasp, 116, 9
\bibitem[Stern et al.(2007)]{stern2007} Stern, D., et al.\ 2007, 
   \apj, 663, 677
\bibitem[Strauss et al.(1999)]{strauss1999} Strauss, M.~A., et al.\ 
   1999, \apjl, 522, L61
\bibitem[Sutherland \& Hodgman(1974)]{sutherland1974} Sutherland, 
   I.~E. \& Hodgman, G.~.W., 1974, Communications of the Association 
   for Computing Machinery, 17
\bibitem[Swaters et al.(2009)]{swaters2009} Swaters, R.~A., Valdes, F., 
   \& Dickinson, M.~E.\ 2009, Astronomical Data Analysis Software and Systems XVIII, 411, 506
\bibitem[Teerikorpi(2004)]{teerikorpi2004} Teerikorpi, P.\ 2004, \aap, 424, 73
\bibitem[Tinney et al.(2005)]{tinney2005} Tinney, C.~G., 
   Burgasser, A.~J., Kirkpatrick, J.~D.,  \& McElwain, M.~W.\ 2005, \aj, 130, 2326
\bibitem[Tinney et al.(2003)]{tinney2003} Tinney, C.~G., 
   Burgasser, A.~J., \& Kirkpatrick, J.~D.\ 2003, \aj, 126, 975
\bibitem[Tinney(1993)]{tinney1993} Tinney, C.~G.\ 1993, \aj, 105, 
   1169
\bibitem[Tody(1986)]{tody1986} Tody, D.\ 1986, \procspie, 627, 
   733
\bibitem[Tokunaga et al.(2002)]{tokunaga2002} Tokunaga, A.~T., 
   Simons, D.~A., \& Vacca, W.~D.\ 2002, \pasp, 114, 180
\bibitem[Tsvetanov et al.(2000)]{tsvetanov2000} Tsvetanov, Z.~I., et 
   al.\ 2000, \apjl, 531, L61
\bibitem[Vacca et al.(2003)]{vacca2003} Vacca, W.~D., Cushing, 
   M.~C., \& Rayner, J.~T.\ 2003, \pasp, 115, 389
\bibitem[van Altena et al.(2001)]{vanaltena2001} van Altena, W.~F., 
   Lee, J.~T., \& Hoffleit, E.~D.\ 2001, VizieR Online Data Catalog, 1238, 0
\bibitem[van Altena et al.(1995)]{vanaltena1995} van Altena, W.~F., 
   Lee, J.~T., \& Hoffleit, D.\ 1995, VizieR Online Data Catalog, 1174, 0
\bibitem[van Dam et al.(2006)]{vandam2006} van Dam, M.~A., et al.\ 
   2006, \pasp, 118, 310
\bibitem[van 
   Leeuwen(2007)]{vanleeuwen2007} van Leeuwen, F.\ 2007, \aap, 474, 653
\bibitem[Vrba et al.(2004)]{vrba2004} Vrba, F.~J., et al.\ 2004, 
   \aj, 127, 2948
\bibitem[Warren et al.(2007)]{warren2007} Warren, S.~J., et al.\ 
   2007, \mnras, 381, 1400
\bibitem[Werner et al.(2004)]{werner2004} Werner, M.~W., et al.\ 
   2004, \apjs, 154, 1
\bibitem[West et al.(2008)]{west2008} West, A.~A., Hawley, 
   S.~L., Bochanski, J.~J., Covey, K.~R., Reid, I.~N., Dhital, S., Hilton, 
   E.~J., \& Masuda, M.\ 2008, \aj, 135, 785
\bibitem[Whitworth \& Zinnecker(2004)]{whitworth2004} Whitworth, A.~P., 
   \& Zinnecker, H.\ 2004, \aap, 427, 299
\bibitem[Wilson et al.(2003)]{wilson2003} Wilson, J.~C., et al.\ 
   2003, \procspie, 4841, 451
\bibitem[Wilson et al.(2003)]{wilson2003b} Wilson, J.~C., Miller, 
   N.~A., Gizis, J.~E., Skrutskie, M.~F., Houck, J.~R., Kirkpatrick, J.~D., 
   Burgasser, A.~J., \& Monet, D.~G.\ 2003, Brown Dwarfs, 211, 197
\bibitem[Wilson et al.(2001)]{wilson2001} Wilson, J.~C., 
   Kirkpatrick, J.~D., Gizis, J.~E., Skrutskie, M.~F., Monet, D.~G., 
   \& Houck, J.~R.\ 2001, \aj, 122, 1989
\bibitem[Wizinowich et al.(2006)]{wizinowich2006} Wizinowich, P.~L., 
   et al.\ 2006, \pasp, 118, 297
\bibitem[Wright et al.(2010)]{wright2010} Wright, E.~L., et al.\ 
   2010, \aj, 140, 1868
\bibitem[Wright et al.(2011)]{wright2011} Wright, E.~L., Mainzer, A.,
   Gelino, C., \& Kirkpatrick, D.\ 
   2010, arXiv:1104.2569
\bibitem[Wroblewski \& Torres(1991)]{wroblewski1991} Wroblewski, H., 
   \& Torres, C.\ 1991, \aaps, 91, 129
\bibitem[York et al.(2000)]{york2000} York, D.~G., et al.\ 2000, 
   \aj, 120, 1579
\bibitem[Zhang et al.(2009)]{zhang2009} Zhang, Z.~H., et al.\ 2009, \aap, 497, 619
\end{thebibliography}
\end{document}